\documentclass[aps,prd,preprint,tightenlines,eqsecnum,showpacs,nofootinbib]{revtex4}
\usepackage{epsfig}

\def\Journal#1#2#3#4{{#1} {\bf #2}, #3 (#4)}

\def\NPB{{\em Nucl. Phys.} B}
\def\PLB{{\em Phys. Lett.}  B}
\def\PRL{\em Phys. Rev. Lett.}
\def\PR{\em Phys. Rev.}
\def\PRD{{\em Phys. Rev.} D}
\def\ZPC{{\em Z. Phys.} C}
\def\FP{\em Fortsch. Phys.}

\def\JMP{\em J. Math. Phys.}
\def\JCP{\em J. Comp. Phys.}
\def\PREP{\em Phys. Rep.}
\def\lsim{\mathrel{\rlap{\lower4pt\hbox{\hskip1pt$\sim$}}
    \raise1pt\hbox{$<$}}}                
\def\gsim{\mathrel{\rlap{\lower4pt\hbox{\hskip1pt$\sim$}}
    \raise1pt\hbox{$>$}}}                

\def\a{\alpha}
\def\b{\beta}

\def\d{\delta}
\def\e{\epsilon}

\def\g{\gamma}

\def\j{\psi}

\def\l{\lambda}
\def\m{\mu}
\def\n{\nu}

\def\p{\pi}
\def\q{\theta}
\def\r{\rho}
\def\s{\sigma}

\def\F{\Phi}
\def\G{\Gamma}
\def\J{\Psi}
\def\L{\Lambda}
\def\O{\Omega}

\def\ve{\varepsilon}

\def\co{{\cal O}}

\def\bo{{\raise.15ex\hbox{\large$\Box$}}}         
\def\sumb{\overline{\sum}}                        
\def\dg{\sp\dagger}                               

\def\slash#1{\rlap{\hbox{$\mskip 1 mu /$}}#1}      
\def\leftrightarrowfill{$\mathsurround=0pt \mathord\leftarrow \mkern-6mu
        \cleaders\hbox{$\mkern-2mu \mathord- \mkern-2mu$}\hfill
        \mkern-6mu \mathord\rightarrow$}       
\def\dvec#1{\vbox{\ialign{##\crcr
        \leftrightarrowfill\crcr\noalign{\kern-1pt\nointerlineskip}
        $\hfil\displaystyle{#1}\hfil$\crcr}}}           
\def\dilog#1{\,{\rm Li_2}#1}                        

\def\be{\begin{equation}}
\def\ee{\end{equation}}

\def\bex{\begin{displaymath}}
\def\eex{\end{displaymath}}

\def\bea{\begin{eqnarray}}
\def\eea{\end{eqnarray}}
\def\NO{\nonumber}

\def\beax{\begin{eqnarray*}}
\def\eeax{\end{eqnarray*}}

\begin{document}

\preprint{ANL-HEP-PR-00-044}
\preprint{RMU-SEMS-020307}
\preprint{FSU-HEP-010130}

\title{THE TWO CUTOFF PHASE SPACE SLICING METHOD}

\author{B. W. Harris}
\email{harris@rmu.edu}
\affiliation{High Energy Physics Division \\
Argonne National Laboratory \\
Argonne, Illinois 60439, USA 
\\ and \\
Robert Morris University \\
Coraopolis, Pennsylvania 15108, USA}

\author{J. F. Owens}
\email{owens@hep.fsu.edu}
\affiliation{Physics Department \\
Florida State University \\
Tallahassee, Florida 32306-4350, USA}

\date{\today}

\begin{abstract}
The phase space slicing method of two cutoffs for
next-to-leading-order Monte-Carlo style QCD corrections has been
applied to many physics processes.  The method is intuitive, simple to
implement, and relies on a minimum of process dependent information.
Although results for specific applications exist in the literature,
there is not a full and detailed description of the method.  Herein
such a description is provided, along with illustrative examples;
details, which have not previously been published, are included
so that the method may be applied to additional hard scattering
processes.
\end{abstract}


\maketitle


\section{Introduction}

Perturbative quantum chromodynamic (QCD) calculations are essential in
the effort to describe large momentum transfer hadronic scattering
processes.  At one time it was sufficient to work at lowest order for
the hard scattering subprocesses and utilize the leading-logarithm
approximation to treat the higher order gluon radiation and
quark-antiquark pair production which give rise to the scale
dependence of the parton distribution and fragmentation functions, and
to the running of the strong coupling $\a_s$.  As the experimental
systematic and statistical errors decreased, the need for increased
precision for the theoretical calculations became apparent, leading to
the widespread use of next-to-leading-order expressions for the hard
scattering subprocesses with the remaining higher order terms being
treated in the next-to-leading-logarithm approximation.  Early
calculations of this type were typically performed with a combination
of analytic and numerical integration techniques.  The phase space
integrations at the parton level were often performed analytically,
and the convolutions with the parton distribution or fragmentation
functions done numerically.  This approach is satisfactory for fully
or singly inclusive cross sections, but information is lost about
quantities over which the integrations have been performed.  Thus, if
cuts are to be placed on two or more partons (or hadrons or jets), the
calculation must be started anew.  Furthermore, for some observables
it is difficult to calculate the appropriate Jacobian for the
transformation from partonic to hadronic variables.  For these reasons
it was recognized that Monte Carlo techniques would be useful for such
calculations.  The Jacobians would be handled by the choice of
histogramming variables and several observables could be histogrammed
simultaneously.  Additionally, it would be simple to define jets and
to implement experimental cuts on the four-vectors of the produced
partons.

In light of the above observations, a method for performing
next-to-leading-logarithm calculations using Monte Carlo techniques
was developed \cite{bergmann}.  Two cutoff parameters
serve to separate the regions of phase space containing the soft
and collinear singularities from the non-singular regions; nowadays 
this is referred to as the phase-space slicing technique.

The usefulness and generality of the method may be appreciated by considering 
the many physics processes to which it has been applied.  
The basic core of the method was first developed to study QCD corrections 
to dihadron production \cite{bergmann}.  It has subsequently been applied to  
direct jet photoproduction \cite{boo1}, hadronic photon--jet \cite{boo2},
direct photon \cite{boo3}, W \cite{breno}, ZZ \cite{oo}, 
WW \cite{ohnemus1}, WZ \cite{ohnemus2}, 
two photon \cite{boo_2photon,cg1}, Z$\g$ \cite{ohnemus3}, and  
W--Higgs \cite{os_whiggs,bbo_whiggs} production,
nonstandard three vector boson couplings in W$\g$ \cite{bho1}, 
WZ \cite{bho2} and WW \cite{bho3} production,
hadronic photon--heavy quark production \cite{bbg1,bbg2}, 
jet photoproduction \cite{ho},
quantum electrodynamic (QED) corrections to hadronic Z production \cite{bks},
QCD corrections to slepton pair production \cite{bhr},
electroweak corrections to W production \cite{bkw},  
single-top-quark production \cite{harris}, and 
dihadron production \cite{Owens:2001rr}.

Despite this usefulness, a full and detailed description of the method
does not exist in the literature.  Here we provide such a description.
Naturally, as the method was applied to the above physics processes,
refinements were made.  We therefore take this opportunity to
modernize and systematize the presentation relative to that given in
\cite{bergmann}, and show details, which have not previously been published.
Searches for signals of new physics often rely on next-to-leading-order Monte 
Carlo-based calculations. It is anticipated that the details provided here 
will prove to be helpful for anyone wanting to apply the method to 
additional processes. 

In the course of a next-to-leading-order calculation 
ultra-violet singularities show up in loop integrals 
where the momenta go to infinity.  They are removed through the 
process of renormalization.  (See, for example, Refs.\ \cite{smith,sterman} 
for a discussion.)$\;$  Soft (infrared) divergences arise 
if the theory includes a massless field like the photon in QED or 
the gluon in QCD.  They are encountered in both loop and phase space 
integrals and are found in the low energy region where the integration 
momenta go to zero.
The soft singularities cancel between the virtual and bremsstrahlung 
processes \cite{kln}.
If the massless field couples to another massless field, or 
to itself, a collinear (mass) singularity may occur in both loop and 
phase space integrals.  Final state mass singularities cancel when summed  
over degenerate (experimentally indistinguishable) final states 
according to the theorem of Kinoshita-Lee-Nauenberg \cite{kln}.
For tagged hadrons there is no final state sum, and the associated mass 
singularities are factorized into fragmentation functions.
Similarly, initial state singularities do not cancel because there is 
typically no sum over degenerate states; they are removed by 
factorization \cite{css,bodwin}.

The goal of the practitioner of next-to-leading-order calculations 
is to organize the
soft and collinear singularity cancellations described above 
without loss of information in terms of observable quantities.  
The phase space slicing method 
provides a relatively simple and robust method to do this.
Several other methods for handling the organization of the cancellations exist 
in the literature and have been used to study a wide variety of high energy 
processes, including some of those listed above. 
The phase space slicing method of one cutoff first developed in 
\cite{ycut1,ycut2} divides the phase space according to 
$s_{ij} =(p_i+p_j)^2 > ys_{12}$ where $p_i$ and $p_j$ label the
momenta of partons $i$ and $j$, and $y$ is a small dimensionless
parameter.  Another variant for jets \cite{gg,ggk} and hadrons and
heavy quarks \cite{kl} partitions phase space according to 
$s_{ij} >s_{\rm min}$ where $s_{\rm min}$ is a small dimension-full parameter.
It is also possible to engineer the singularity cancellation using
plus distributions, commonly referred to as the subtraction method,
which has been applied to jets 
\cite{ert,kn,ks,eks1,eks2,Frixione:1995ms,Frixione:1997np,Nagy:1996bz} 
and heavy quark final states \cite{mnr,fmnr,hs,ho2}. 
The subtraction method taken together
with factorization formulae that interpolate between the soft and
collinear approximations to the matrix elements is known as 
the dipole method \cite{dipole}.  The dipole method was 
originally developed for jet and light hadron cross sections 
where it has seen extensive application 
\cite{Nagy:1997yn,Weinzierl:1999yf,Nagy:2001xb,Campbell:1999ah,Ellis:1998fv,
Beenakker:1999xh,Maina:1996ep,Catani:1999nf,Kramer:1999bf}.
It has recently been extended to handle massive fermions and partons 
\cite{ditt,Phaf:2001gc,Catani:2002hc} and is finding many applications
\cite{harris,Denner:2000bj,Dittmaier:2001ay,Dittmaier:2000tc} in 
that domain as well. 
A brief comparison of the slicing and
subtraction methods is given in Appendix A\@.  Properly implemented, all
methods should give identical physics predictions.

This paper proceeds as follows.  
In Section II we give details of the phase space slicing method with two 
cutoffs.  We examine the soft and collinear regions of phase 
space and see how to arrive at a finite cross section.  
In Section III the process of electron-positron annihilation 
into quarks is studied for massive, massless, and tagged final 
states.  After that, the examples of lepton pair production and 
single particle production in hadronic 
collisions are given.  We conclude in Section IV.  
As mentioned above, Appendix A contains a brief comparison with the 
subtraction method.
Appendix B contains angular integrals useful in the soft analysis.
A discussion of terms that vanish like the ratio of the two cutoffs is 
given in Appendix C.
Appendix D explains how to improve numerical convergence.

\section{The Method}

This section contains the main derivations for jet, fragmentation, and 
heavy quark final states, as well as a discussion of initial state mass 
factorization.  Before getting into too many details, it will be helpful to 
outline the procedure first. The typical calculation involves lowest order 
two-to-two subprocesses which have two-body final states and higher order 
two-to-three subprocesses which lead to both two- and three-body 
final states.  In addition, the one-loop virtual 
corrections also contribute to the two-body final states.

We begin by decomposing the three-body phase space used to calculate the 
two-to-three contribution to the partonic cross section into two regions 
which we call soft, S, and hard, H, by writing
\footnote{Implicit in the cross section is a 
measurement function which serves to implement the jet algorithm and/or
define the experimentally visible portion of phase space (the cuts).
For the cancellation of singularities to take place, the measurement
function is required to be infrared-safe\cite{sw}.}
\be
\s = \frac{1}{2\F} \int \sumb |M_3|^2 d\G_3 
   = \frac{1}{2\F} \int_{\rm S} \sumb |M_3|^2 d\G_3 
   + \frac{1}{2\F} \int_{\rm H} \sumb |M_3|^2 d\G_3 \, ,
\ee
where $\F=\l^{1/2}(s,m_1^2,m_2^2)$ is the usual flux factor which depends
on the partonic center-of-momentum energy squared $s$ and the 
incident particle masses 
$m_1$ and $m_2$, $\sumb |M_3|^2$ is the two-to-three body squared 
matrix element averaged (summed)
over initial (final) degrees of freedom, and $d\Gamma_3$ 
is the three-body phase space. The partitioning of phase space into S and H 
depends on a parameter $\d_s$ in a manner to be described below. 
Within S the double pole (eikonal) approximation to the matrix 
elements is made and then analytically integrated over the unobserved 
degrees of freedom in $n$ space-time dimensions.  The result, 
depending on the masses of the partons involved, may 
contain double and/or single poles in $n-4$, and accompanying 
double and/or single logarithms in the soft cutoff $\d_s$.  
We always work in the approximation where the cutoffs are small, so 
terms of order $\d_s$ may be neglected.  Just how small is needed will 
be studied below.

Next, if there are collinear singularities present,
the hard region is further decomposed into collinear 
C, and non-collinear ${\rm \overline{C}}$, regions as follows:
\be
\frac{1}{2\F} \int_{\rm H} \sumb |M_3|^2 d\G_3 = 
\frac{1}{2\F} \int_{\rm HC} \sumb |M_3|^2 d\G_3 + 
\frac{1}{2\F} \int_{\rm H\overline{C}} \sumb |M_3|^2 d\G_3 \, . 
\ee
This partitioning depends on a second cutoff $\d_c$.  Within HC the
leading collinear pole approximation to the squared matrix element
is made.  As explained below, exact collinear kinematics may be used to
define the integration domain of HC when $\d_c \ll \d_s$.  The
integrations over the unobserved degrees of freedom are performed
analytically in $n$ space-time dimensions giving a factorized result
where single poles in $n-4$, and single logarithms in both cutoffs
$\d_c$ and $\d_s$, multiply splitting functions and lower-order
squared matrix elements.  This is in the approximation where terms of
order $\d_c$ and $\d_s$ are neglected.

The cancellation of poles in $n-4$ is based on the
Kinoshita-Lee-Nauenberg theorem \cite{kln}, or mass factorization
\cite{css,bodwin}, depending on the situation.  For experimentally
degenerate final states, the soft and final state hard collinear
singularities cancel upon addition of the interference of the leading
order diagrams with the renormalized one-loop virtual diagrams.  The
remaining initial state collinear singularities are factorized and
absorbed into the parton distribution functions. The result is finite
in $n=4$ dimensions, but depends logarithmically on the cutoffs.  For
tagged hadrons there is no final state sum, and the associated mass
singularities are factorized into fragmentation functions.

The integration over the hard non-collinear ${\rm H\overline{C}}$ 
portion of the phase space is performed using standard Monte Carlo 
techniques.  The result is finite by construction and the expressions may be 
evaluated in four dimensions. At this stage, the calculation yields a set of 
two-body weights which have explicit logarithmic dependence on the two 
cutoffs and the three-body weights for which a logarithmic dependence 
on the cutoffs develops as the Monte Carlo integration is performed. 
When all of the contributions are combined at the histogramming stage,  
the cutoff dependence cancels for suitably defined infrared-safe 
observables.
In the following subsections we look at each of these steps in detail.

\subsection{Soft}
\label{sec:soft}
In this subsection we describe the procedure for extracting soft gluon
singularities.  When one of the gluons is soft, the phase space is
greatly simplified and the eikonal (double pole) approximation of the 
matrix element is valid \cite{gy,cs,bcm,fmr,cg}.  The cross section
is simple enough to be analytically integrated over the
unobserved degrees of freedom in $n$ space-time dimensions.  The
required integrals are well known \cite{hs,been,willy}.  The result,
depending on the masses of the partons involved, contains double
and/or single poles in $n-4$, and accompanying double and/or single
logarithms in the soft cutoff $\d_s$.  We work in the
approximation where terms of order $\d_s$ are neglected.

Generically, we begin by writing the two-to-three body contribution 
to the partonic cross section
\be
\s = \frac{1}{2\F} \int \sumb |M_3|^2 d\G_3 \, ,
\ee
as the sum of soft, ${\rm S}$, and hard, ${\rm H}$, terms
\be
\s = \s_{\rm S} + \s_{\rm H} \, ,
\label{eqn:softhard}
\ee
where
\be
\s_{\rm S} = \frac{1}{2\F} \int_{\rm S} \sumb |M_3|^2 d\G_3 \, ,
\label{eqn:soft}
\ee
and
\be
\s_{\rm H} = \frac{1}{2\F} \int_{\rm H} \sumb |M_3|^2 d\G_3 \, .
\label{eqn:hard}
\ee
In this section we examine $\s_{\rm S}$ in detail.  Further 
evaluation of $\s_{\rm H}$ is deferred to the following subsections.

Let the particles in the scattering be labeled by their four-momenta 
$p_1+p_2 = p_3+p_4+p_5$ and define the Mandelstam invariants 
$s_{ij}=(p_i+p_j)^2$ and $t_{ij}=(p_i-p_j)^2$.
Consider the case when parton 5 is a soft gluon.  
The soft region ${\rm S}$ is defined in terms of the gluon energy $E_5$ in the 
$p_1+p_2$ rest frame by $0 \leq E_5 \leq \d_s \sqrt{s_{12}}/2$.  
The hard region ${\rm H}$ is the complement: $E_5 > \d_s \sqrt{s_{12}}/2$.
The gluon energy can be calculated from the other 
invariants in the problem as follows. Start with 
$p_1+p_2-p_5 = p_3+p_4$ which, after squaring both sides, yields  
$(p_1+p_2)^2 -2p_5 \cdot (p_1+p_2) = (p_3+p_4)^2$.
In the $p_1+p_2$ rest frame $p_1+p_2=\sqrt{s_{12}}(1,0,0,0)$, so 
$s_{12}-2E_5\sqrt{s_{12}}=s_{34}$.
Solving for the gluon energy gives
\be
E_5=\frac{s_{12}-s_{34}}{2\sqrt{s_{12}}} \, .
\label{eqn:e5}
\ee
This expression for $E_5$ and the definition of the soft region are 
independent of the masses of the other particles in the reaction.  
Now that we have defined the boundaries of the soft portion of phase 
space, we examine the approximations that can be made.

The three-body phase space in $n$ dimensions is given by
\be
d\G_3 = \frac{d^{n-1}p_3}{2p^0_3(2\p)^{n-1}}
        \frac{d^{n-1}p_4}{2p^0_4(2\p)^{n-1}}
        \frac{d^{n-1}p_5}{2p^0_5(2\p)^{n-1}}
        (2\p)^n \d^n(p_1+p_2-p_3-p_4-p_5) \, .
\label{eqn:ps3}
\ee
The divergence in the integral of the matrix element over phase space 
will be at worst logarithmic.  Therefore, up to corrections 
of $\co(\d_s)$, we can set $p_5^\m=0$ in the delta 
function and regroup terms, yielding
\be
d\G_3|_{\rm soft} = \left[ \frac{d^{n-1}p_3}{2p^0_3(2\p)^{n-1}}
        \frac{d^{n-1}p_4}{2p^0_4(2\p)^{n-1}}
        (2\p)^n \d^n(p_1+p_2-p_3-p_4) \right]
        \frac{d^{n-1}p_5}{2p^0_5(2\p)^{n-1}} \, .
\ee
This is simply
\be
d\G_3|_{\rm soft} = d\G_2 \frac{d^{n-1}p_5}{2p^0_5(2\p)^{n-1}} \, ,
\ee
where
\be
d\G_2= \frac{d^{n-1}p_3}{2p^0_3(2\p)^{n-1}}
        \frac{d^{n-1}p_4}{2p^0_4(2\p)^{n-1}}
        (2\p)^n \d^n(p_1+p_2-p_3-p_4) \, ,
\label{eqn:ps2}
\ee
is the two-body phase space of partons 3 and 4.
Likewise, up to corrections of $\co(\d_s)$, we can parameterize 
the gluon's $n-$momentum in the $p_1+p_2$ rest frame as
\be
p_5=E_5(1, \ldots, \sin\q_1 \sin\q_2, \sin\q_1 \cos\q_2, \cos\q_1) \, ,
\label{eqn:p5_soft}
\ee
where the dots indicate the $n-4$ unspecified momentum components which may be 
trivially integrated over using
\bea
d^{n-1}p_5 &=& d|\vec{p_5}| |\vec{p_5}|^{n-2} d\O_{n-2} \NO \\
           &=& dE_5 E_5^{n-2} \sin^{n-3}\!\q_1\,d\q_1\sin^{n-4}\!\q_2\,d\q_2 
               \O_{n-4} \, .
\eea
The angular volume element
\be
\O_{n-4} = \frac{ 2 \p^{(n-3)/2} }{ \G[(n-3)/2] } \, ,
\ee
may be rewritten using 
\be
\G[(n-3)/2] = \sqrt\p 2^{2\e} \frac{\G(1-2\e)}{\G(1-\e)} \, ,
\ee
where we have set $n=4-2\e$.  The final result for the phase space volume 
approximated in the soft region is
\be
d\G_3|_{\rm soft} = d\G_2 \left[ \left( \frac{4\p}{s_{12}} \right)^\e
         \frac{\G(1-\e)}{\G(1-2\e)} \frac{1}{2(2\p)^2} \right] dS \, ,
\label{eqn:soft_PS3}
\ee
with
\be
dS =  \frac{1}{\p} 
     \left( \frac{4}{s_{12}} \right)^{-\e} \int_0^{\d_s\sqrt{s_{12}}/2} 
     dE_5 E_5^{1-2\e} \sin^{1-2\e}\!\q_1\,d\q_1\sin^{-2\e}\!\q_2\,d\q_2 \, .
\label{eqn:soft_PS}
\ee
Once we have the corresponding soft approximation to the matrix element
this integral can be performed, yielding the advertised singularity
structure.  But before proceeding, we pause to note that choosing 5 to
be the soft gluon is not special.  A similar analysis holds when
parton 3 or 4 is a gluon.  At this order in perturbation theory, only
one gluon may be soft at a time.

Soft photon emission in QED is characterized by the factorization of
an eikonal current from the scattering amplitude.  The structure of
multiple soft photon emission has been studied by Grammer and Yennie
\cite{gy}.  In QCD the process is different because gluons carry color
charge and can therefore radiate during the scattering.  Fortunately, when 
QCD amplitudes are decomposed into color sub-amplitudes they enjoy
the same factorization properties as QED amplitudes \cite{cs,bcm,fmr,cg}.

Let parton 5 be the soft gluon and take it to have 
color index $a$ ($=1, \ldots, N^2-1$) and 
Lorentz index $\mu$.  The matrix element factorizes as 
\be
M^a_3|_{\rm soft} \simeq g \m_r^{\e} \ve^{\m}(p_5){\bf J}_{\m}^a(p_5) 
{\bf M}_2 \, .
\label{eqn:me_soft}
\ee
The mass dimensions of the strong coupling have been isolated into the 
parameter $\m_r$ which is identified with the renormalization scale, leaving 
the dimensionless coupling $g$.  The soft gluon's polarization vector, 
denoted by $\ve^{\m}(p_5)$, is Lorentz contracted with the non-abelian 
eikonal current given by
\be
{\bf J}^a_{\m}(p_5) = \sum_{f=1}^4 {\bf T}^a_f \frac{p_f^{\m}}{p_f 
\cdot p_5} \, ,
\ee
which itself is color contracted with the color sub-amplitude ${\bf M}_2$.  
The sum corresponds to the soft gluon being emitted from each 
external line in turn.  The $SU(N)$ color charge associated with the 
emitting parton $f$ is denoted by ${\bf T}_f$.
Squaring and summing Eq.\ (\ref{eqn:me_soft}) over the soft gluon 
polarizations gives
\be
|M_3|^2|_{\rm soft} \simeq -g^2 \m_r^{2\e} \sum_{f,{f^\prime}=1}^4 
\frac{p_f \cdot p_{f^\prime}}{p_f \cdot p_5 \; p_{f^\prime} 
\cdot p_5} M^0_{f{f^\prime}} \, ,
\label{eqn:me2_soft}
\ee
where
\be
M^0_{f{f^\prime}} = ( {\bf T}^a_f {\bf M}_2 ) ( {\bf T}^a_{f^\prime} 
    {\bf M}_2 )^{\dg} 
  = \left[ M_{c_1 \ldots b_f \ldots b_{f^\prime} \ldots c_4} \right]^* 
    T_{b_f d_f}^a T_{b_{f^\prime} d_{f^\prime}}^a 
    M_{c_1 \ldots d_f \ldots d_{f^\prime} \ldots c_4} \, ,
\label{eqn:me0_cc}
\ee
is the square of the color connected Born amplitude.  If the emitting 
parton is a final state quark or initial state anti-quark, the color 
charge is in the fundamental representation 
$(T^a)_{ij}=t^a_{ij}$ ($i,j = 1, \ldots, N$).  For a final state 
anti-quark or initial state quark $(T^a)_{ij}=-t^a_{ij}$.  
If the emitting parton is a gluon, the color charge is 
in the adjoint representation $(T^a)_{bc}=-if_{abc}$. 

Substituting (\ref{eqn:soft_PS3}) and (\ref{eqn:me2_soft}) into 
(\ref{eqn:soft}) gives
\be
d\s_{\rm S} =
         \left[ \frac{\a_s}{2\p} \frac{\G(1-\e)}{\G(1-2\e)} 
         \left( \frac{4\p\mu_r^2}{s_{12}} \right)^\e \right] 
         \sum_{f,{f^\prime}=1}^4 d\s_{f{f^\prime}}^{0} \int
         \frac{-p_f \cdot p_{f^\prime}}{p_f \cdot p_5 \; p_{f^\prime} 
         \cdot p_5} dS \, ,
\label{eqn:soft_final}
\ee
where 
\be
d\s^0_{f{f^\prime}} = \frac{1}{2\F} \sumb M^{0}_{f{f^\prime}} d\G_2 \, .
\ee
The integration over the eikonal factors depends on the 
masses of the particles in the reaction.  We leave the integrals to be 
performed on a case-by-case basis, although all possible 
mass combinations may be worked out.  Specific applications of 
Eq.\ (\ref{eqn:soft_final}) are given below.

From a practical point of view, 
the presentation just made requires only the calculation of the 
eikonal factors and the color connected Born amplitudes 
in $n=4-2\e$ dimensions.  There is no need to 
evaluate the full two-to-three body matrix element in $n$ dimensions.
Depending on the complexity of the process, this may be 
a simplification.
However, if the full two-to-three body matrix element in $n=4-2\e$ 
dimensions is at hand, setting $p_5^\m=0$ 
everywhere in the numerator and retaining only the leading singular 
terms as $p_5^\m \rightarrow 0$ will reproduce Eq.\ (\ref{eqn:me2_soft})
directly, this being known as the double pole approximation \cite{bergmann}.

\subsection{Collinear}
\label{sec:collinear}
We now return to the further evaluation of the hard portion of the 
cross section which was separated out in Sec.\ \ref{sec:soft}.  
The phase space is greatly simplified in the limit where two of the partons 
are collinear.  In the same limit, the leading pole approximation 
of the matrix element is valid. The cross section is simple 
enough to be analytically integrated over the unobserved 
degrees of freedom in $n$ space-time dimensions.
The result contains single poles in $n-4$, and accompanying 
logarithms of the soft $\d_s$ and collinear $\d_c$ 
cutoffs.  Terms of order $\d_s$ and $\d_c$ are neglected.

To this end 
we further decompose $\s_{\rm H}$ given in Eq.\ (\ref{eqn:hard}) 
into a sum of hard-collinear ${\rm HC}$ and hard-non-collinear 
${\rm H\overline{C}}$ terms
\be
\s_{\rm H} = \s_{\rm HC} + \s_{\rm H\overline{C}} \, ,
\ee
with
\be
\s_{\rm HC} = \frac{1}{2\F} \int_{\rm HC} \sumb |M_3|^2 d\G_3 \, ,
\label{eqn:sig_HC}
\ee
and
\be
\s_{\rm H\overline{C}} = \frac{1}{2\F} \int_{\rm H\overline{C}} \sumb 
|M_3|^2 d\G_3 \, .
\label{eqn:sig_HNC}
\ee
The ${\rm HC}$ regions of phase space are those where any invariant
($s_{ij}$ or $t_{ij}$) becomes smaller in magnitude than $\d_c s_{12}$, 
the collinear condition, while at the same time all gluons
remain hard.  The complementary ${\rm H\overline{C}}$ pieces are
finite and may be evaluated numerically in four dimensions using
standard Monte-Carlo techniques \cite{vegas}.

The piece containing the collinear singularities, $\s_{\rm HC}$, 
is treated according to whether the singularities are initial or 
final state in origin.  
For the former, factorization provides the formalism 
for removing the singularities.
In the latter case, we distinguish between experimentally 
degenerate and tagged final states, and rely on either the 
Kinoshita-Lee-Nauenberg theorem or factorization to dispose of the 
singularities.   We discuss the final state cases first, then 
return to the initial state.

\subsubsection{Indistinguishable Final States}
\label{sec:jets}

Consider the case when there is a sum over experimentally 
degenerate final states, such as a jet or total 
cross section.   Let partons 4 and 5 be massless and 
collinear to each other, $0 \le s_{45} \le \d_c s_{12}$.  
If we define $p_{45}=p_4+p_5$,
then for fixed $p_5$ we have $d^{n-1}p_{45}=d^{n-1}p_4$.
The three body phase space Eq.\ (\ref{eqn:ps3}) may be written as 
\be
d\G_3|_{\rm coll} = \left[ \frac{d^{n-1}p_3}{2p^0_3(2\p)^{n-1}}
        \frac{d^{n-1}p_{45}}{2p^0_{45}(2\p)^{n-1}}
        (2\p)^n \d^n(p_1+p_2-p_3-p_{45}) \right]
        \frac{d^{n-1}p_5}{2p^0_5(2\p)^{n-1}} \frac{p^0_{45}}{p^0_4} \, .
\ee
This is simply
\be
d\G_3|_{\rm coll} = d\G_2 \frac{d^{n-1}p_5}{2p^0_5(2\p)^{n-1}}
\frac{p^0_{45}}{p^0_4} \, ,
\label{eqn:collps}
\ee
where $d\G_2$ is the two-body phase space of the particles 3 and 45.
In the collinear limit ($p_t \rightarrow 0$ with $z$ fixed) we can write
\bea
p_{45} &=& (P,0,0,P) \NO \\
p_4 &\simeq& ( zP+\frac{p_t^2}{2zP},\vec{p}_t,zP) \NO \\
p_5 &\simeq& ( (1-z)P+\frac{p_t^2}{2(1-z)P},-\vec{p}_t,(1-z)P) \, .
\label{eqn:coll}
\eea
Then $p_4+p_5=p_{45}+\co(p_t^2)$ and
\be
s_{45} = 2 p_4 \cdot p_5 \simeq \frac{p_t^2}{z(1-z)} \, .
\ee
Now using $d^{n-2}p=dpp^{n-3}d\O_{n-3}$ and
$z(1-z)ds_{45} = dp_t^2$ we find
\be
\frac{d^{n-1}p_5}{2p^0_5(2\p)^{n-1}} \frac{p^0_{45}}{p^0_4} =
\frac{(4\p)^\e}{16\p^2\G(1-\e)} dz ds_{45} [s_{45}z(1-z)]^{-\e} \, .
\label{eqn:dp5}
\ee
The corresponding approximation to the matrix element 
is obtained by imposing collinear kinematics on the portion of the 
two-to-three matrix element proportional to the leading collinear 
singularity.  This is
known as the leading pole or collinear approximation.
As a consequence of the factorization theorems \cite{css,bodwin}, 
the squared matrix element factors into the product 
of a splitting kernel and a leading order squared matrix element 
\cite{bergmann,ert,ks,mnr,ap}.

As above, let partons 4 and 5 be collinear.  The matrix element 
factorizes as 
\be
\sumb |M_3(1+2 \rightarrow 3+4+5)|^2 \simeq \sumb 
|M_2(1+2 \rightarrow 3+4')|^2
P_{44'}(z,\epsilon) g^2 \mu_r^{2 \epsilon} \frac{2}{s_{45}} \, ,
\label{eqn:lpole}
\ee
where the $P_{ij}(z,\e)$ are the unregulated ($z<1$)
splitting functions calculated in $n=4-2\e$ dimensions related to the
usual Altarelli-Parisi splitting kernels \cite{ap}.  
We label as $4'$ the parton
which splits into the 45 collinear pair.  Generally, Eq.\ (\ref{eqn:lpole}) 
contains an additional term that vanishes after
integration over the azimuthal angles in $n$ dimensions \cite{ks,mnr}.
Such a term does not contribute to our result.

Substituting Eqs.\ (\ref{eqn:collps}), (\ref{eqn:dp5}), 
and (\ref{eqn:lpole}) into Eq.\ (\ref{eqn:sig_HC}) gives
\bea
d\s_{\rm HC}^{1+2\rightarrow 3+4+5} &=&  d\s_0^{1+2\rightarrow 3+4'} 
\left[ \frac{\a_s}{2\p} \frac{\G(1-\e)}{\G(1-2\e)} 
                 \left( \frac{4\p\mu_r^2}{s_{12}} \right)^\e \right] \NO \\
&\times& \int_0^{\d_c s_{12}} \frac{ds_{45}}{s_{45}} 
  \left( \frac{s_{45}}{s_{12}} \right) ^{-\e}
\int dz [z(1-z)]^{-\e}  P_{44'}(z,\e) \, ,
\label{eqn:dsHC}
\eea
where we have used
\be
\frac{1}{\G(1-\e)} = \frac{\G(1-\e)}{\G(1-2\e)} + {\cal O}(\e^2) \, .
\ee
The collinear condition ($0 \le s_{45} \le \d_c s_{12}$) 
sets the $s_{45}$ integration limits.
The hard condition sets the $z$ integration limits which depend 
additionally on the splitting function involved, and also on the mass of the 
parton recoiling against the 45 pair (3 in this case). 
This later dependence enters through the threshold condition, 
as discussed below.  

First take parton 3 to be  massless.
For $q \rightarrow qg$ splitting the hard region is defined by 
$\d_s \sqrt{s_{12}}/2 \le E_5 \le \sqrt{s_{12}}/2$, 
assuming 4 labels $q$ and 5 labels $g$.  From Eq.\ (\ref{eqn:coll}) we have
$s_{34}=(p_3+p_4)^2=2p_3 \cdot p_4 \simeq (2p_3 \cdot p_{45})z$ and 
$s_{12}=(p_3+p_{45})^2 = s_{45} + 2 p_3 \cdot p_{45} 
\simeq 2 p_3 \cdot p_{45}$ which together yield 
$s_{34} \simeq zs_{12}$.  Using Eq.\ (\ref{eqn:e5}) the hard condition 
becomes
\be
0 \le z \le 1-\d_s \, .
\label{eqn:hard_limit}
\ee
For the $g \rightarrow gg$ splitting it is required that both gluons be hard, 
{\it i.e.,}\ $E_4$ and $E_5 \ge \d_s \sqrt{s_{12}}/2$.  $z$ then satisfies 
the relation   
$\d_s \le z \le 1 - \d_s$.  For the $g \rightarrow q \overline q$ splitting 
there are no soft singularities, so $0 \le z \le 1$ may be taken.
In all of these cases the $z$ integration limits are independent of 
$s_{45}$ by virtue of the approximation $s_{45}=0$ implicit in 
$s_{12} \simeq 2 p_3 \cdot p_{45}$ (this point is discussed 
further at the end of Sec.\ \ref{sec:massless_example} and in Appendix C).  
The outermost integration over $s_{45}$ may therefore be performed 
giving the result
\be
d\s_{\rm HC}^{1+2\rightarrow 3+4+5} =  d\s_0^{1+2\rightarrow 3+4'} 
\left[ \frac{\a_s}{2\p} \frac{\G(1-\e)}{\G(1-2\e)} 
                 \left( \frac{4\p\mu_r^2}{s_{12}} \right)^\e \right]
\left( -\frac{1}{\e}\right) \d_c^{-\e} \int 
dz z^{-\e}(1-z)^{-\e} 
                 P_{44'}(z,\e) \, .
\label{eqn:dsHC1}
\ee
For $z<1$ the $n-$dimensional unregulated splitting functions may be 
written as $P_{ij}(z,\e)=P_{ij}(z)+\e P^{\prime}_{ij}(z)$.  Explicitly,
\bea
P_{qq}(z) &=& C_F \frac{1+z^2}{1-z} \label{eqn:ap_unreg1} \\
P_{qq}^{\prime}(z) &=& -C_F(1-z) \\
P_{gq}(z) &=& C_F \frac{1+(1-z)^2}{z} \\
P_{gq}^{\prime}(z) &=& -C_Fz \\
P_{gg}(z) &=& 2N\left[ \frac{z}{1-z}+\frac{1-z}{z}+z(1-z)\right] \\
P_{gg}^{\prime}(z) &=& 0 \\
P_{qg}(z) &=& \frac{1}{2} \left[ z^2+(1-z)^2 \right] \\
P_{qg}^{\prime}(z) &=& -z(1-z) \, ,
\label{eqn:ap_unreg2}
\eea
where $N=3$ and $C_F=(N^2-1)/2N=4/3$ for QCD.
Expanding the integrand in Eq.\ (\ref{eqn:dsHC1}) to $\co(\e)$
and integrating over $z$ yields the final state hard-collinear 
terms
\be
d\s_{HC}^{1+2\rightarrow 3+4+5} =  d\s_0^{1+2\rightarrow 3+4'} 
\left[ \frac{\a_s}{2\p} \frac{\G(1-\e)}{\G(1-2\e)} 
                 \left( \frac{4\p\mu_r^2}{s_{12}} \right)^\e \right]
                 \left( \frac{A_1^{4' \rightarrow 45}}{\e} 
                        + A_0^{4' \rightarrow 45} \right) \, ,
\label{eqn:HC_final}
\ee
where
\bea
A_1^{q \rightarrow qg} &=& C_F \left( 3/2+2\ln\d_s \right) \label{eqn:A1q} \\
A_0^{q \rightarrow qg} &=& C_F \left[ 7/2 - \p^2/3 - \ln^2\d_s 
        - \ln\d_c \left(3/2+2\ln\d_s \right) \right] \label{eqn:A0q} \\
A_1^{g \rightarrow q\overline{q}} &=& -n_f/3 \\
A_0^{g \rightarrow q\overline{q}} &=& n_f/3 \left( \ln\d_c-5/3 \right) \\
A_1^{g \rightarrow gg} &=& N \left( 11/6 + 2 \ln\d_s \right) \\
A_0^{g \rightarrow gg} &=& N \left[ 67/18 - \p^2/3 - \ln^2\d_s 
    - \ln\d_c \left( 11/6 + 2 \ln\d_s \right) \right] \, . \label{eqn:A0g}
\eea
where $n_f$ denotes the number of active flavors.

When the mass of the parton recoiling against the 45 pair, $m_3$, is 
retained, the factorization of the phase
space and matrix element is unaffected.  Likewise, the collinear
condition $0 \le s_{45} \le \d_c s_{12}$ remains unchanged.  It is
only the boundries of the hard region that are modified.  
The full kinematic range of
the invariant $s_{34}$ is $m_3^2 \le s_{34} \le s_{12}$.  The
threshold for producing a particle of mass $m_3$ sets the lower limit.
In terms of $s_{34}$ the hard region is 
$m_3^2 \le s_{34} \le (1-\d_s) s_{12}$.  
This implies the hard condition Eq.\ (\ref{eqn:hard_limit}) 
becomes $0 \le z \le 1-\d_s/(1-m_3^2/s_{12})$.  A similar analysis 
follows for the $g \rightarrow gg$ splitting case wherein 
$\d_s/(1-m_3^2/s_{12}) \le z \le 1-\d_s/(1-m_3^2/s_{12})$.  
We therefore immediately see that Eqs.\ (\ref{eqn:A1q})--(\ref{eqn:A0g}) 
are valid with the replacement $\d_s \rightarrow \d_s/(1-m_3^2/s_{12})$.

\subsubsection{Tagged Final States}
\label{sec:hadrons}

Next, consider a process where a particular type of hadron is identified in 
the final state. This necessitates the introduction of a fragmentation 
function $D_{h/c}(z)$ which gives the probability density for finding a 
hadron $h$ which carries a fraction $z$ of the momentum of the parent 
parton $c$. Consider the case where parton 4 fragments into a hadron $h$, 
for which the lowest order cross section is
\be
d\s_0^{1+2\rightarrow 3+h} = d\s_0^{1+2\rightarrow 3+4} D_{h/4}(z)dz \, .
\label{eqn:lo_f}
\ee
The hard collinear cross section expression in Eq.\ (\ref{eqn:dsHC1}) becomes
\bea
d\s_{\rm HC}^{1+2\rightarrow 3+h+5} &=&  d\s_0^{1+2\rightarrow 3+4'} 
\left[ \frac{\a_s}{2\p} \frac{\G(1-\e)}{\G(1-2\e)} 
                 \left( \frac{4\p\mu_r^2}{s_{12}} \right)^\e \right]
\left( -\frac{1}{\e}\right) \d_c^{-\e} \NO \\
&\times&\int dy y^{-\e}(1-y)^{-\e} 
                 D_{h/4}(x) dx P_{44'}(y,\e) \d(xy-z)dz \, .
\label{eqn:ds_frag}
\eea
The delta function insures that the hadron $h$ carries a momentum fraction 
$z$ of the parent parton's momentum (parton $4'$ in this example). 
Here there is a splitting $4' \rightarrow 4 5$ followed by 
parton 4 fragmenting to hadron $h$. When all possible $2 \rightarrow 3$ 
subprocesses are considered, there will be several contributions of this same 
form, corresponding to a sum over the parton 4. For example, if $4'$ is a 
gluon, there can be $g \rightarrow g g$ followed by $g \rightarrow h$ or 
$g \rightarrow q \overline q$ followed by $q \rightarrow h$ or 
$\overline q \rightarrow h$. Similarly, if $4'$ is a quark $q$, there  
can be $q \rightarrow q g$ followed by $q \rightarrow h$ or by 
$g \rightarrow h$.  Furthermore, the limits of integration on $y$ depend on 
the splitting function as in the case discussed in the previous 
subsection.

The collinear singularity, evidenced by the pole in $\e$ in 
Eq.\ (\ref{eqn:ds_frag}), must 
be factorized and absorbed into the bare fragmentation function. To do 
this, we introduce a scale dependent parton fragmentation function 
\be
D_{h/c}(z,M_f)=D_{h/c}(z)+\left( -\frac {1}{\e}\right) 
\left[ \frac{\a_s}{2\p} \frac{\G(1-\e)}{\G(1-2\e)} 
                 \left( \frac{4\p\mu_r^2}{M_f^2} \right)^\e \right]
\int_z^1 \frac{dy}{y} D_{h/c'}(z/y) P^+_{c'c}(y) \, .
\ee
In this expression there is an implied sum over the index $c'$ corresponding 
to the sum over the different fragmentation possibilities referred to above. 
The final state factorization scale has been denoted by $M_f$.
Notice, too, that the integration over $y$ extends from $z$ to 1.
This form for the scale dependent fragmentation function  corresponds 
to the $\overline{\rm MS}$ convention.  The regulated ($x \le 1$) 
splitting functions \cite{ap} are given by
\bea
P^+_{qq}(x) &=& C_F \left[ \frac{1+x^2}{(1-x)_+} 
+ \frac{3}{2} \d(1-x) \right] \label{eqn:pqq} \\
P^+_{gq}(x) &=& C_F \left[ \frac{1+(1-x)^2}{x} \right] \\
P^+_{gg}(x) &=& 2N \left[ \frac{x}{(1-x)_+} + \frac{1-x}{x} + x(1-x) \right] 
+ \left( \frac{11}{6} N - \frac{1}{3} n_f \right) \d(1-x) \\
P^+_{qg}(x) &=& \frac{1}{2} \left[ x^2 + (1-x)^2 \right] \, .
\label{eqn:pqg}
\eea
Next, we rewrite the bare fragmentation function in Eq.\ (\ref{eqn:lo_f}) 
in terms of the 
scale dependent expression given above, yielding to $\co(\alpha_s)$
\bea
d\s_0^{1+2\rightarrow 3+h} &=& d\s_0^{1+2\rightarrow 3+4} \NO \\
&\times& \left\{ D_{h/4}(z,M_f) + 
\left( \frac {1}{\e}\right) \left[ \frac{\a_s}{2\p} 
\frac{\G(1-\e)}{\G(1-2\e)} 
                 \left( \frac{4\p\mu_r^2}{M_f^2} \right)^\e \right]
\int_z^1 \frac{dy}{y} D_{h/c'}(z/y) P^+_{c'4}(y)\right\} dz \, . \NO \\
\eea
The second term is sometimes refered to as the mass factorization
counterterm.  When $d\s_0$ and $d\s_{\rm HC}$ are added together,
there is a cancellation between the two singular expressions. Note,
however, that this cancellation is not complete since the limits of
the $y$ integration in the two expressions differ.

After the cancellation, the resulting $\co(\alpha_s)$ expression for the 
fragmentation contribution is 
\bea
\label{eqn:frag_twobody}
d\s_{frag}^{1+2\rightarrow 3+h} &=& d\s_0^{1+2\rightarrow 3+4'}
\left[ \frac{\a_s}{2\p} \frac{\G(1-\e)}{\G(1-2\e)} 
                 \left( \frac{4\p\mu_r^2}{s_{12}} \right)^\e \right] \NO \\
&\times&\left\{ \widetilde{D}_{h/4'}(z,M_f) 
+ \left[ \frac{A_1^{sc}(4'\rightarrow 4+5)}
{\e} + A_0^{sc}(4'\rightarrow 4+5)\right] D_{h/4}(z,M_f) 
\right\}dz \, . \NO \\
\eea
The soft collinear factors $A_i^{sc}$ result from the mismatch in the $y$ 
integrations in the fragmentation and subtraction pieces, mentioned above. 
They are given by 
\bea
A_0^{sc} &=& A_1^{sc} \ln \left( \frac{s_{12}}{M_f^2} \right) 
\label{eqn:sc_0} \\
A_1^{sc}(q\rightarrow qg) &=& C_F(2 \ln \d_S + 3/2 ) \label{eqn:sc_1q} \\
A_1^{sc}(g\rightarrow gg) &=& 2N \ln \d_s + (11N-2 n_f)/6 \, .
\label{eqn:sc_1g}
\eea
The modified fragmentation function $\widetilde{D}_{h/c}(z, M_f)$ is given by
\be
\label{eqn:Dtilde}
\widetilde{D}_{h/c}(z,M_f) = \sum_{c'}\int_z^{1-\d_s \d_{c'c}} \frac{dy}{y} 
D_{h/c'}(z/y,M_f) \widetilde{P}_{c'c}^{\rm frag} \, ,
\ee
where 
\be
\label{eqn:Ptilde}
\widetilde{P}_{c'c}^{\rm frag} = P_{c'c}(y) \ln \left[ 
\frac{y(1-y)\d_c s_{12}}{M_f^2} \right] - P'_{c'c}(y) \, .
\ee
$P(y)$ and $P'(y)$ are the $n=4$ and $\co (\e)$ pieces, respectively, 
of the unregulated splitting kernels given in 
Eqs.\ (\ref{eqn:ap_unreg1})--(\ref{eqn:ap_unreg2}).
The $\widetilde{D}$ functions contain an explicit logarithm of $\d_c$ as well 
as logarithmic dependences on $\d_s$ which are built up by the integration 
on $y$ when $c'=c$. 
In Appendix D it is shown how to make the $\d_s$ dependence 
explicit, thereby improving the convergence of the Monte Carlo integration. 

Comparing with the previous subsection, we see that when going to 
the fragmentation case, the hard collinear 
terms, Eqs.\ (\ref{eqn:A1q})--(\ref{eqn:A0g}), for the fragmenting 
parton are replaced by a combination of the $\widetilde{D}$ 
function and soft collinear factors $A_i^{sc}$.  Nevertheless, a careful 
comparison of the two cases shows that the poles in 
$\e$ cancel and the final results for physics observables 
are independent of the cutoffs. This 
will be illustrated by several examples to follow.

\subsubsection{Initial State}
\label{sec:initial}

The treatment of the initial state collinear singularities is much the 
same as that for the previous case of final state fragmentation. The 
collinear singularities are absorbed into the bare parton distribution 
functions leaving a finite remainder which is written in terms of 
modified parton distribution functions. In addition, there are accompanying 
soft collinear factors as in the fragmentation case. 
However, some of the details are different, so a brief summary of the 
derivation is given here. In order not to unnecessarily complicate the 
discussion, only the details for one of the incoming partons will be shown. 

Consider a process which involves a parton on leg 2 coming from an incoming 
hadron $B$, so that in lowest order 
\be
d\s_0^{1+B\rightarrow 3+4}= G_{2/B}(x)dx d\hat\s_0^{1+2\rightarrow 3+4} \, ,
\label{eqn:lo_initial}
\ee
where $G_{2/B}(x)dx$ denotes the probability of getting parton 2 from 
hadron $B$ with a momentum fraction between $x$ and $x+dx$. The 
hat symbol $\hat{}$ is used here to label a purely partonic subprocess.
We are interested in the next-to-leading-order corrections coming from 
the various possible parton splittings which can occur on leg 2. The hard 
collinear contribution Eq.\ (\ref{eqn:sig_HC})  
is calculated by applying the collinear approximation 
to the appropriate three-body matrix elements as follows:
\be
\sumb\vert M_3(1+2\rightarrow 3+4+5)\vert^2 \simeq \sumb \vert M_2(
1+2'\rightarrow 3+4)\vert^2 P_{2'2}(z,\e)g^2 \mu_r^{2\e} 
\frac{-2}{z t_{25}} \, ,
\ee
where $z$ denotes the fraction of parton 2's momentum carried by parton $2'$ 
with parton 5 taking a fraction $(1-z)$. Using the approximation 
$p_2-p_5 \simeq z p_2$, the three-body phase space may be written as 
\be
d\G_3|_{coll}=\left[\frac{d^{n-1}p_3}{2p_3^0(2\pi)^{n-1}}\frac{d^{n-1}p_4}
{2p_4^0(2\pi)^{n-1}}(2\pi)^n \d^n(p_1+zp_2-p_3-p_4)\right] 
\frac{d^{n-1}p_5}{2p_5^0(2\pi)^{n-1}} \, .
\ee
The square bracketed portion is just two-body phase space evaluated at a 
squared parton-parton energy of $z s_{12}$. The $p_5$ dependent part may be 
rewritten as  
\be
\frac{d^{n-1}p_5}{2p_5^0(2\pi)^{n-1}} = 
\frac{(4\pi)^{\e}}{16 \pi^2 \G(1-\e)}dz dt_{25}[-(1-z)t_{25}]^{-\e} \, .
\ee
The allowed range for $t_{25}$ is given by the collinear condition
$0 < -t_{25} < \d_c s_{12}$.  The $t_{25}$ integration yields
\be
\int_0^{\d_c s_{12}} -dt_{25} (-t_{25})^{-1-\e} = -\frac{1}{\e} 
(\d_c s_{12})^{-\e} \, .
\ee
Using these results, the three-body cross section in the hard collinear 
region may be written as 
\bea 
d\s_{HC}^{1+B\rightarrow 3+4+5} & = &G_{2/B}(y)dy d\hat\s_0^
{1+2'\rightarrow 3+4} (zs_{12},t_{13},t_{14})
\left[ \frac{\alpha_s}{2\pi} \frac{\G(1-\e)}{\G(1-2\e)}
\left(\frac{4 \pi \mu_r^2}{s_{12}}\right)^{\e}\right] \NO \\
&\times& \left(-\frac{1}{\e}\right)
\d_c^{-\e}P_{2'2}(z,\e)dz (1-z)^{-\e} \d(yz-x)dx \, .
\eea
Note that a factor of $1/z$ has been absorbed into the flux factor for the 
two-body subprocess. The delta function insures 
that the fraction of hadron $B's$ momentum carried by parton $2'$ into the 
two-body subprocess is $x$ in order to be able to combine this result with the
lowest order contribution. The delta function may be used to perform the $y$ 
integration, but one point must first be made. $s_{12}$ is related to 
the square of the overall hadronic squared center-of-mass energy $S$ by 
$s_{12}=y S$. On the other hand, in the lowest order subprocess the relation 
is $s_{12}=x S$. It is convenient to rewrite the above expression using this 
latter definition for $s_{12}$. Therefore, after the $y$ integration, 
we obtain
\bea 
d\s_{HC}^{1+B\rightarrow 3+4+5} & = & G_{2/B}(x/z) d\hat\s_0^
{1+2'\rightarrow 3+4} (s_{12},t_{13},t_{14})
\left[ \frac{\alpha_s}{2\pi} \frac{\G(1-\e)}{\G(1-2\e)}
\left(\frac{4\pi\m_r^2}{s_{12}}\right)^{\e}\right] \NO \\ 
&\times& \left(-\frac{1}{\e}\right)
\d_c^{-\e}P_{2'2}(z,\e)\frac{dz}{z} \left[ \frac {(1-z)}{z} \right]^{-\e}
dx \, .
\label{eqn:HC_initial}
\eea
Comparing with the corresponding result for final state fragmentation, we see 
that a factor of $[z(1-z)]^{-\e}$ has been changed to $[(1-z)/z]^{-\e}$.

In order to factorize the collinear singularity into the parton distribution 
function, we introduce a scale dependent parton distribution function using 
the $\overline{\rm MS}$ convention:
\be
G_{b/B}(x,\m_f)=G_{b/B}(x)+\left(-\frac{1}{\e}\right) \left[
\frac{\alpha_s}{2\pi} 
\frac{\G(1-\e)}{\G(1-2\e)} \left(\frac{4\pi \m_r^2}{\m_f^2}\right)^{\e}\right]
\int_z^1 \frac{dz}{z} P_{bb'}(z)G_{b'/B}(x/z) \, .
\ee
Next, using this definition, we replace $G_{2/B}(x)$ in the lowest order 
expression (\ref{eqn:lo_initial}) and combine the result with 
the hard collinear contribution (\ref{eqn:HC_initial}). 
The resulting $\co(\alpha_s)$ expression 
for the initial state collinear contribution is 
\bea
d\s_{coll}^{1+B\rightarrow 3+4+5} &=& d\hat\s_0^{1+2'\rightarrow 3+4}
\left[ \frac{\alpha_s}{2\pi} \frac{\G(1-\e)}{\G(1-2\e)}
\left(\frac{4 \pi \m_r^2}{s_{12}}\right)^{\e}\right] \NO \\
&\times& \left\{
\widetilde{G}_{2'/B}(z,\m_f) + 
\left[ \frac{A_1^{sc}(2\rightarrow 2'+5)}{\e} + A_0^{sc}
(2\rightarrow 2'+5)\right] G_{2/B}(z,\m_f)\right\}dz \, .
\label{eqn:sig_coll}
\eea
Note that in this expression the soft collinear factors (given in 
Eqs.\ (\ref{eqn:sc_0})--(\ref{eqn:sc_1g})) 
depend on the initial state factorization scale
$\m_f$.  The $\widetilde{G}$ functions are given by
\be
\widetilde{G}_{c/B}(x,\m_f) = \sum_{c'}  \int_x^{1-\d_s\d_{cc'}} \frac{dy}{y} 
                       G_{c'/B}(x/y,\m_f) \widetilde{P}_{cc'}(y)
\label{eqn:g_tilde}
\ee
with 
\be
\widetilde{P}_{ij}(y) = P_{ij}(y)\ln\left(\d_c\frac{1-y}{y}
\frac{s_{12}}{\m_f^2}\right) - P_{ij}^{\prime}(y) \, .
\ee
The $n=4$ and $\co (\e)$ pieces of the unregulated splitting kernels, 
$P(y)$ and $P'(y)$, are given in 
Eqs.\ (\ref{eqn:ap_unreg1})--(\ref{eqn:ap_unreg2}).
An example of a hadron-hadron process will be given in the next section.

As in the final state hadron case, the $\widetilde{G}$ functions contain an
explicit logarithm of $\d_c$ as well as logarithmic dependences on
$\d_s$ which are built up by the integration on $y$. In Appendix D it
is shown how to make the $\d_s$ dependence explicit, thereby improving
the convergence of the Monte Carlo integration.

\section{Examples}

In this section we provide five illustrative examples applying the
method developed in the previous section.  The results
are shown to be in complete agreement with those available in the
literature.  We begin by calculating the QCD corrections to
electron-positron annihilation into a massive quark pair.  
The quark mass serves to regulate any would be collinear singularities. 
There are only final state soft singularities and, hence, only the 
soft cutoff is required.  Next, the
QCD corrections to electron-positron annihilation into a massless
quark pair are considered.  In this case final state soft and collinear
singularities are encountered, necessitating the use of both soft and 
collinear cutoffs. The example of inclusive photon production in
hadronic final states of electron-positron annihilation is then
presented, illustrating the use of fragmentation functions.  Finally,
we close the examples section by showing how to calculate the QCD
corrections to lepton pair and single particle inclusive 
production in hadron-hadron collisions.
Both examples contain initial state soft and collinear singularities,
necessitating the use of scale dependent parton distribution functions. 
Furthermore, the single particle inclusive cross section calculation 
also requires the use of scale dependent fragmentation functions.

\subsection{Electron-positron annihilation to massive quark pair}

\begin{figure}
\centerline{\hbox{\epsfig{figure=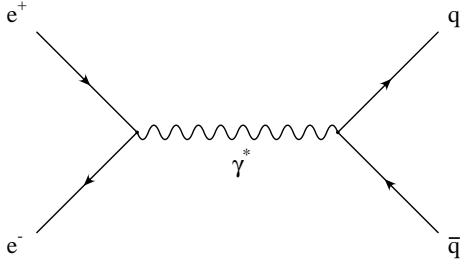,width=2.5in}}}
\caption[]{Leading order contribution to electron-positron annihilation 
via photon exchange.}
\label{ee_born}
\end{figure}
Electron-positron annihilation into a massive quark pair has a particularly 
simple singularity structure, that of soft singularities in the final state 
only.  It will therefore be used as a first example of the method described 
in the section above.

Working in the single photon exchange approximation, the leading order 
Feynman diagram is shown in Fig.\ \ref{ee_born}.  Neglecting the electron mass 
and denoting the quark mass by $m$, the leading order cross section 
\be
d\s^0 = \frac{1}{2s_{12}} \sumb |M_2|^2 d\G_2 \, ,
\ee
calculated in $n=4-2\e$ dimensions is expressed in terms of 
the (summed and averaged) matrix element squared 
\be
\sumb |M_2|^2 = 2 N e^4 Q_q^2 \left( 
    \frac{{t_{13}^\prime}^2+{t_{23}^\prime}^2}{s_{12}^2} 
    + \frac{2m^2}{s_{12}} - \e \right) \, ,
\ee
and the two-body phase space 
\be
d\G_2 = \frac{2^{2\e}}{16\p} \left( \frac{4\p}{s_{12}} \right)^\e \b^{1-2\e}
        \frac{1}{\G(1-\e)} \int_0^\p \sin^{1-2\e} \! \q \, d\q \, .
\ee
The center-of-mass scattering angle is denoted by $\q$ and 
$\b=\sqrt{1-4m^2/s_{12}}$.  $Q_q$ is the quark charge 
in units of $e$ and $N=3$ is the
number of colors.  When masses are present, it is often convenient to
define primed Mandelstam invariants which are the ones defined
previously minus some combination of squared masses.  In this case, 
$s_{ij}^\prime \equiv s_{ij} - m^2$ and $t_{ij}^\prime \equiv t_{ij}-m^2$.  
Performing the integration over $\q$ we obtain the 
well known $\e=0$ result,
\be
\s^0 = \frac{4\p\a^2}{3s_{12}}Q_q^2N(1+\frac{2m^2}{s_{12}})\b \, ,
\ee
with $\a=e^2/4\p$.

Because the quark mass regulates any would-be final state collinear
singularity, the appropriate decomposition of the two-to-three 
contribution to the cross section is simply given by 
Eq.\ (\ref{eqn:softhard}).  For the $\co (\a_s)$ QCD corrections we
therefore need the soft cross section (\ref{eqn:soft}), the hard cross
section (\ref{eqn:hard}), and the virtual corrections.

\begin{figure}
\centerline{\hbox{\epsfig{figure=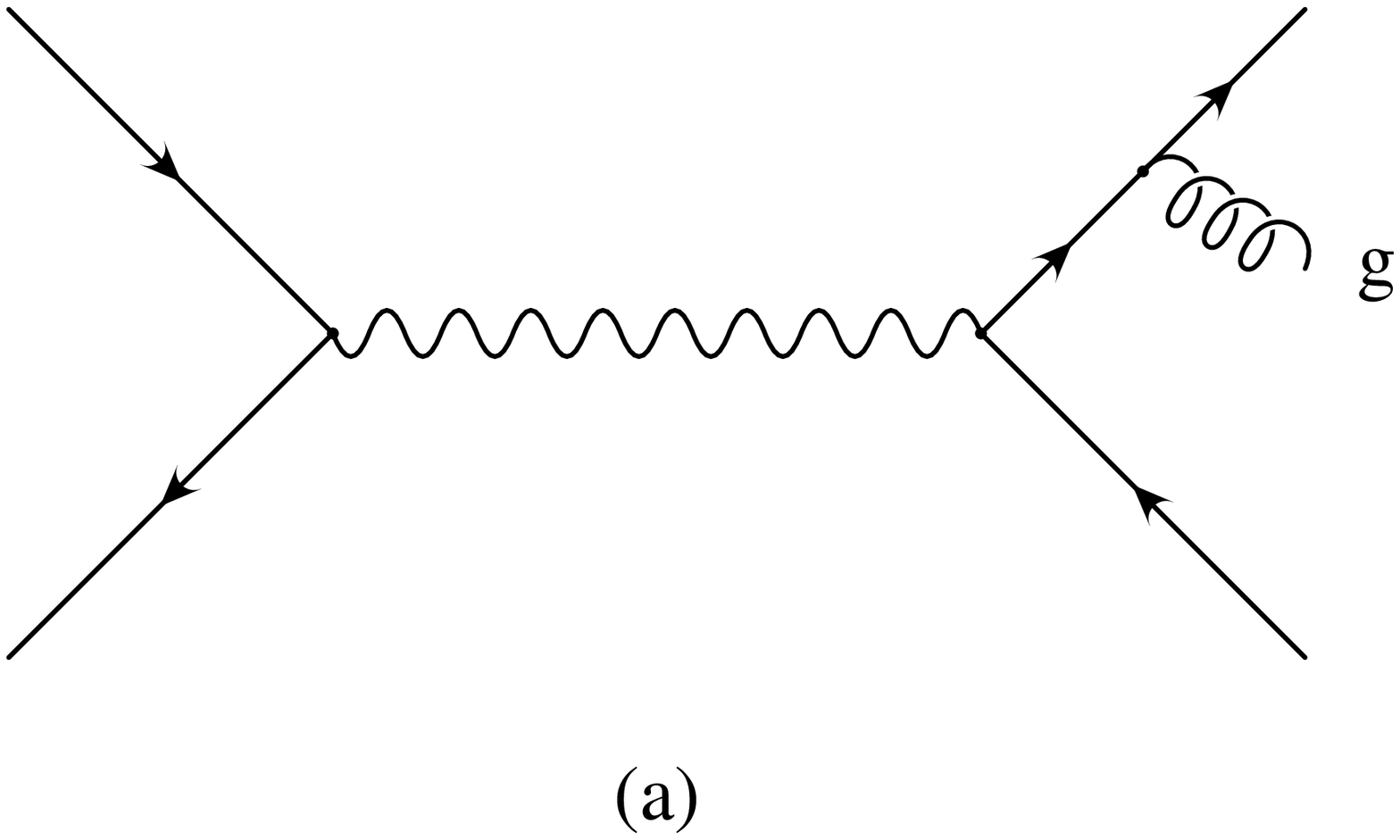,width=2.5in}
                  \epsfig{figure=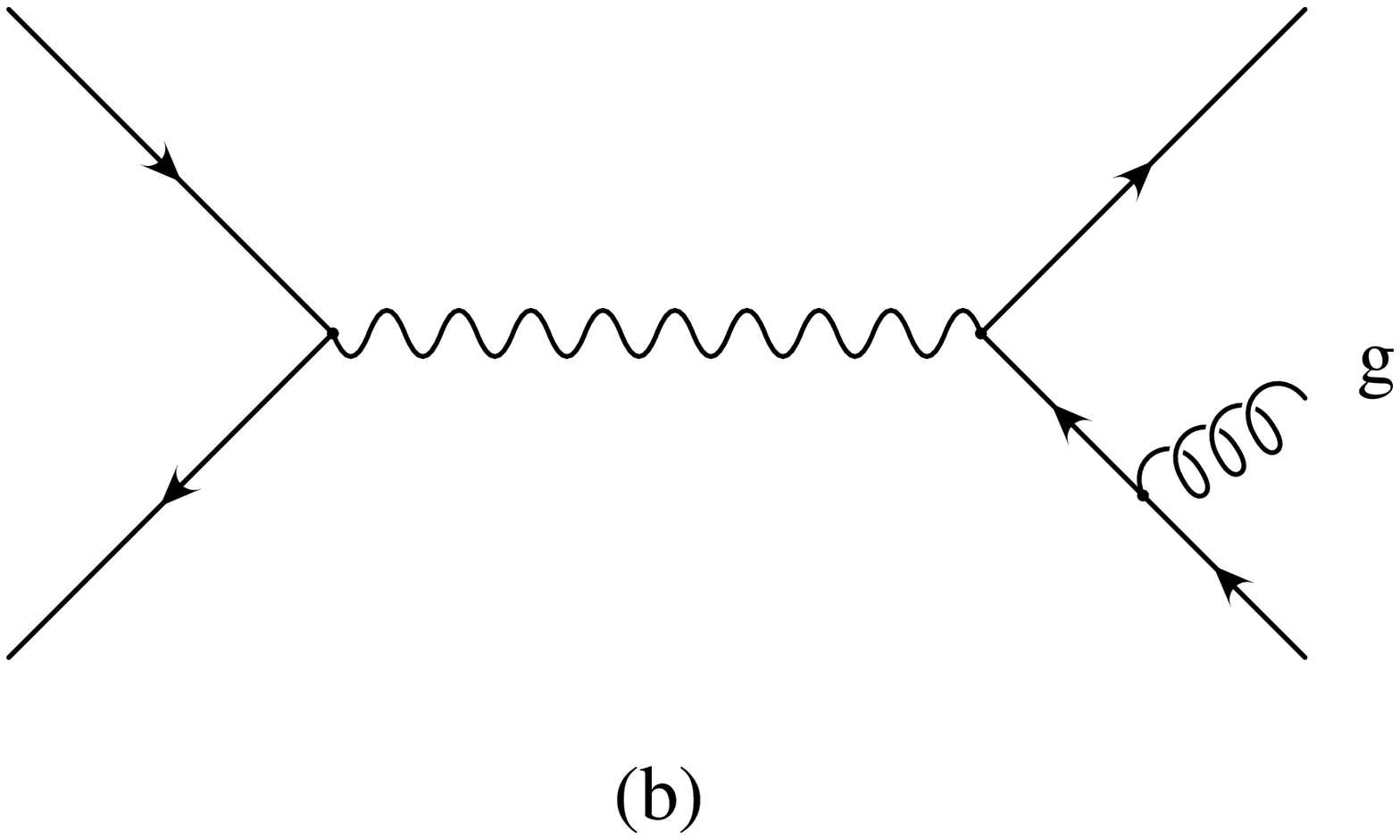,width=2.5in}}}
\caption[]{Real emission contribution to electron-positron annihilation.}
\label{ee_real}
\end{figure}

The real emission diagrams that give $|M_3|^2$ 
are shown in Fig.\ \ref{ee_real}.
If we define the (summed and averaged) squared matrix element as 
\be
\sumb |M_3|^2 = \frac{1}{4} e^4 Q_q^2 g^2 \frac{1}{s_{12}^2} N C_F \J \, ,
\label{eqn:2to3}
\ee
then
\bea
\J &=& {\rm tr} \left[\slash{p}_2\g^\m\slash{p}_1\g^\nu\right] \NO \\
&\times& \left\{ {\rm tr} \left[ 
(\slash{p}_3+m)\g_\m(\slash{p}_4+\slash{p}_5-m)
\g^\s(\slash{p}_4-m)\g_\s(-\slash{p}_4-\slash{p}_5+m)\g_\n\right]/
{s_{45}^\prime}^2 \right. \NO \\
& & + 2\, {\rm tr} \left[ (\slash{p}_3+m)\g^\s(\slash{p}_3+\slash{p}_5+m)\g^\m
(\slash{p}_4-m)\g_\s(\slash{p}_4+\slash{p}_5-m)\g_\n\right]/s_{35}^\prime
s_{45}^\prime \NO \\
&  & + \left. {\rm tr} \left[ (\slash{p}_3+m)\g^\s(\slash{p}_3+\slash{p}_5+m)
\g^\m(-\slash{p}_4+m)\g_\n(\slash{p}_3+\slash{p}_5+m)\g_\s\right]/
{s_{35}^\prime}^2 \right\} \, ,
\label{eqn:psi}
\eea
where the strong coupling is denoted by $g$ and $C_F=(N^2-1)/2N=4/3$.  To
obtain the hard contribution, $\s_{\rm H}$, the traces may be
evaluated in four space-time dimensions.

There is a soft singularity when the energy of the gluon 
in Fig.\ \ref{ee_real} goes to zero.  The corresponding soft 
contribution to the cross section, $\s_{\rm S}$, is given by 
Eq.\ (\ref{eqn:soft_final}).   In this case, the sum in 
(\ref{eqn:soft_final}) is taken over the 
final state quark legs (labeled 3 and 4) and 
the color connected Born cross sections 
are related to the leading order cross section by 
$d\s_{33}^0=-d\s_{34}^0=d\s_{44}^0=C_F d\s^0$.  We find
\be
d\s_{\rm S} =  d\s^0
         \left[ \frac{\a_s}{2\p} \frac{\G(1-\e)}{\G(1-2\e)} 
         \left( \frac{4\p\mu_r^2}{s_{12}} \right)^\e \right] 
         C_F \int \left[ -\frac{m^2}{(p_3 \cdot p_5)^2}
         -\frac{m^2}{(p_4 \cdot p_5)^2} +
         \frac{s-2m^2}{p_3 \cdot p_5 \, p_4 \cdot p_5} \right] dS \; .
\label{eqn:soft_massive}
\ee
The poles need to be integrated over the soft phase space 
according to Eq.\ (\ref{eqn:soft_PS}) to extract the singularities 
in dimensional regularization.  Define
\bea
I(s_{35}^\prime) &=& \int \frac{1}{{s_{35}^\prime}^2} dS \label{eqn:I1} \\
I(s_{45}^\prime) &=& \int \frac{1}{{s_{45}^\prime}^2} dS \label{eqn:I2} \\
I(s_{35}^\prime s_{45}^\prime) &=& \int \frac{1}{s_{35}^\prime s_{45}^\prime} 
dS \, . \label{eqn:I3}
\eea
In terms of the $p_1p_2$ center-of-momentum 
scattering angle $\q$, $p_3$ and $p_4$ may be written as 
\bea
p_3 &=& \frac{\sqrt{s_{12}}}{2} (1, 0, \ldots, 0, \b\sin\q,\b\cos\q) \NO \\
p_4 &=& \frac{\sqrt{s_{12}}}{2} (1, 0, \ldots, 0, -\b\sin\q,-\b\cos\q) \, .
\label{eqn:p3p4}
\eea
Using these together with Eq.\ (\ref{eqn:p5_soft}) we find
\bea
s_{35}^\prime &=& 2p_3\cdot p_5 = \sqrt{s_{12}}E_5
      (1-\b\sin\q\sin\q_1\cos\q_2-\b\cos\q\cos\q_1)
\NO \\
s_{45}^\prime &=& 2p_4\cdot p_5 = \sqrt{s_{12}}E_5
      (1+\b\sin\q\sin\q_1\cos\q_2+\b\cos\q\cos\q_1) \, .
\label{eqn:s35s45}
\eea
The gluon energy integrals in Eqs.\ (\ref{eqn:I1})--(\ref{eqn:I3})
may be performed trivially
\be
\left( \frac{4}{s_{12}} \right)^{-\e} 
           \int_0^{\d_s\sqrt{s_{12}}/2} 
           dE_5 E_5^{1-2\e} \frac{1}{s_{12}E_5^2} = 
        \frac{1}{s_{12}} \left( -\frac{1}{2\e} \right) 
        \d_s^{-2\e} \, .
\label{eqn:energy_int}
\ee
The remaining angular integrals are well know and are tabulated 
in the Appendix B.  The complete results are
\be
I(s_{35}^\prime) = I(s_{45}^\prime) = \frac{1}{2m^2} \left( -\frac{1}{2\e} 
       + \ln \d_s - \frac{1}{2\b} \ln \frac{1+\b}{1-\b} \right)
\ee
\be
I(s_{35}^\prime s_{45}^\prime) = \frac{1}{s_{12}\b} 
  \left( - \frac{1}{2\e} \ln 
  \frac{1+\b}{1-\b} - \dilog\frac{2\b}{1+\b} - \frac{1}{4} \ln^2 
  \frac{1+\b}{1-\b} + \ln \d_s \ln \frac{1+\b}{1-\b} \right) \, .
\ee
\begin{figure}
\centerline{\hbox{\epsfig{figure=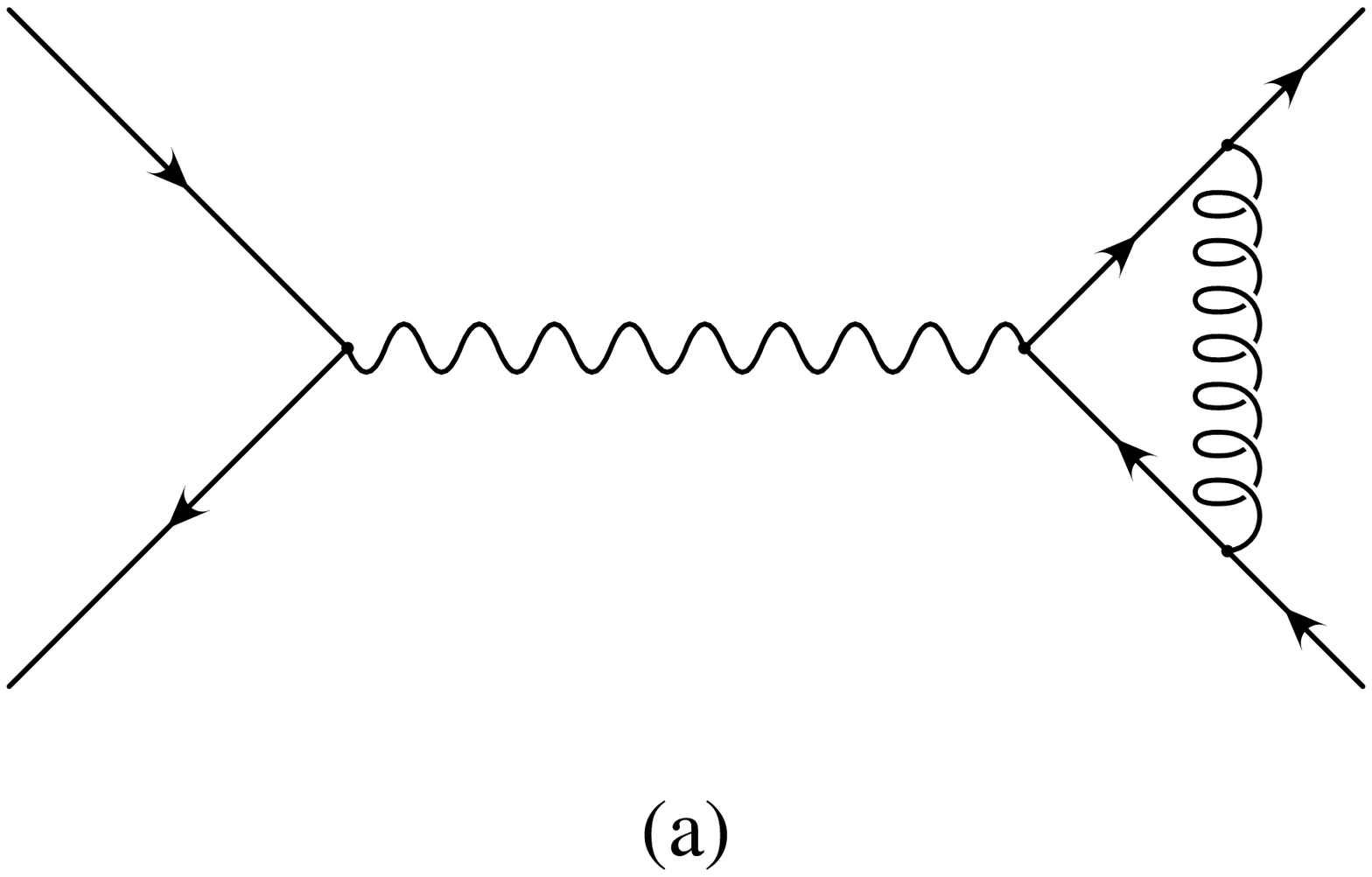,width=2.5in}
                  \epsfig{figure=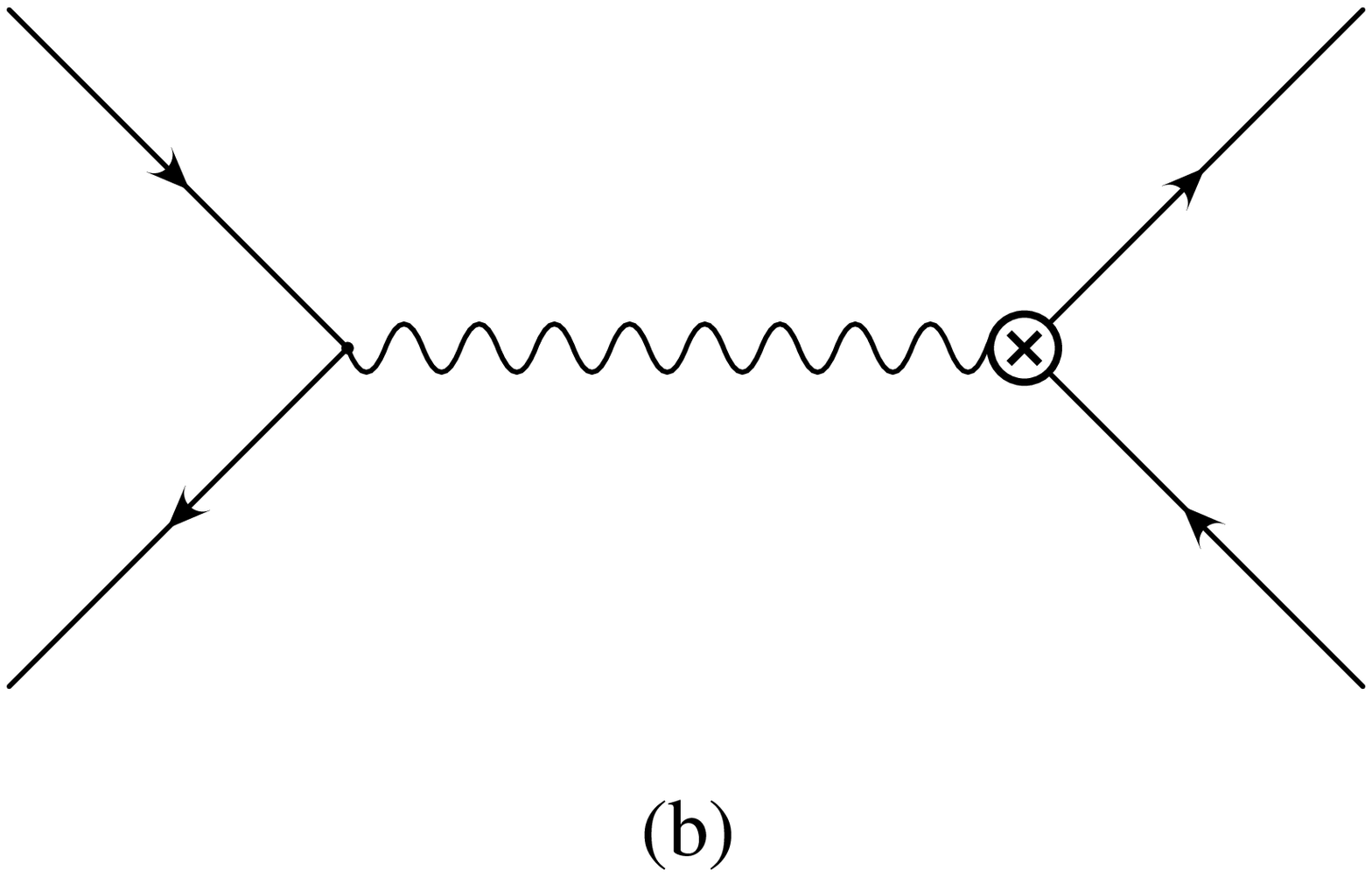,width=2.5in}}}
\centerline{\hbox{\epsfig{figure=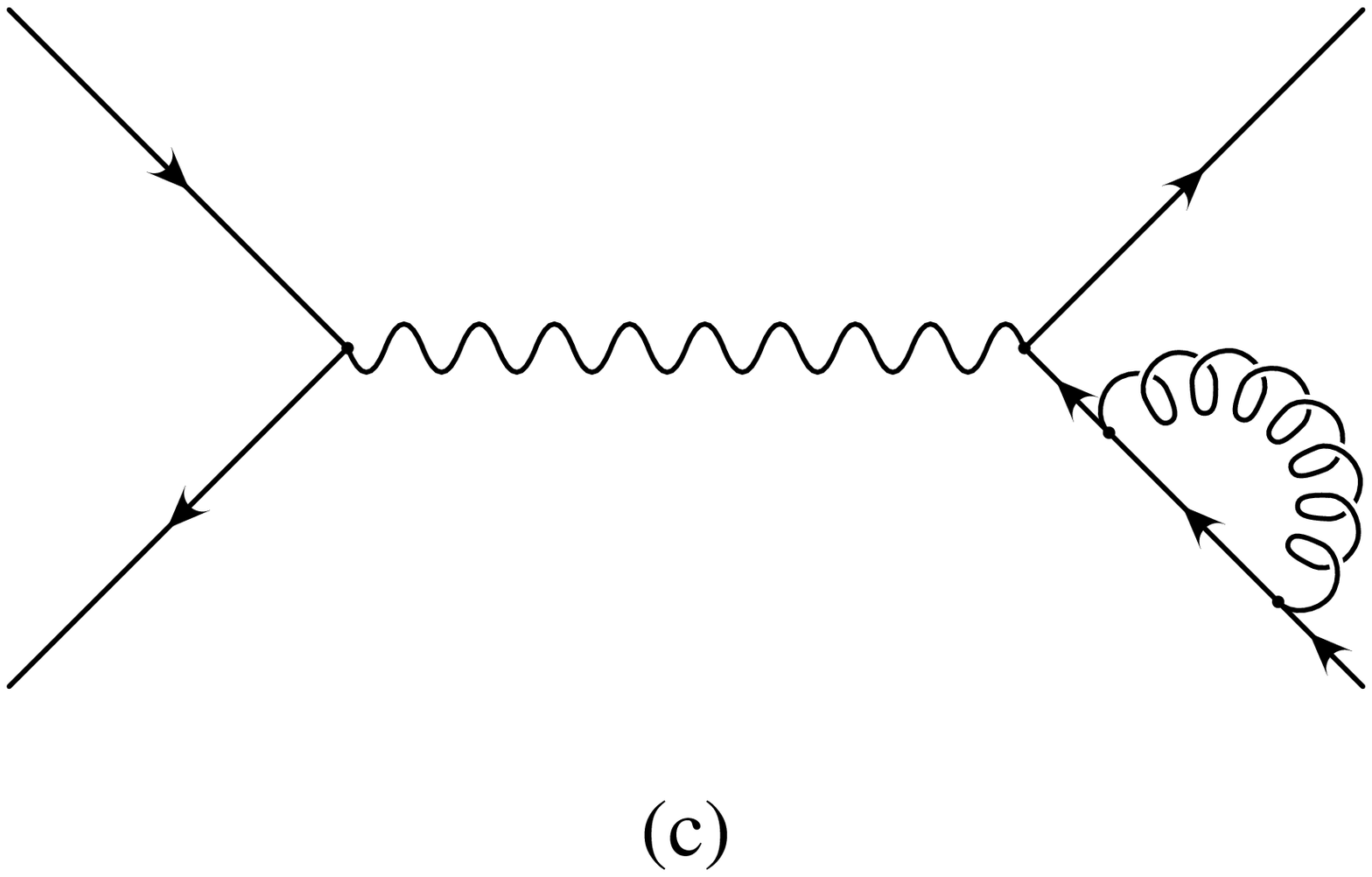,width=2.5in}
                  \epsfig{figure=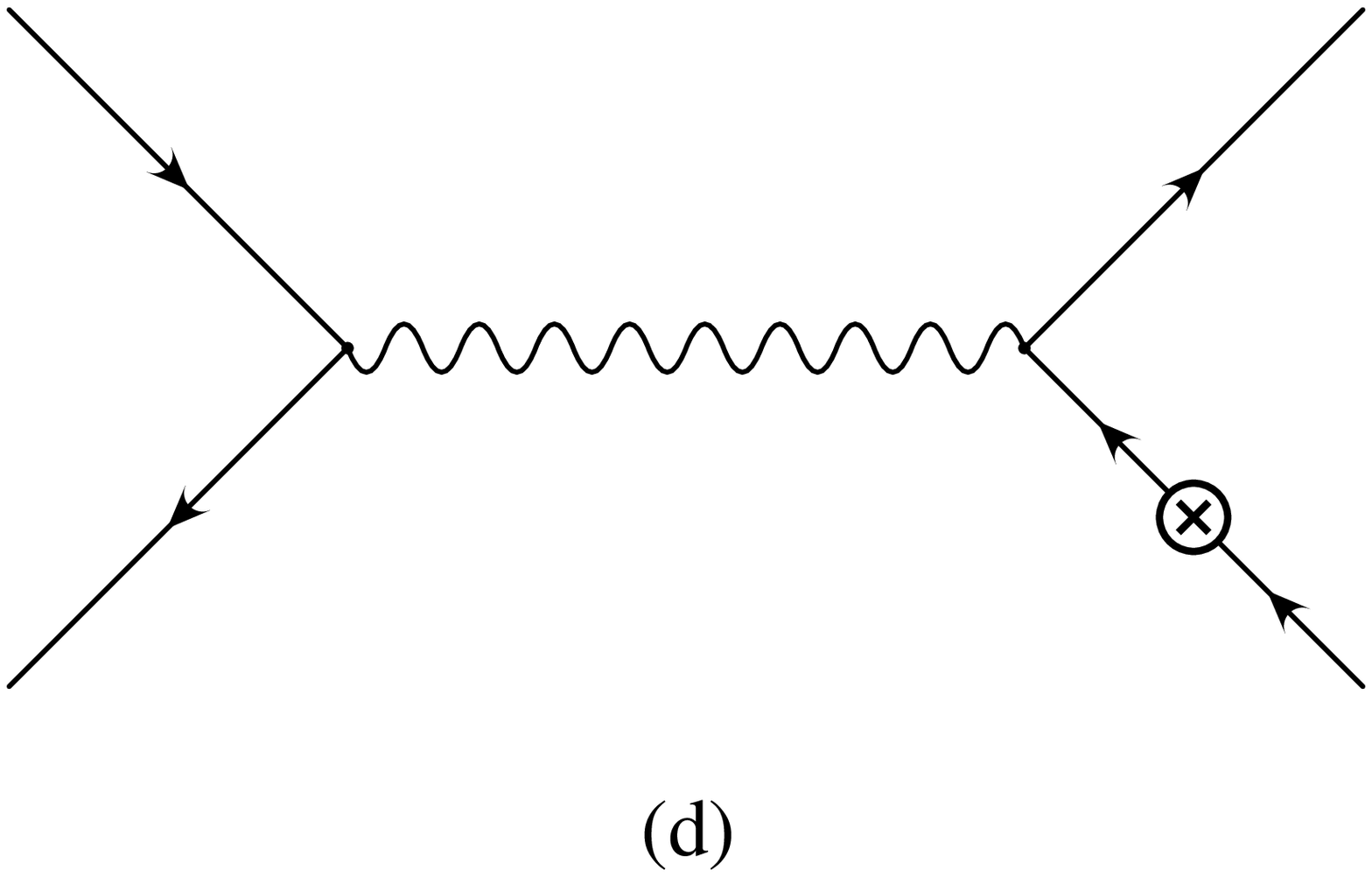,width=2.5in}}}
\caption[]{Loop and counterterm corrections to electron-positron annihilation 
via photon exchange.}
\label{ee_oneloop}
\end{figure}
We may therefore write the final expression for the soft cross section as
\be
d\s_{\rm S} = d\s^0 \left[ \frac{\a_s}{2\p}
\frac{\G(1-\e)}{\G(1-2\e)} \left( \frac{4\p\mu_r^2}{s_{12}} \right)^\e \right]
\left( \frac{A_1^s}{\e}+A_0^s \right) \, ,
\ee
where
\bea
A_1^s &=& 2 C_F \left( 1 - \frac{1+\b^2}{2\b} \ln \frac{1+\b}{1-\b} 
      \right) \NO \\
A_0^s &=& 4 C_F \left[ -\ln \d_s + \frac{1}{2\b} \ln \frac{1+\b}{1-\b} 
          \right. \NO \\
      &-& \left. \frac{1+\b^2}{2\b} \left( \dilog\frac{2\b}{1+\b} 
       + \frac{1}{4} \ln^2 \frac{1+\b}{1-\b} - \ln \d_s \ln 
       \frac{1+\b}{1-\b}\right) \right] \, .
\eea

\begin{figure}
\centerline{\hbox{\epsfig{figure=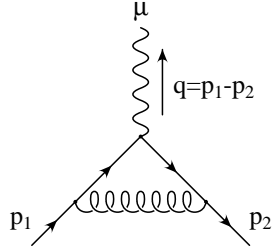,width=1.5in}}}
\caption[]{The vertex correction.}
\label{vertex}
\end{figure}

The virtual contribution is obtained from the one-loop 
diagrams shown in Fig.\ \ref{ee_oneloop}.  In the on-shell 
renormalization scheme diagrams $(c)$ and $(d)$ cancel exactly.
The vertex correction needed in Fig.\ \ref{ee_oneloop}$(a)$ is shown 
separately in Fig.\ \ref{vertex}.  After performing the loop integrals 
the result for the vertex valid for $q^2 > 4m^2$ is
\be
\G_\mu = \left( -ieQ_q \right) \d_{ij} \left( g^2 C_F C_\e \right) 
   \overline{u}(p_2) \left[ A \frac{\left(p_1+p_2\right)_\mu}{2m} 
   + B \g_\mu \right] u(p_1) \, ,
\ee
with
\bea
A &=& \frac{\b^2-1}{\b} \ln \frac{1-\b}{1+\b}  \\
B &=& \frac{1}{\e} \left( 1 + \frac{1+\b^2}{\b} \ln \frac{1+\b}{1-\b} 
      \right) + 3\b\ln \frac{1+\b}{1-\b} \NO \\
  &+& \frac{1+\b^2}{\b} \left( -\frac{1}{2} \ln^2 \frac{1-\b}{1+\b} 
      +2 \ln \frac{1-\b}{1+\b} \ln \frac{2\b}{1+\b} 
      + 2 \dilog\frac{1-\b}{1+\b} + \frac{2\p^2}{3} \right) \, ,
\eea
and
\be
C_\e = \frac{\p^{2-\e}}{(2\p)^{4-2\e}} \G(1+\e) 
\left( \frac{\m_r^2}{m^2} \right)^{\e} \, .
\ee
The vertex counterterm, $Z_1$, implicit in Fig.\ \ref{ee_oneloop}($b$)
is fixed in the on-shell renormalization scheme by the condition that
the renormalized vertex through $\co (g^2)$ evaluated at zero momentum
transfer equals the leading contribution ($-ie\L^\mu(q=0)=-ie\g^\mu$).  
This results in
\be
Z_1 = 1 - g^2 C_F C_\e \left( \frac{3}{\e} + 4 \right) \, .
\ee
The interference of the leading order diagram with 
the one loop renormalized diagrams yields 
\be
d\s_{\rm V} = d\s^0  \left[ \frac{\a_s}{2\p}
\frac{\G(1-\e)}{\G(1-2\e)} \left( \frac{4\p\mu_r^2}{s_{12}} \right)^\e \right]
\left( \frac{A_1^v}{\e}+A_0^v \right) + d\widetilde{\s}^v \, ,
\ee
with
\bea
A_1^v &=& -2 C_F \left( 1 - \frac{1+\b^2}{2\b} \ln \frac{1+\b}{1-\b} 
      \right) \NO \\
A_0^v &=& C_F \left[ -2 \left( 1 - \frac{1+\b^2}{2\b} \ln \frac{1+\b}{1-\b} 
      \right) \ln\left(\frac{s_{12}}{m^2}\right) + 3\b\ln\frac{1+\b}{1-\b}
      -4 \right. \NO \\ 
&+& \left. \frac{1+\b^2}{\b} \left( -\frac{1}{2}\ln^2\frac{1-\b}{1+\b}+2
\ln\frac{1-\b}{1+\b}\ln\frac{2\b}{1+\b}+2\dilog\frac{1-\b}{1+\b}
+\frac{2}{3}\pi^2 \right) \right] \, ,
\eea
and
\be
d\widetilde{\s}^v = \frac{8\p^2\a^2}{s_{12}} Q_q^2 
\left( \frac{\a_s}{2\pi} \right)
\left[ 4NC_F \frac{\b^2-1}{\b} \ln\frac{1-\b}{1+\b}\left(\frac{m^2}{s_{12}}
-\frac{t_{13}^{\prime}t_{14}^{\prime}}{s_{12}^2} \right) \right] d\G_2 \, .
\ee
Observe that the sum of the soft and ultra-violet 
renormalized virtual terms is finite,
$A_1^s + A_1^v = 0$, as required \cite{kln}.
We are therefore free to return to 4 dimensions with the finite remainders 
$A_0^s+A_0^v$.  The final result for the $\co (\a_s)$ correction 
consists of two contributions to the cross section: a two-body 
term $\s^{(2)}$ and a three-body term $\s^{(3)}$ where
\be
\s^{(2)}= \int \left[ d\s_0 \left( \frac{\a_s}{2\pi} \right) 
\left(A_0^s +A_0^v\right) +d\widetilde \s^v\right] \, ,
\ee
and
\be
\s^{(3)} = \s_H = \frac{1}{2s_{12}}\int_H \sumb |M_3|^2 d\G_3 \, .
\ee
\begin{figure}
\centerline{\hbox{\epsfig{figure=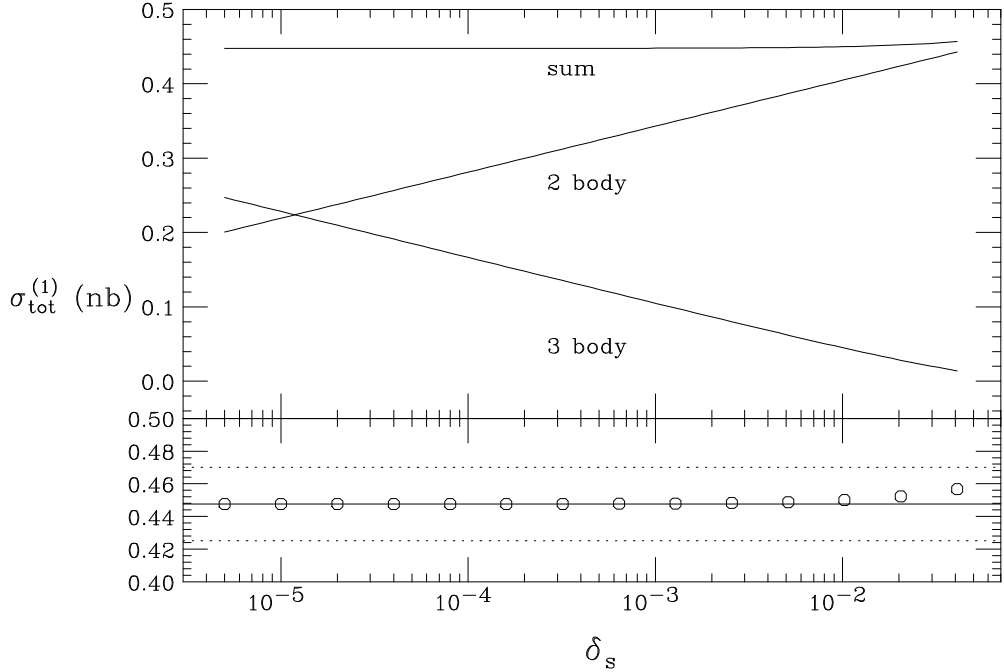,width=3.5in,angle=90}}}
\caption[]{The next-to-leading order contribution to the total cross 
section for producing a massive quark pair in 
electron-positron annihilation via single photon exchange.  The two- and 
three-body contributions together with their sum are shown as a function 
of the soft cutoff $\d_s$.  
The bottom enlargement shows the sum (open circles) 
relative to $\pm5\%$ (dotted lines) of the analytical result (solid line) 
given in Eq.\ (\protect{\ref{eqn:anal_massive}}).}
\label{fig:delta_massive}
\end{figure}
As a check, these results may be integrated to give a total 
rate and compared against the known next-to-leading order 
result \cite{zerwas} taken from \cite{rn}:
\bea
\s_{\rm tot}^{(1)} 
& = & \s_0 \frac{\a_s}{4\p} C_F 
      \left[ (32-8\r^2)\dilog{(t)} + (16-4\r^2){\rm F}_3(t) \right. \NO \\
& + & (2+\r)\sqrt{1-\r}\;{\rm F}_4(t) + (8-2\r^2)\ln(t)\ln(1+t) \NO \\
& + & \left. (-12+2\r+7\r^2/4)\ln(t) + (3+9\r/2)\sqrt{1-\r} \right] \, ,
\label{eqn:anal_massive}
\eea
where $\s_0$ is the leading order cross section for producing a pair of 
massless quarks given by
\be
\s_0 = \frac{4\p\a^2}{3s_{12}} N Q_q^2 \, ,
\label{eqn:lo_massless}
\ee
and we have also defined 
\be
t = \frac{1-\sqrt{1-\r}}{1+\sqrt{1-\r}} \, ,
\ee
with $\r=4m^2/s_{12}$ and
\bea
{\rm F}_3(t) & = & \dilog{(-t)}+\ln(t)\ln(1-t) \NO \\
{\rm F}_4(t) & = & 6\ln(t)-8\ln(1-t)-4\ln(1+t) \, .
\eea
The next-to-leading order corrections are shown in 
Fig.\ \ref{fig:delta_massive} for $\sqrt{s}=11$ GeV and $m=m_b=5$ GeV.  
The two- and three-body contributions together with their sum are 
shown as a function of the soft cutoff $\d_s$.  The bottom enlargement 
shows the sum (open circles) relative to $\pm5\%$ (dotted lines) of the 
analytical result (solid line) given in Eq.\ (\ref{eqn:anal_massive}).
The result quickly converges to the known result.

It is satisfying that the fully inclusive rate from the slicing method 
agrees with that from Ref.\ \cite{rn}.  Having made this necessary 
check, the results may be used to histogram a wide variety observables and 
to study various physics issues.  We refrain from any such studies here, 
and instead pass to our second example.

\subsection{Electron-positron annihilation to massless quark pair}
\label{sec:massless_example}

The process to be studied in this section is similar to that of the 
last section, but with one key difference: the quarks are considered 
massless from the beginning.  Therefore, in addition to the final state 
soft singularities there are final state collinear singularities.
The leading order cross section
\be
d\s^0 = \frac{1}{2s_{12}} \sumb |M_2|^2 d\G_2 \, ,
\ee
is expressed in terms of the (summed and averaged) 
matrix element squared 
\be
\sumb |M_2|^2 = 2 N e^4 Q_q^2 \left( 
    \frac{t_{13}^2+t_{23}^2}{s_{12}^2} 
    - \e \right) \, ,
\ee 
calculated from Fig.\ \ref{ee_born}, 
and the two body phase space
\be
d\G_2 = \frac{2^{2\e}}{16\p} \left( \frac{4\p}{s_{12}} \right)^\e 
        \frac{1}{\G(1-\e)} \int_0^\p \sin^{1-2\e} \! \q \, d\q \, .
\ee
In four space-time dimensions, integration over the phase space 
produces the result shown previously in Eq.\ (\ref{eqn:lo_massless}). 

For the QCD corrections we need 
the soft cross section (\ref{eqn:soft}), 
the hard-collinear cross section (\ref{eqn:sig_HC}), 
the hard-non-collinear cross section (\ref{eqn:sig_HNC}), 
and the virtual contribution.
The real emission diagrams are shown in Fig.\ \ref{ee_real} where   
the quark lines are to be interpreted as massless.  The two-to-three 
body matrix element squared $|M_3|^2$ needed to evaluate the 
hard-non-collinear cross section Eq.\ (\ref{eqn:sig_HNC}) follows 
directly from Eq.\ (\ref{eqn:2to3}) of the previous example 
by setting $m=0$ and evaluating the traces in four space-time 
dimensions.
The soft cross section Eq.\ (\ref{eqn:soft}) may also be obtained 
from the results of the last example by setting $m=0$ in 
Eq.\ (\ref{eqn:soft_massive}) giving 
\be
d\s_{\rm S} =  d\s^0
         \left[ \frac{\a_s}{2\p} \frac{\G(1-\e)}{\G(1-2\e)} 
         \left( \frac{4\p\mu_r^2}{s_{12}} \right)^\e \right] 
         C_F \int \left(
         \frac{s}{p_3 \cdot p_5 \, p_4 \cdot p_5} \right) dS \; .
\ee
The pole needs to be integrated over the soft phase space 
according to Eq.\ (\ref{eqn:soft_PS}).  To this end, define 
\be
I(s_{35} s_{45}) = \int \frac{1}{s_{35} s_{45}} dS \, .
\ee
Using the massless ($\b=1$) form of Eq.\ (\ref{eqn:s35s45}), 
the energy integral Eq.\ (\ref{eqn:energy_int}), 
and the angular integrals given in Appendix B, the result is
\be
I(s_{35} s_{45}) = \frac{1}{2s_{12}} \left( \frac{1}{\e^2} 
- \frac{2}{\e}\ln \d_s +2\ln^2\d_s \right) \, .
\ee
We may therefore write the final expression for the soft cross section as
\be
d\s_{\rm S} = d\s^0 \left[ \frac{\a_s}{2\p}
\frac{\G(1-\e)}{\G(1-2\e)} \left( \frac{4\p\mu_r^2}{s_{12}} \right)^\e \right]
\left( \frac{A_2^s}{\e^2}+\frac{A_1^s}{\e}+A_0^s \right) \, ,
\ee
with
\bea
A_2^s &=& 2C_F \NO \\
A_1^s &=& -4C_F \ln\d_s \NO \\
A_0^s &=&4C_F \ln^2\d_s \, .
\eea

The final state hard collinear cross section was 
derived in Sec.\ \ref{sec:jets}.
The relevant splitting is $q \rightarrow qg$.  From 
Eq.\ (\ref{eqn:HC_final}) we have
\be
d\s^{q \rightarrow qg}_{\rm HC} = d\s^0 \left[ \frac{\a_s}{2\p}
\frac{\G(1-\e)}{\G(1-2\e)} \left( \frac{4\p\mu_r^2}{s_{12}} \right)^\e \right]
\left( \frac{A_1^{q \rightarrow qg}}{\e}+A_0^{q \rightarrow qg} \right) \, ,
\ee
with
\bea
A_1^{q \rightarrow qg} &=& C_F \left( 3/2 + 2 \ln\d_s \right) \NO \\
A_0^{q \rightarrow qg} &=& C_F \left[ 7/2 - \p^2/3 - \ln^2\d_s 
      - \ln\d_c \left( 3/2 + 2 \ln\d_s \right) \right] \, .
\eea

The interference of the one-loop diagrams in Fig.\ \ref{ee_oneloop} 
with the leading order diagram yields the virtual contribution.
In Fig.\ \ref{ee_oneloop}, diagrams 
$(b)$ and $(d)$ add to zero via the Ward identity.  Diagram $(c)$ 
vanishes for massless quarks.  This leaves diagram $(a)$, 
comprised of the vertex shown in Fig.\ \ref{vertex} 
evaluated for massless quarks.  The result for the vertex is
\be
\G_\mu = \left( -ieQ_q \right) \d_{ij} 
   \overline{u}(p_2) \g_\mu u(p_1) \g(q^2) \, ,
\label{eqn:vertex_massless}
\ee
with
\be
\g(q^2) = -\frac{\a_s}{2\p}C_F\left( \frac{4\p\mu_r^2}{-q^2} \right)^\e 
\frac{\G(1-\e)}{\G(1-2\e)} \left( \frac{1}{\e^2} + \frac{3}{2\e} + 4
+ \frac{\p^2}{6} \right) \, .
\ee
We may therefore write the final expression for the virtual 
contribution to the cross section as
\be
d\s_{\rm V} = d\s^0 \left[ \frac{\a_s}{2\p}
\frac{\G(1-\e)}{\G(1-2\e)} \left( \frac{4\p\mu_r^2}{s_{12}} \right)^\e \right]
\left( \frac{A_2^v}{\e^2}+\frac{A_1^v}{\e}+A_0^v \right) \, ,
\ee
with
\bea
A_2^v &=& -2C_F \NO \\
A_1^v &=& -3C_F \NO \\
A_0^v &=& -2C_F(4-\p^2/3) \, .
\eea

\begin{figure}
\centerline{\hbox{\epsfig{figure=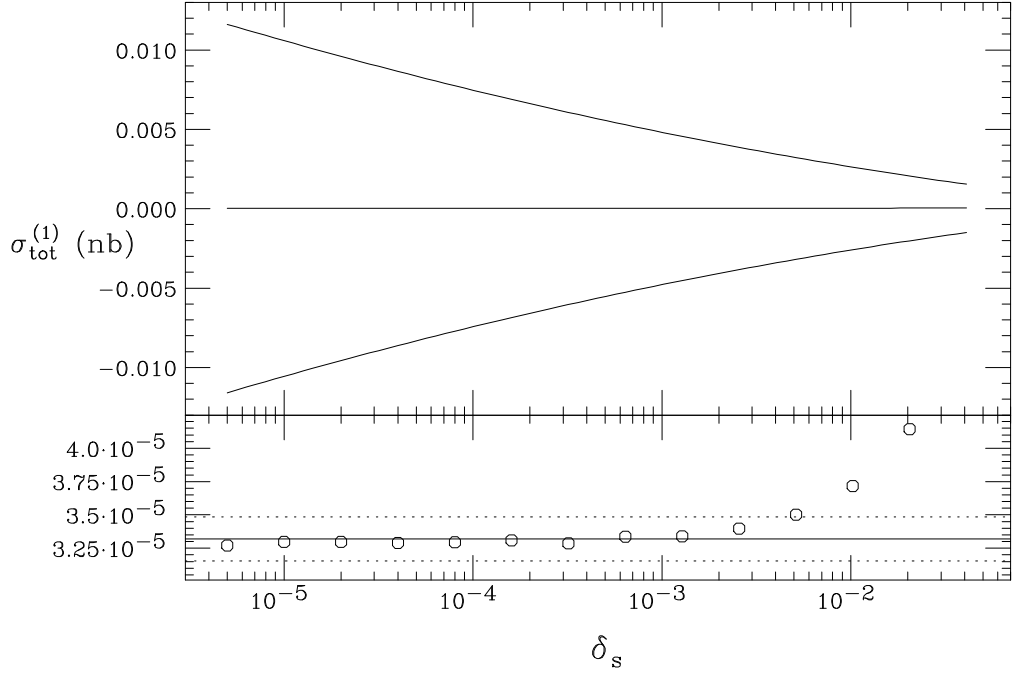,width=3.5in,angle=90}}}
\caption[]{The next-to-leading order contribution to the total cross 
section for producing a massless quark pair in electron-positron 
annihilation via single photon exchange.  The two-body (negative) 
and three-body (positive) contributions together with their sum are shown 
as a function of the soft cutoff $\d_s$ with the collinear cutoff 
$\d_c=\d_s/300$.  The bottom enlargement shows 
the sum (open circles) relative to $\pm5\%$ (dotted lines) of the 
analytical result (solid line) given in 
Eq.\ (\protect{\ref{eqn:anal_massless}}).}
\label{fig:delta_massless}
\end{figure}

\begin{figure}
\centerline{\hbox{\epsfig{figure=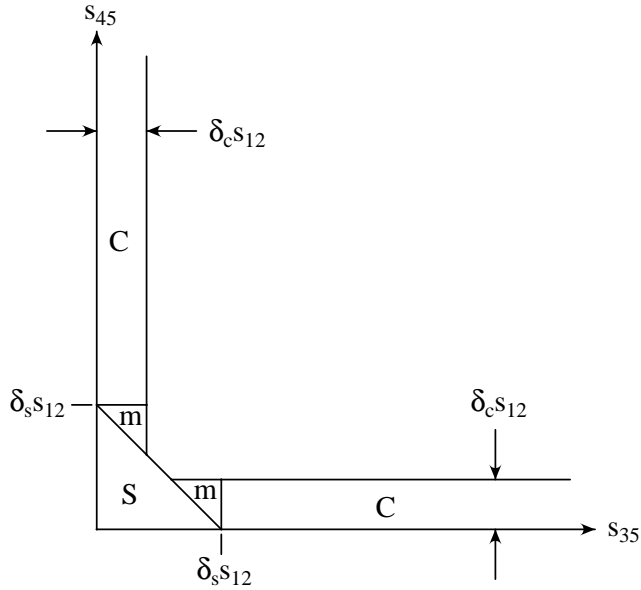,width=3.5in}}}
\caption[]{The $s_{35}-s_{45}$ plane for electron-positron annihilation 
to massless quarks showing the delineation into soft S and collinear 
C regions.  The triangles marked m give vanishing contribution for 
$\d_c \ll \d_s$.}
\label{slice}
\end{figure}

The full two-body weight is given by the sum $d\s_{\rm S}+d\s_{\rm V} + 
2 d\s^{q\rightarrow qg}_{\rm HC}$.  The factor of two occurs since 
there are two quark legs, either of which can emit a gluon.
At this point we have a finite result since $A_2^s+A_2^v=0$ and 
$A_1^s+A_1^v+2A_1^{q \rightarrow qg}=0$ as required \cite{kln}.
The finite two-body weight is given by 
\be
\s^{(2)} = \int d\s_0 \left( \frac{\a_s}{2\pi} \right) 
\left(A_0^s+A_0^v+2A_0^{q \rightarrow qg}\right) \,
\ee
while the three-body contribution is given by 
\be
\s^{(3)}= \s_{H\overline C}=\frac{1}{2s_{12}}\int_{H\overline C} \sumb 
|M_3|^2 d\G_3 \, .
\ee

A necessary check may be made by integrating these 
results and comparing with the known analytic answer.
The contributions from $\s^{(2)}$ (negative) and $\s^{(3)}$ (positive) 
and their sum are shown in Fig.\ \ref{fig:delta_massless} for 
$\sqrt{s}=500$ GeV as a function of the soft cutoff $\d_s$ with 
the collinear cutoff $\d_c=\d_s/300$.  The known result may be found in 
\cite{sterman}, for example, and is given by
\be
\s_{\rm tot}^{(1)} = \s_0 \frac{\a_s}{4\p} 3C_F  \, ,
\label{eqn:anal_massless}
\ee
where $\s_0$ is given in Eq.\ (\ref{eqn:lo_massless}).  
The bottom enlargement shows the sum (open circles) relative to 
$\pm5\%$ (dotted lines) of the known result (solid line) given 
in Eq.\ (\ref{eqn:anal_massless}).  Very good agreement is found 
below $\d_s \sim 2 \times 10^{-3}$.

Before proceeding further, it is instructive to examine some issues related to 
the cutoff dependence of this technique.  As shown in 
Fig.\ \ref{fig:delta_massless}, the answer converges to the known result 
for $\d_s < 10^{-3}$ when $\d_c=\d_s/300$.  We have imposed the 
requirement $\d_c \ll \d_s$ which may be understood by examining 
the nature of the three-body phase space for this case.  
Neglecting both initial and final 
state masses, four-momentum conservation yields $s_{12}=s_{34}+s_{35}+s_{45}$. 
The soft region is defined by $E_5 < \d_s \sqrt{s_{12}}/2$ which, 
taken with Eq.\ (\ref{eqn:e5}), can be recast as 
$s_{45} < \d_s s_{12} - s_{35}$.   This is shown as the region S in the 
plot of $s_{45}$ versus $s_{35}$ in Fig.\ \ref{slice}.  Two collinear 
regions defined by the constraints $s_{35}$ or $s_{45} < \d_c s_{12}$ are 
shown as the regions labeled C in Fig.\ \ref{slice}.  There are two 
small regions labeled ``m'' which are properly included in the collinear 
regions C.  However, using a fixed upper limit of $1-\d_s$ in 
calculating the hard collinear contributions 
(cf.\ Eq.\ (\ref{eqn:hard_limit})) these regions are excluded.  
They are also not included in the 
hard-non-collinear three-body integrations.  With some effort, it is 
possible to analytically evaluate the required integrals (\ref{eqn:dsHC}) 
over the m regions.  The result (derived in Appendix C) is 
that occurrences of $\ln\d_c\ln\d_s$ in Eq.\ (\ref{eqn:A0q}) 
are to be replaced by $\ln\d_c\ln\d_s - \dilog{(\d_c/\d_s)}$.  
From the properties of the dilogarithm 
function  we note that the correction 
term vanishes like $\d_c/\d_s$ in the limit of small $\d_c$.
Accordingly, the contributions from the regions denoted by m in 
Fig.\ \ref{slice} may be made negligible by 
requiring $\d_c \ll \d_s$.
Of course, as $\d_c$ and $\d_s$ become smaller the 
statistical errors on the sum of the two- and three-body weights increase.  
In practice, one must compromise between the errors induced by larger cutoffs 
and the statistical errors. 
For many calculations it has been found that choosing 
$\d_c$ to be 50 - 100 times smaller than $\d_s$ is sufficient for answers 
accurate to a few percent. Acceptable ranges for $\d_s$ must be determined 
on a case by case basis, as illustrated by the examples shown here. 
Furthermore, the sign of the deviations as $\d_s$ grows differs from 
process to process.

\subsection{Electron-positron annihilation to photons}

In this section we consider an example fragmentation process, 
inclusive photon production in hadronic final states of 
electron-positron annihilation, 
calculated to leading order in the electromagnetic coupling $\a$.
For pedagogical purposes only the radiation of photons from the final state
quark or antiquark will be included, {\it i.e.},\ initial state radiation 
will be neglected.
This process is different from the previous two examples in 
that there are final state collinear singularities only, and they 
are removed through the factorization procedure.

\begin{figure}
\centerline{\hbox{\epsfig{figure=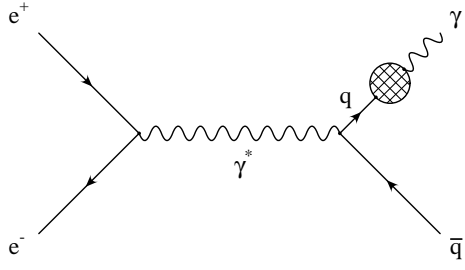,width=2.5in}}}
\caption[]{Leading order non-perturbative contribution 
to inclusive photon production via photon exchange.}
\label{photon_had}
\end{figure}
The diagram for the leading order non-perturbative contribution 
is shown in Fig.\ \ref{photon_had}.  The cross section from 
Eq.\ (\ref{eqn:lo_f}) is
\be
d\s^{e^+e^- \rightarrow \gamma X}_0 =\sum_q dz D_{\g/q}(z) 
d\s^{e^+e^- \rightarrow q \overline{q}}_0 \, .
\ee
To simplify notation it is helpful to write the 
Born-level total cross section for $e^+e^- \rightarrow q \overline{q}$ 
in terms of that for $e^+e^- \rightarrow \m^+ \m^-$
\be
\s^{e^+e^- \rightarrow q \overline{q}}_0=NQ_q^2 
\s^{e^+e^- \rightarrow \m^+ \m^-}_0 \, .
\ee
We further denote $\s^{e^+e^- \rightarrow \gamma X}_0$ by $\s$,
and $\s^{e^+e^- \rightarrow \m^+ \m^-}_0$ by $\s_{\m\m}$. 
Taking into account $D_{\g/q}(z) =  D_{\g/\overline{q}}(z)$, and 
letting $f$ denote the quark flavor, we arrive at the 
result for the leading order non-perturbative contribution
\be
\label{eqn:photonNP}
\frac{1}{\s_{\m\m}} \frac{d\s}{dz} = 2N\sum_{f=1}^{n_f} 
Q_f^2 D_{\g/f}(z) \, .
\ee

\begin{figure}
\centerline{\hbox{\epsfig{figure=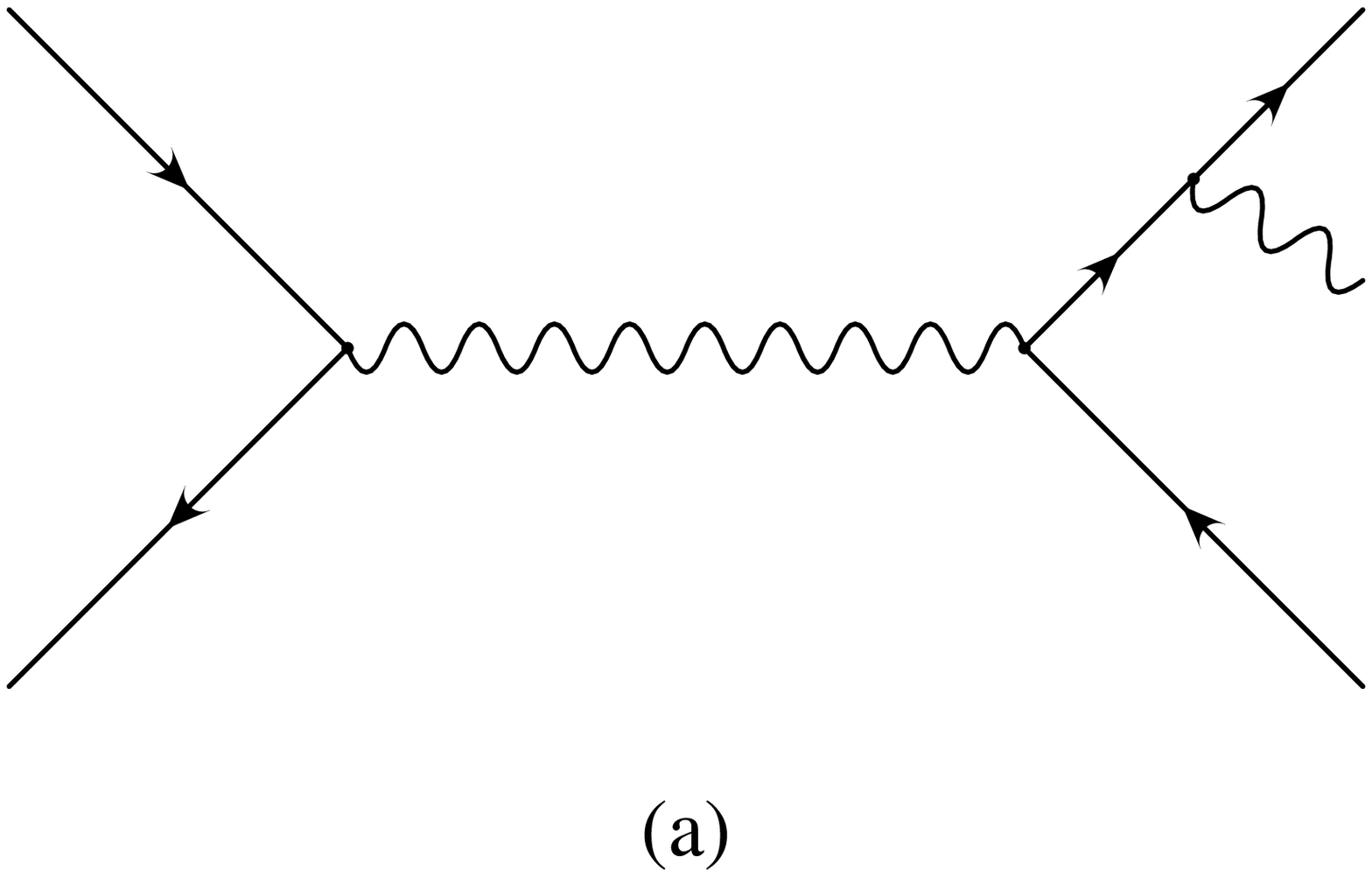,width=2.5in}
                  \epsfig{figure=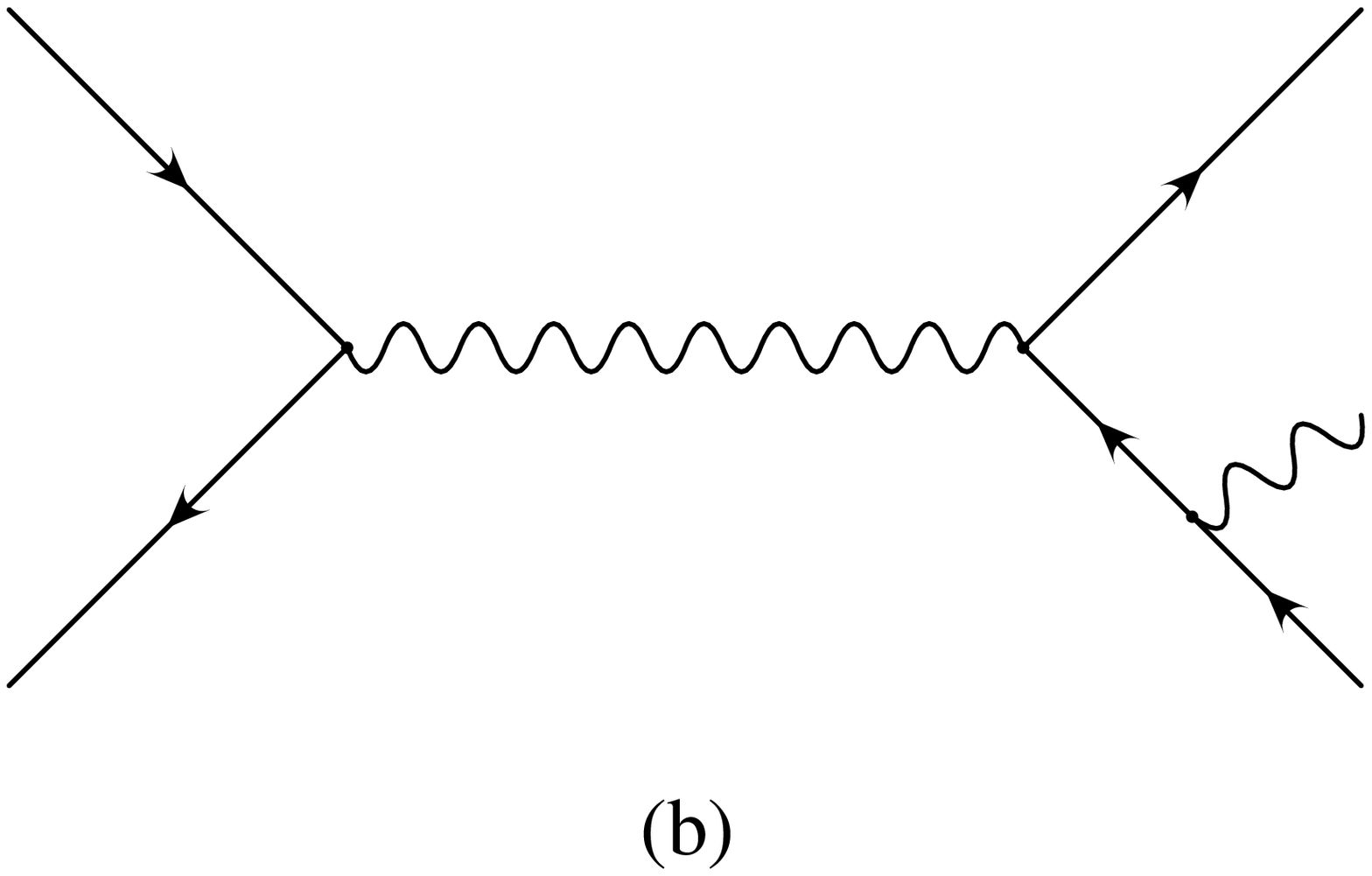,width=2.5in}}}
\caption[]{Leading order perturbative contribution 
to inclusive photon production via photon exchange.}
\label{photon_real}
\end{figure}

Additionally, there are two- and three-body pieces that 
make up the leading order perturbative contribution.
The Feynman diagrams are shown in Fig.\ \ref{photon_real}.
Because there are only final state collinear singularities 
present, the relevant decomposition of the two-to-three 
contribution to the cross section is into collinear and 
non-collinear terms, $\s=\s_{\rm C}+\s_{\rm \overline{C}}$. 
The collinear term $\s_{\rm C}$ is handled as 
discussed in Sec.\ \ref{sec:hadrons}.  In this case
there are no soft singularities so the soft-collinear terms $A^{sc}$ 
are not present.  The 
two-body piece follows from Eq.\ (\ref{eqn:frag_twobody})
\be
\frac{1}{\s_{\m\m}} \frac{d\s^{(2)}}{dz} = 2N \frac{\a}{2\p}
\sum_{f=1}^{n_f} Q_q^4 \widetilde{D}_{\g/f}(z,M_f) \, ,
\ee
once the replacement $\a_s \rightarrow \a$ is made.
$\widetilde{D}_{\g/f}(z,M_f)$ is as given in Eq.\ (\ref{eqn:Dtilde}) 
and may be expanded using $D_{\g/\g}(x)=\d(1-x)+{\cal O}(\a^2)$ and 
$D_{\g/i}(x)={\cal O}(\a)$ for $i=g,q$.  The leading term in $\a$ is therefore
\be
\widetilde{D}_{\g/q}(z,M_f) = \int_z^1 \frac{dy}{y} \d(1-z/y) 
\widetilde{P}^{\rm frag}_{\g q}(y) =  \widetilde{P}^{\rm frag}_{\g q}(z) \, ,
\ee
where $\widetilde{P}^{\rm frag}$ is as given in Eq.\ (\ref{eqn:Ptilde}) with
\be
P_{\g q}(z,\e)=\frac{1+(1-z)^2-\e z^2}{z} \, .
\ee
The final result for the two-body piece of the leading order 
perturbative contribution is
\be
\label{eqn:photon2}
\frac{1}{\s_{\m\m}} \frac{d\s^{(2)}}{dz} = 2N \frac{\a}{2\p}
\sum_{f=1}^{n_f} Q_f^4 \left\{ \frac{1+(1-z)^2}{z} \ln \left[
\frac{z(1-z)\d_cs}{M_f^2} \right] + z \right\} \, .
\ee

The complementary non-collinear piece $\s_{\rm \overline{C}}$ follows 
from the matrix element represented in Fig.\ \ref{photon_real}
\be
\s^{(3)} = \frac{1}{2s} \int_{\rm \overline{C}} \sumb |M_3|^2 d\G_3 
= \frac{4e^6}{3s^2} N \sum_{f=1}^{n_f} Q_f^4
\int_{\rm \overline{C}} d\G_3 \left( \frac{2s_{34}s}{s_{35}s_{45}}
+ \frac{s_{35}}{s_{45}} + \frac{s_{45}}{s_{35}} \right) \, .
\ee
We could now study the cutoff dependence numerically as in the 
previous examples.  However, the integration over phase space 
may be performed analytically and rather straightforwardly so 
we take this route to demonstrating the $\d_c$ independence of 
the full result.  To this end, consider, in the virtual photon rest frame, 
the three-body phase space
\be
d\G_3 = \frac{1}{8} \frac{1}{(2\p)^5} \frac{1}{E_3} \frac{d^3p_4}{E_4} 
\frac{d^3p_5}{E_5} \d(\sqrt{s}-E_3-E_4-E_5) \, .
\ee
Let $q^{\m}$ denote the virtual photon four-momentum.
Taking $\vec{p}_5$ along the $z$ axis and defining $x_4=2E_4/\sqrt{s}$, 
$z=2E_5/\sqrt{s}$ we may write the four-momenta as
\bea
q &=& \sqrt{s} (1,0,0,0) \\
p_5 &=& z \frac{\sqrt{s}}{2} (1,0,0,1) \\
p_4 &=& x_4 \frac{\sqrt{s}}{2} (1,\sin\q,0,\cos\q) \, .
\eea
Momentum conservation $\vec{q}=\vec{p}_3+\vec{p}_4+\vec{p}_5$ gives
\be
p_3 = (E_3,-x_4\frac{\sqrt{s}}{2}\sin\q,0,
-x_4\frac{\sqrt{s}}{2}\cos\q-z\frac{\sqrt{s}}{2}) \, .
\ee
The mass-shell condition $p_3^2=0$ can be used to fix $E_3$ as
\be
E_3 = \frac{\sqrt{s}}{2} \sqrt{ x_4^2+z^2+2x_4z\cos\q} \, .
\ee
Writing $d^3p_4=2\p\,d\!\cos\q E_4^2 dE_4$ and $d^3p_5=4\p E_5^2 dE_5$
the phase space delta function may be used to perform 
the $\cos\q$ integral.  The invariants $s_{34}$, $s_{35}$, and $s_{45}$ 
may then be written in terms of $x_4$ and $z$
\bea
s_{35} &=& s(1-x_4) \NO \\
s_{45} &=& s(z+x_4-1) \NO \\
s_{34} &=& s(1-z) \, .
\eea
The three-body piece is now
\be
\frac{1}{\s_{\m\m}} \frac{d\s^{(3)}}{dz} = N \frac{\a}{2\p}
\sum_{f=1}^{n_f} Q_f^4 \int_{\rm \overline{C}} dx_4 
\left[ \frac{2(1-z)}{(1-x_4)(z+x_4-1)}
+ \frac{1-x_4}{z+x_4-1} + \frac{z+x_4-1}{1-x_4} \right] \, .
\ee
This is to be integrated over the non-collinear ${\rm \overline{C}}$ 
region defined by $s_{45}>\d_c s$ and $s_{35}>\d_c s$ which is equivalent
to $1 - z + \d_c \le x_4 \le 1 - \d_c$.  The integral may easily be performed.
Dropping terms of ${\cal O}(\d_c)$, the 
final result for the three-body piece of the leading order 
perturbative contribution is
\be
\label{eqn:photon3}
\frac{1}{\s_{\m\m}} \frac{d\s^{(3)}}{dz} = 2N \frac{\a}{2\p}
\sum_{f=1}^{n_f} Q_f^4 \left[ \frac{1+(1-z)^2}{z} \ln \left(
\frac{z}{\d_c} \right) - z \right] \, .
\ee

Adding the non-perturbative (\ref{eqn:photonNP}), perturbative two-body 
(\ref{eqn:photon2}), and perturbative three-body (\ref{eqn:photon3}) 
contributions 
we obtain the well known result \cite{photon1,photon2}
\be
\frac{1}{\s_{\m\m}} \frac{d\s}{dz} = 2N \sum_{f=1}^{n_f} Q_f^2 \left\{
D_{\g/f}(z) + \frac{\a}{2\p} Q_f^2 \frac{1+(1-z)^2}{z} \ln \left[ 
\frac{z^2(1-z)s}{M_f^2} \right] \right\} \, ,
\ee
which is independent of $\d_c$.

\subsection{Drell-Yan}

Our next example is that of the QCD
corrections to lepton pair production in hadron-hadron collisions 
which illustrates the method for handling initial state collinear 
singularities developed in Sec.\ \ref{sec:initial}.

\begin{figure}
\centerline{\hbox{\epsfig{figure=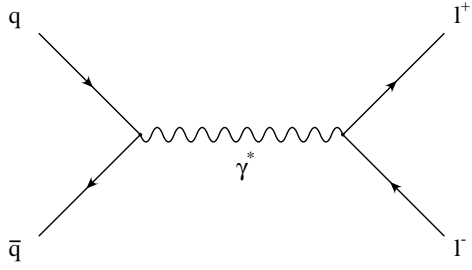,width=2.5in}}}
\caption[]{Leading order contribution to Drell-Yan production of a 
lepton pair via photon exchange.}
\label{dy_born}
\end{figure}

The leading order contribution mediated by a virtual photon is 
shown in Fig.\ \ref{dy_born}.  
The leading order partonic cross section 
\be
d\hat{\s}^0 = \frac{1}{2s_{12}} \sumb |M_2|^2 d\G_2 \, ,
\ee
is expressed in terms of the (summed and averaged) 
matrix element squared calculated in $n=4-2\e$ dimensions 
\be
\sumb |M_2|^2 = e^4 Q_f^2 \frac{2}{N} \left( 
    \frac{t_{13}^2+t_{23}^2}{s^2} - \e \right) \, ,
\ee
and the two-body phase space 
\be
d\G_2 = \frac{2^{2\e}}{16\p} \left( \frac{4\p}{s_{12}} \right)^\e 
        \frac{1}{\G(1-\e)} \int_0^\p \sin^{1-2\e} \! \q \, d\q \, .
\ee

For the QCD corrections there are five pieces to consider: 
the finite hard-non-collinear partonic cross section 
Eq.\ (\ref{eqn:sig_HNC}), the mass factorized 
hard-collinear cross section Eq.\ (\ref{eqn:sig_coll}), 
which consists of two pieces, the soft part of the initial 
state factorization counterterms and the mass factorization 
residuals (the $\widetilde{G}$ functions), 
the soft cross section Eq.\ (\ref{eqn:soft_final}), and 
the virtual corrections.

\begin{figure}
\centerline{\hbox{\epsfig{figure=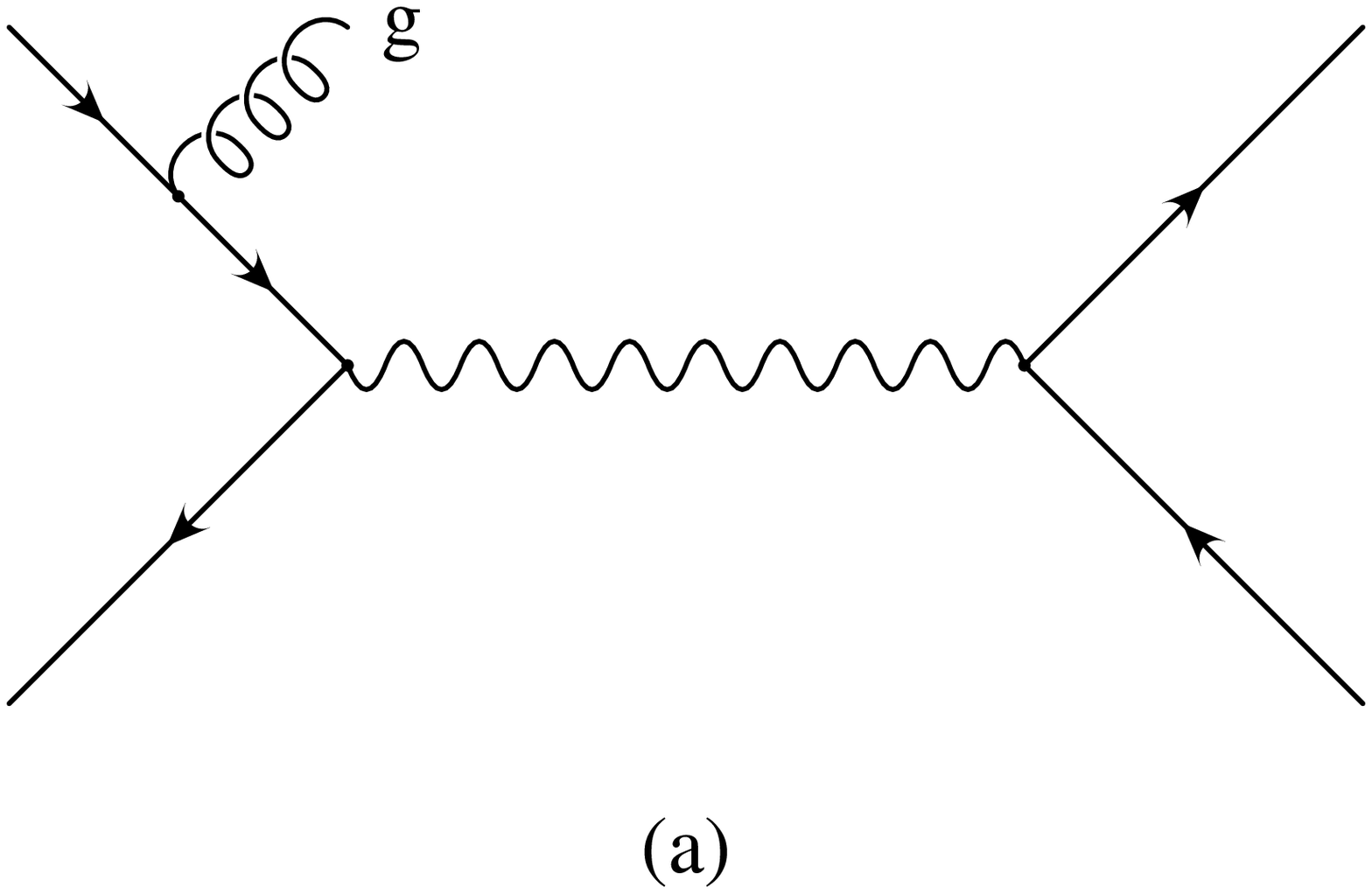,width=2.5in}
                  \epsfig{figure=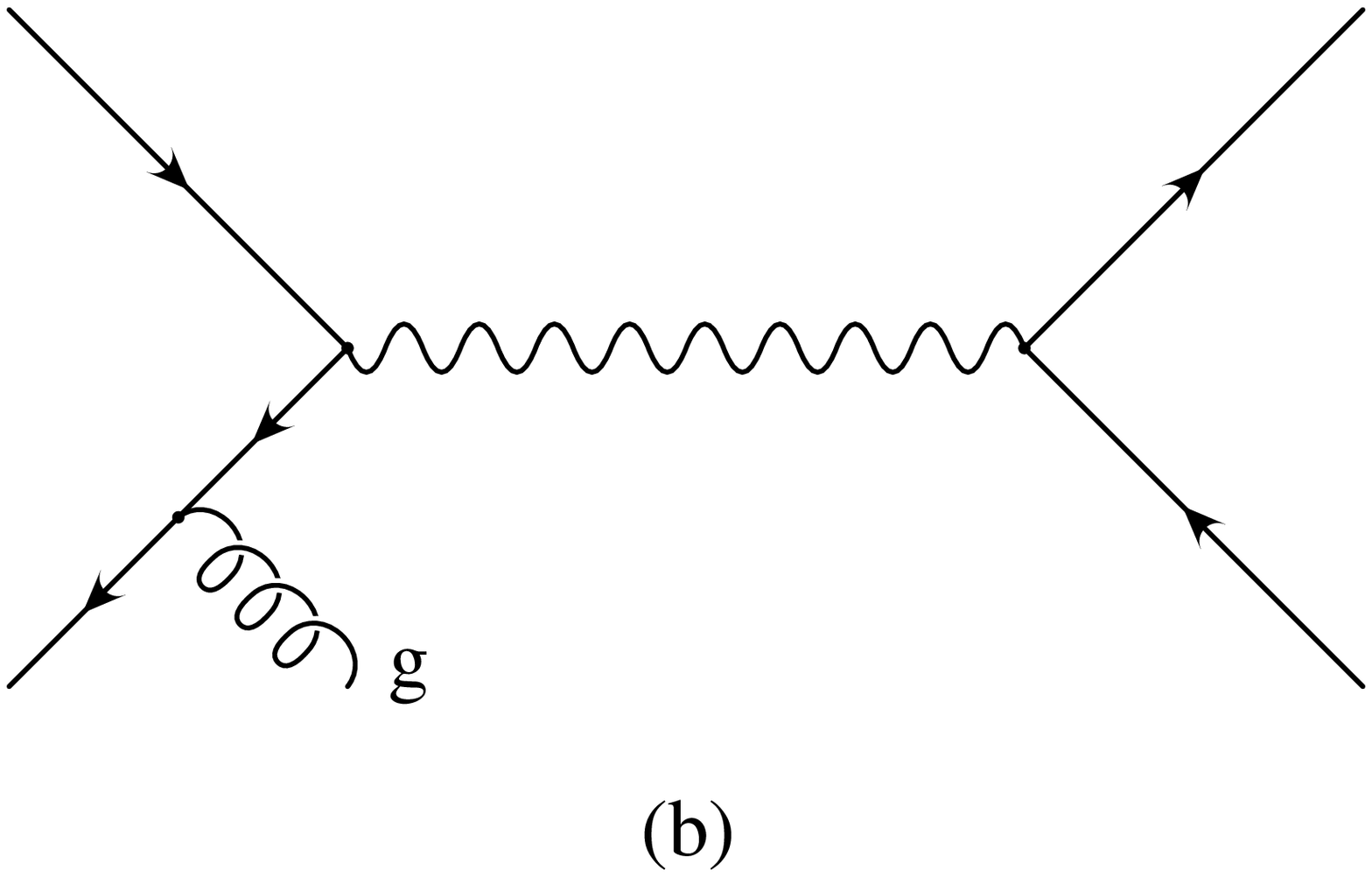,width=2.5in}}}
\caption[]{Real emission contribution to Drell-Yan production of a 
lepton pair via photon exchange.}
\label{dy_real}
\end{figure}

Shown in Fig.\ \ref{dy_real} are the real emission diagrams 
that give $|M_3|^2$.  Defining the 
(summed and averaged) matrix element squared as  
\be
\sumb |M_3|^2 = \frac{1}{4} e^4 Q_f^2 g^2 \frac{1}{s_{34}^2} 
\frac{C_F}{N} \J \, .
\label{2to3_massless_dy}
\ee
we find 
\bea
\J &=& {\rm tr} \left[\slash{p}_3\g^\m\slash{p}_4\g^\nu\right] \NO \\
&\times& \left\{ -{\rm tr} \left[ \slash{p}_2\g_\m(\slash{p}_1-\slash{p}_5)
\g^\s\slash{p}_1\g_\s(\slash{p}_1-\slash{p}_5)\g_\n\right]/
t_{15}^2 \right. \NO \\
& & + 2\, {\rm tr} \left[ \slash{p}_2\g^\s(\slash{p}_2-\slash{p}_5)\g^\m
\slash{p}_1\g_\s(\slash{p}_1-\slash{p}_5)\g_\n\right]/t_{15}t_{25} 
\NO \\
&  & - \left. {\rm tr} \left[ \slash{p}_2\g^\s(\slash{p}_2-\slash{p}_5)
\g^\m\slash{p}_1\g_\n(\slash{p}_2-\slash{p}_5)\g_\s\right]/
t_{25}^2 \right\} \, .
\label{eqn:psi_massless_dy}
\eea
The hard-non-collinear partonic cross section is obtained by 
evaluating the traces in four space-time dimensions.

There is a soft singularity when the gluon's energy goes to zero 
in the real emission diagrams.  This contributes to the 
soft cross section presented in Eq.\ (\ref{eqn:soft_final}).  
The sum runs over the initial state quark lines (labeled by 1 and 2). 
The color connected Born cross section $d\s_{12}^0=-C_Fd\s^0$. 
We find
\be
d\s_{\rm S} =  d\hat{\s}^0
         \left[ \frac{\a_s}{2\p} \frac{\G(1-\e)}{\G(1-2\e)} 
         \left( \frac{4\p\mu_r^2}{s_{12}} \right)^\e \right] 
         C_F \int \left(
         \frac{s}{p_1 \cdot p_5 \, p_2 \cdot p_5} \right) dS \; .
\ee
The integration of the pole over the soft phase space 
measure (\ref{eqn:soft_PS}) is written in terms of 
\be
I(t_{15} t_{25}) = \int \frac{1}{t_{15} t_{25}} dS \, .
\ee
In the $p_1p_2$ center-of-momentum system we take 
\bea
p_1 &=& \frac{\sqrt{s_{12}}}{2} (1, 0, \ldots, 0, 1) \NO \\
p_2 &=& \frac{\sqrt{s_{12}}}{2} (1, 0, \ldots, 0, -1) \, .
\eea
Using these together with Eq.\ (\ref{eqn:p5_soft}) we find
\bea
t_{15} &=& -2p_1\cdot p_5 = -\sqrt{s_{12}}E_5
      (1-\cos\q_1)
\NO \\
t_{25} &=& -2p_2\cdot p_5 = -\sqrt{s_{12}}E_5
      (1+\cos\q_1) \, .
\eea
Using the energy integral Eq.\ (\ref{eqn:energy_int}) and 
the angular integrals given in Appendix B, we find 
\be
I(t_{15} t_{25}) = \frac{1}{2s_{12}} \left( \frac{1}{\e^2} 
- \frac{2}{\e}\ln \d_s +2\ln^2\d_s \right) \, .
\ee
We may therefore write the final expression for the soft cross section as
\be
d\s_{\rm S} = d\hat{\s}^0 \left[ \frac{\a_s}{2\p}
\frac{\G(1-\e)}{\G(1-2\e)} \left( \frac{4\p\mu_r^2}{s_{12}} \right)^\e \right]
\left( \frac{A_2^s}{\e^2}+\frac{A_1^s}{\e}+A_0^s \right) \, ,
\ee
with
\bea
A_2^s &=& 2C_F \NO \\
A_1^s &=& -4C_F \ln\d_s \NO \\
A_0^s &=&4C_F \ln^2\d_s \, .
\eea

\begin{figure}
\centerline{\hbox{\epsfig{figure=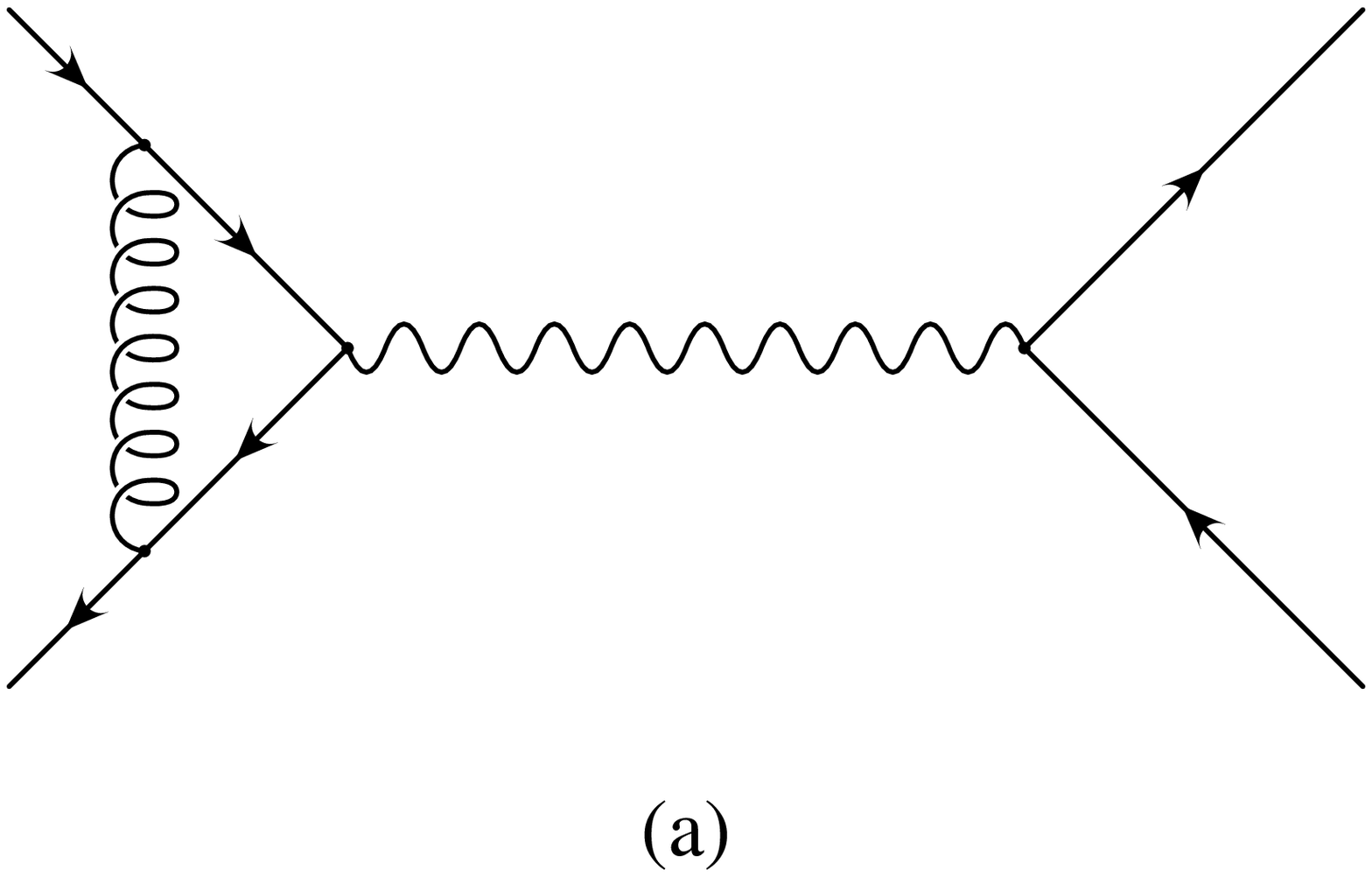,width=2.5in}
                  \epsfig{figure=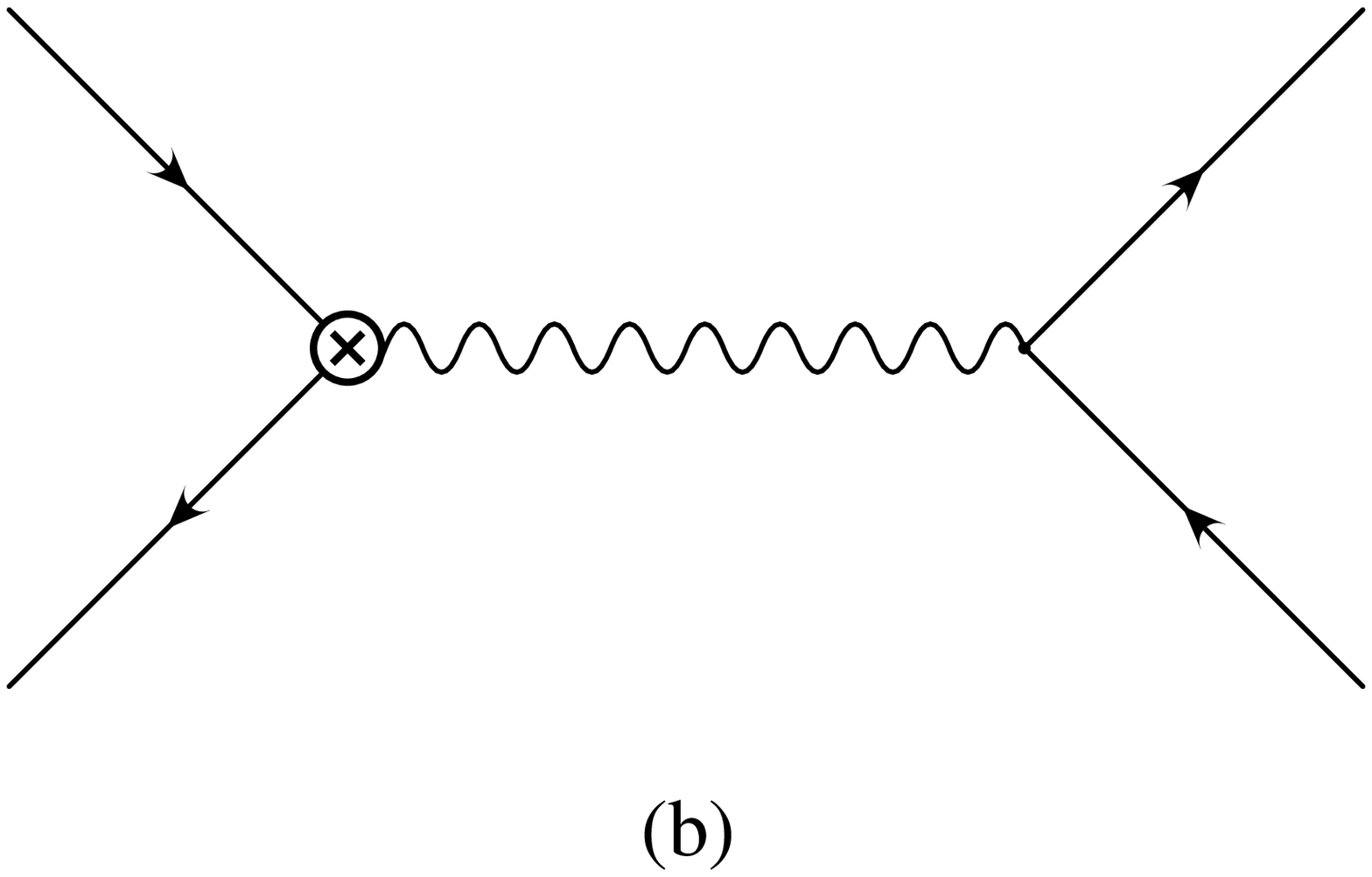,width=2.5in}}}
\centerline{\hbox{\epsfig{figure=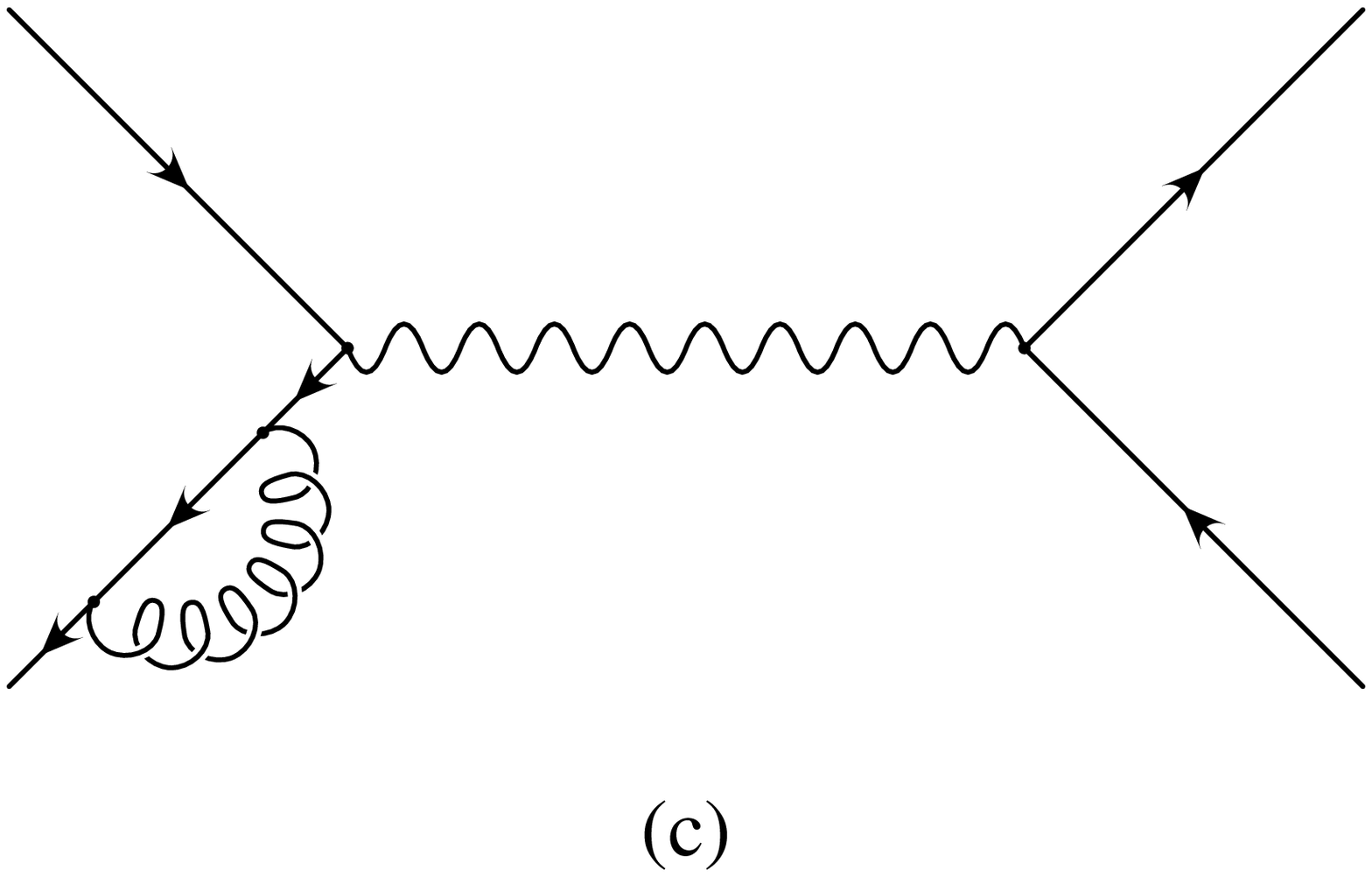,width=2.5in}
                  \epsfig{figure=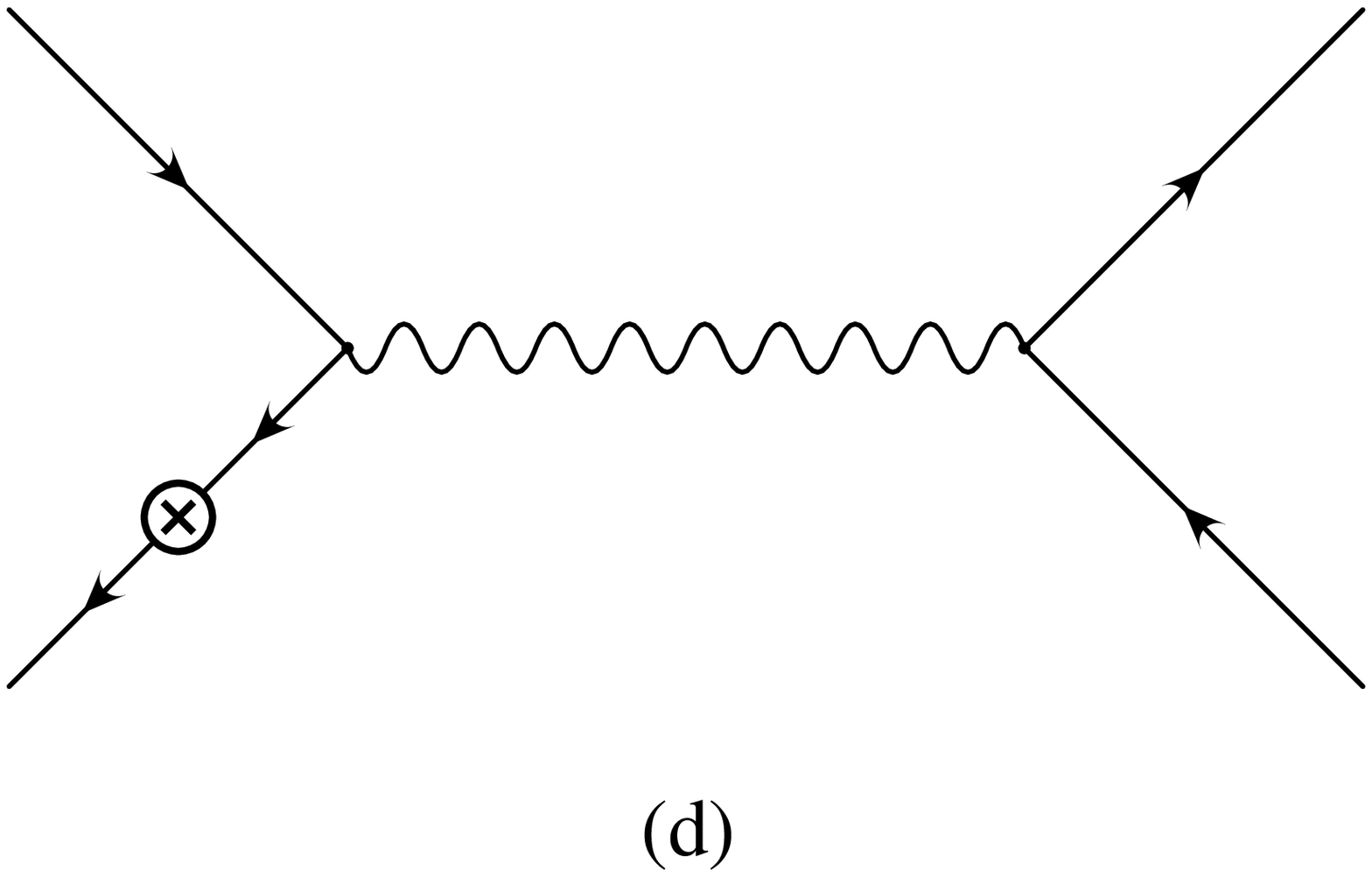,width=2.5in}}}
\caption[]{Loop and counterterm corrections to Drell-Yan production of a 
lepton pair via photon exchange.}
\label{dy_oneloop}
\end{figure}

Because the quarks are massless, there is a collinear 
singularity when the gluon becomes collinear to either of the 
initial state quark lines.  This singularity is removed through 
the factorization program described in Sec.\ \ref{sec:initial}.  
The soft-collinear pieces of the initial state 
factorization counterterms are given in Eq.\ (\ref{eqn:sig_coll}) as
\be
d\s^{\rm coll} = d\s^0 \left[ \frac{\a_s}{2\p}
\frac{\G(1-\e)}{\G(1-2\e)} \left( \frac{4\p\mu_r^2}{s_{12}} \right)^\e \right]
\left( \frac{A_1^{\rm sc}}{\e}+A_0^{\rm sc} \right) \, ,
\ee
with
\bea
A_1^{\rm sc} &=& C_F \left( 2 \ln \d_s + 3/2 \right) \NO \\
A_0^{\rm sc} &=& C_F \left( 2 \ln \d_s + 3/2 \right) 
\ln\frac{s_{12}}{\mu^2_f} \, .
\eea

The one-loop virtual diagrams are shown in Fig.\ \ref{dy_oneloop}.  
As in the case for electron-positron annihilation to a massless quark pair, 
diagrams $(b)$ and $(d)$ add to zero via the Ward identity, and diagram $(c)$ 
vanishes for massless quarks.  This leaves diagram $(a)$ for which 
the vertex shown in Fig.\ \ref{vertex} is needed, for massless quarks. 
The result is given in Eq.\ (\ref{eqn:vertex_massless}).
The final expression for the virtual cross section is
\be
d\s_{\rm V} = d\s^0 \left[ \frac{\a_s}{2\p}
\frac{\G(1-\e)}{\G(1-2\e)} \left( \frac{4\p\mu_r^2}{s_{12}} \right)^\e \right]
\left( \frac{A_2^v}{\e^2}+\frac{A_1^v}{\e}+A_0^v \right) \, ,
\ee
with
\bea
A_2^v &=& -2C_F \NO \\
A_1^v &=& -3C_F \NO \\
A_0^v &=& -2C_F(4-\p^2/3) \, .
\eea
At this point we pause to note that the two-body weight is 
finite: $A_2^s+A_2^v=0$ and
$A_1^s+A_1^v+2A_1^{\rm sc}=0$.  The factor of two occurs since 
there are two quark legs, either of which can emit a gluon.

\begin{figure}
\centerline{\hbox{\epsfig{figure=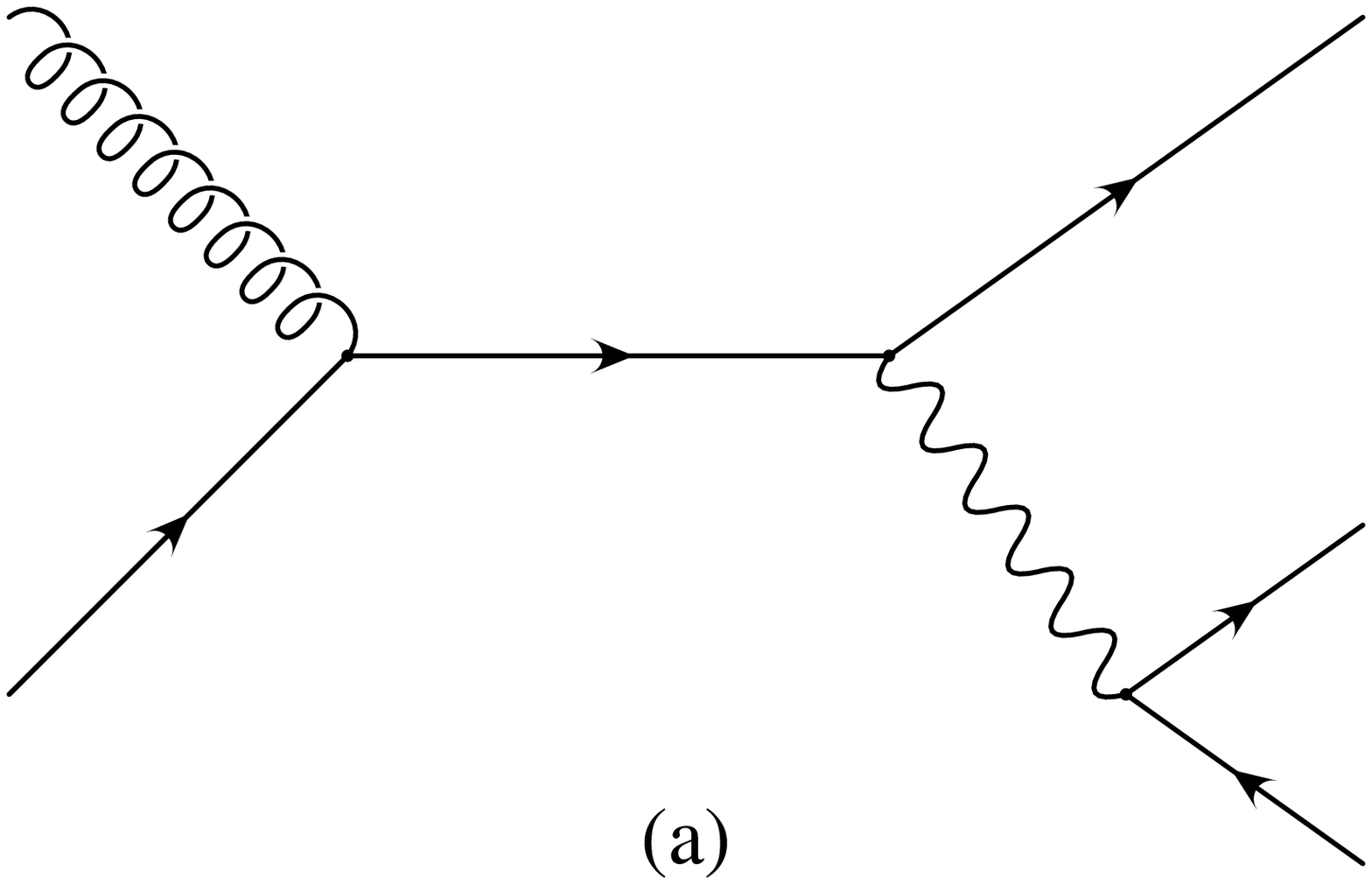,width=2.5in}
                  \epsfig{figure=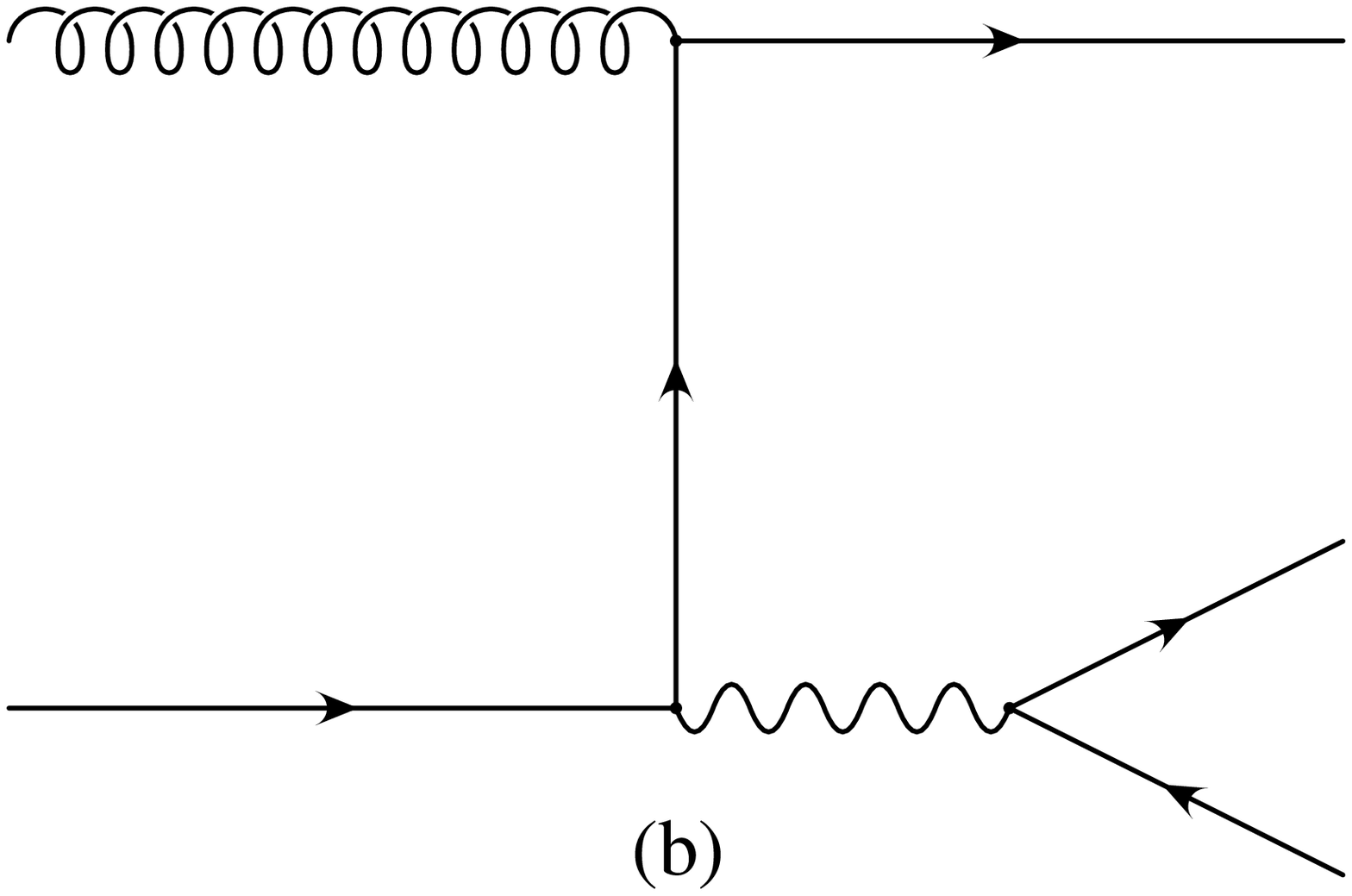,width=2.5in}}}
\caption[]{Quark-gluon initiated contribution to Drell-Yan production of a 
lepton pair.}
\label{dy_compton}
\end{figure}

In addition to the $q\overline{q}$ initiated processes, there are also
$qg$ initiated processes at this order of perturbation theory, as
shown in Fig.\ \ref{dy_compton}.  The singularities are initial state
collinear only in origin and arise from the $P_{qg}$ splitting in 
diagram $(b)$.  They are removed by factorization.  As the $P_{qg}(z)$ kernel 
is finite for $z=1$ there are no soft singularities.  
This implies the $A^{\rm sc}$ terms in Eq.\
(\ref{eqn:sig_coll}) are not present; only the finite $\widetilde{G}$
terms remain.

The final finite two-body cross section is given by the sum of the residual 
$\widetilde{G}$ terms from both the $q\overline{q}$ and $qg$ initiated 
processes 
and the finite two body weights from the $q\overline{q}$ process, 
$A_0^s+A_0^v+2A_0^{\rm sc}$.  The result, summed over all parton flavors, is
\bea
\s^{(2)} &=&\left( \frac{\a_s}{2\pi} \right)
\sum_f\int dx_A dx_B d\hat{\s}_0 \left[ G_{f/A}(x_A,\m_f) 
G_{\overline{f}/B}(x_B,\m_f) 
\left(A_0^s+A_0^v+2A_0^{\rm sc}\right) \right. \NO \\
&+& \left. G_{f/A}(x_A,\m_f) \widetilde{G}_{\overline{f}/B}(x_B,\m_f) 
+ \widetilde{G}_{f/A}(x_A,\m_f) G_{\overline{f}/B}(x_B,\m_f) 
+ (x_A \leftrightarrow x_B)\right] \, .
\eea
The $\widetilde{G}$ functions are given in Eq.\ (\ref{eqn:g_tilde}).
The three-body contribution is given by 
\be
\s^{(3)} = \sum_{i,j=q,g} \int dx_A dx_B G_{i/A}(x_A,\m_f) G_{j/B}(x_B,\m_f) 
d\hat{\s}_{ij} \, ,
\ee
with the hard-non-collinear partonic cross section given by
\be 
d\hat{\s}_{ij} = \frac{1}{2s_{12}}\int_{H\overline C} 
\sumb |M_3^{(ij)}|^2 d\G_3 \, .
\ee
Physical predictions follow from the sum $\s^{(2)}+\s^{(3)}$ which 
is cutoff independent for sufficiently small cutoffs.
The results may be integrated to obtain the total rate for 
$Q^2 > Q^2_{\rm min}$ and 
checked against the known $\co (\a_s)$ corrections
\bea
& & \s = \int^{S}_{Q^2_{\rm min}} dQ^2 
\int^1_{Q^2/S} dx_A \int^1_{Q^2/Sx_A} dx_B \Biggl\{
\sum_{ij=q,\bar{q}} G_{i/A}(x_A,\m_f) G_{j/B}(x_B,\m_f)
{d\hat{\s}_{qq}
\over d Q^2}
 \\ \nonumber \\ \nonumber
& & \quad +\sum_{i=q,\bar{q}}\Bigl[ G_{i/A}(x_A,\m_f) G_{g/B}(x_B,\m_f)
+G_{g/A}(x_A,\m_f )G_{i/B}(x_B,\m_f)\Bigr] {d\hat{\s}_{qg}
\over d Q^2}\Biggr\} \, ,
\eea
where $Q^2$ is the square of the lepton pair invariant mass and 
$S$ is the hadron-hadron center 
of mass energy squared which is related to $s_{12}$, 
the parton-parton center of mass energy squared, via $s_{12}=x_A x_B S$.  
Defining $z=Q^2/s_{12}$, the $\co (\a_s)$ hard scattering partonic 
subprocess cross sections are given by \cite{aem}
\begin{figure}
\centerline{\hbox{\epsfig{figure=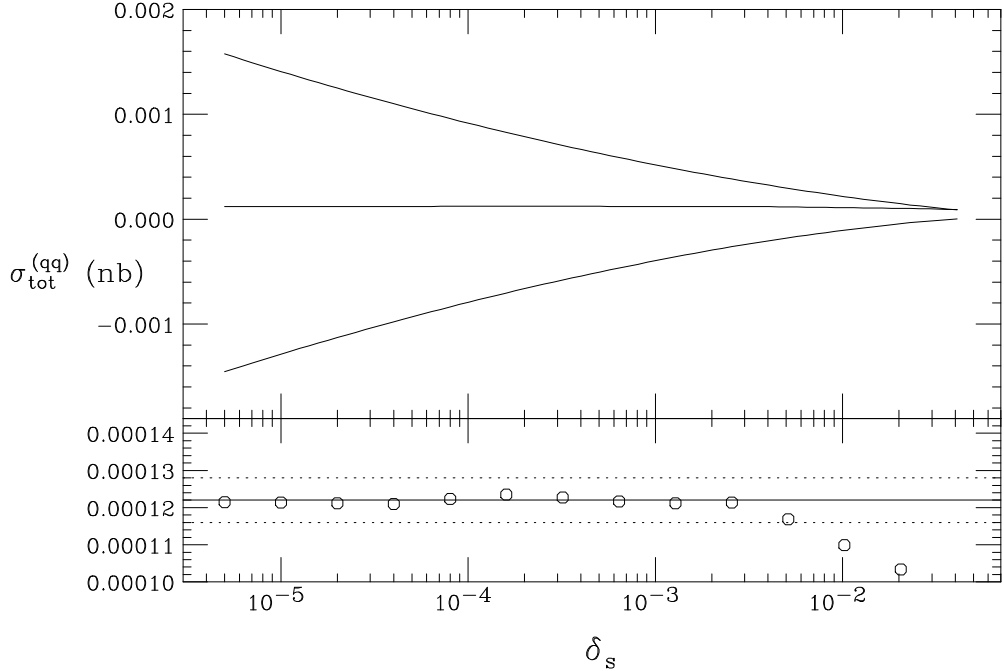,width=3.5in,angle=90}}}
\caption[]{The next-to-leading order quark-quark contribution to the 
Drell-Yan cross section.  The two-body (negative) and three-body (positive) 
contributions together with their sum are shown as a function of the soft 
cutoff $\d_s$ with the collinear cutoff $\d_c=\d_s/50$.
The bottom enlargement shows the sum (open circles) relative to $\pm5\%$ 
(dotted lines) of the analytic 
result (solid line) given in Eq.\ (\protect{\ref{eqn:analqq_dy}}).}
\label{fig:deltaqq_dy}
\end{figure}
\bea
{d\hat{\s}_{qq}\over dQ^2}
&=& \s_0 {\a_s\over 2\p} C_F \Biggl\{
4(1+z^2)\Biggl[{\ln (1-z)\over 1-z}\Biggr]_+ -2{(1+z^2)\over 1-z}\ln z
\NO \\
& & \quad\quad\quad\quad +\Bigl({2\p^2\over 3}-8\Bigr)\d (1-z)
+{3\over 2}P^+_{qq}(z)\ln {Q^2\over \m_f^2} \Biggr\} \, ,
\label{eqn:analqq_dy}
\eea
and
\bea
{d\hat{\s}_{qg}\over dQ^2}
&=& \s_0 {\a_s\over 2\p} \frac{1}{2} \Biggl\{ \frac{3}{2} + z 
- \frac{3}{2} z^2 + 2 P^+_{qg}(z) \Biggl[ \ln\frac{(1-z)^2}{z}-1 + 
\ln {Q^2\over \m_f^2} \Biggr] \Biggr\} \, ,
\label{eqn:analqg_dy}
\eea
where $\sigma_0$
\be
\s_0 = \frac{4\p\a^2Q_q^2}{3Ns_{12}Q^2} \, .
\ee
$P^+_{qq}(z)$ and $P^+_{qg}(z)$ are the splitting kernels given in 
Eqs.\ (\ref{eqn:pqq}) and (\ref{eqn:pqg}).

We show numerical results on the cutoff (in)dependence for 
proton-proton collisions at $\sqrt{S}=28.28$ GeV with $Q_{min}=10$ GeV.
Hard scales are set to the lepton pair mass $\m_f=\m_r=Q$ and the 
number of flavors taken to be $n_f=3$.

Shown in Fig.\ \ref{fig:deltaqq_dy} is 
the next-to-leading order quark-quark contribution to the 
Drell-Yan cross section.
The two- and three-body contributions to the cross section 
(negative and positive, respectively), and their 
sum are shown as a function of the soft cutoff $\d_s$.
The collinear cutoff $\d_c=\d_s/50$.
The bottom portion of the figure shows the sum (open circles) 
relative to $\pm5\%$ (dotted lines) of the analytic 
result (solid line) given in Eq.\ (\ref{eqn:analqq_dy}).

\begin{figure}
\centerline{\hbox{\epsfig{figure=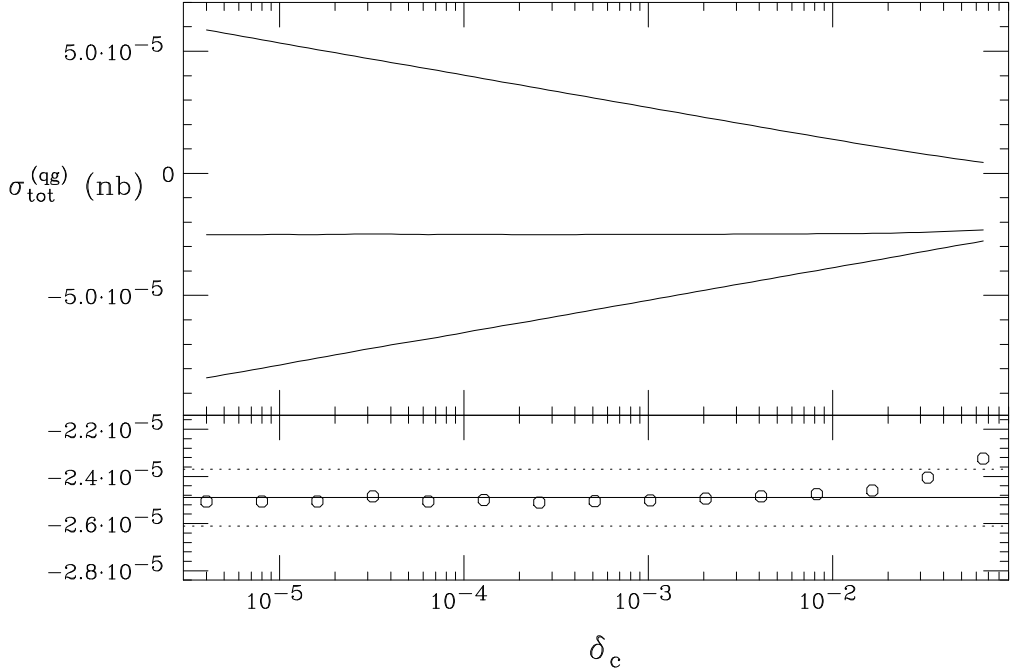,width=3.5in,angle=90}}}
\caption[]{The next-to-leading order quark-gluon contribution to the 
Drell-Yan cross section.  The two-body (negative) and three-body (positive) 
contributions together with their sum are shown as a function of the 
collinear cutoff $\d_c$.
The bottom enlargement shows the sum (open circles)
relative to $\pm5\%$ (dotted lines) of the analytic 
result (solid line) given in Eq.\ (\protect{\ref{eqn:analqg_dy}}).}
\label{fig:deltaqg_dy}
\end{figure}

Finally, we show the next-to-leading order quark-gluon contribution to the 
Drell-Yan cross section in Fig.\ \ref{fig:deltaqg_dy}.
The two- and three-body contributions and their 
sum are shown as a function of the 
collinear cutoff $\d_c$.
The bottom enlargement shows the sum (open circles)
relative to $\pm5\%$ (dotted lines) of the analytic 
result (solid line) given in Eq.\ (\ref{eqn:analqg_dy}).

In both cases, nice agreement is seen with the known analytic 
result, providing a cross check on the use 
of the two cutoff phase space slicing method.

\subsection{Single Particle Inclusive Cross Section}

Our final example is that of the single particle inclusive cross section 
in hadron-hadron collisions. The input needed for this calculation includes 
the squared matrix elements for the $2 \rightarrow 3$ subprocesses \cite{es} 
and the results for the ${\cal O}(\alpha_s^3)$ one-loop contributions to the 
$2 \rightarrow 2$ subprocesses \cite{ks,es}. For the purpose of this 
example, the notation of \cite{ks} will be used, since much of the input 
needed can be found in the appendices of that paper. The partons are 
labelled as $A + B \rightarrow 1 + 2$ \ and $ A + B \rightarrow 1 + 2 +3$ \ 
for the $2 \rightarrow 2$ \ and $2 \rightarrow 3$ subprocesses, respectively. 
A flavor label $a_A$ is used to denote the flavor of parton $A$, and 
similarly for the other partons. 

The lowest-order contribution to the inclusive cross section for producing 
a hadron $h$ in a collision of hadrons of types $A$\ and $B$ can be written  
as 
\begin{eqnarray} 
d\sigma^B &=& \frac{1}{2 x_A x_B s} \sum_{a_A, a_B, a_1, a_2}
G_{a_A/A}(x_A) G_{a_B/B}(x_B) D_{h/a_1}(z_1) dx_A\, dx_B\, dz_1\, \NO \\
& \times & \frac{(4 \pi \alpha_s)^2}{w(a_A) w(a_B)} \psi^{(4)}(\vec a,\vec p) 
d \Gamma_2 
\label{eqn:1pIBorn}
\end{eqnarray}
where $\vec a =\{a_A, a_B, a_1, a_2\}$\ and  $\vec p=\{p_A^{\mu},p_B^{\mu},
p_1^{\mu},p_2^{\mu}\}$ denote the sets of flavor indices and parton 
four-vectors, respectively. The factors appearing in the spin/color 
averaging are given by 
\[ w(a) = \left\{ \begin{array}{ll}
2(1-\epsilon) V & \mbox{a=gluon} \\
2 N & \mbox{a=quark or antiquark} 
\end{array}
\right. \]   
with $N=3$\ and $V=N^2-1.$ The factor $d \Gamma_2$ is the differential 
two-body phase space element from Eq.\ (\ref{eqn:ps2}).  
Eq.\ (\ref{eqn:1pIBorn}) 
gives the contribution where parton 1 fragments into the hadron $h$. Care must 
be taken to explicitly include in the sum over $\vec a$ those terms 
corresponding to the case where parton 2 fragments into $h$. For compactness, 
these terms will not be explicitly written. The squared matrix elements for 
the various subprocesses, denoted by $\psi^{(4)}(\vec a, \vec p)$, may be 
found in Ref.\ \cite{ks}.
 
Next, consider the one-loop virtual corrections to the $2 \rightarrow 2$ 
subprocesses. These take the form
\begin{eqnarray}
d\sigma^v &=& \frac{1}{2x_A x_B s} \sum_{a_A, a_B, a_1, a_2} 
G_{a_A/A}(x_A) G_{a_B/B}(x_B) D_{h/a_1}(z_1) dx_A\, dx_B\, dz_1 \NO \\
& \times &\frac{(4\pi \alpha_s)^2}{w(a_A) w(a_B)}\left[ 
\frac{\alpha_s}{2 \pi} 
\left(\frac{4 \pi \mu^2_R}{2 p_A \cdot p_B}\right)^{\epsilon}
\frac{\Gamma(1-\epsilon)}{\Gamma(1-2\epsilon)}\right]
\psi^{(6)}(\vec a, \vec p) d\Gamma_2
\end{eqnarray}
where
\begin{eqnarray}
\psi^{(6)}(\vec a, \vec p) &=& \psi^{(4)}(\vec a, \vec p )\left[ -\frac{1}
{\epsilon^2}\sum_n C(a_n) - \frac{1}{\epsilon }\sum_n \gamma(a_n) 
\right] \NO \\
& + & \frac{1}{2 \epsilon}\sum_{m,n \atop m\ne n}
\ln \left( \frac{p_m\cdot p_n}{p_A \cdot p_B}\right)
\psi^{(4,c)}_{m,n}(\vec a, \vec p) \NO \\
& - & \frac{\pi^2}{6}\sum_n \psi^{(4)}(\vec a, \vec p) +\psi^{(6)}_{NS}
(\vec a, \vec p) + {\cal O}(\epsilon ).
\end{eqnarray}
This expression for $\psi^{(6)}$ differs slightly from Eq.\ (35) in Ref.\ 
\cite{ks} because we have chosen to extract a different $\epsilon$ dependent 
overall factor: a factor of $\Gamma(1+\epsilon ) \Gamma(1-\epsilon )
\approx 1 + \epsilon^2 \frac{\pi^2}{6}$ has 
been absorbed into the above expression for $\psi^{(6)}$. Furthermore, the 
arbitrary scale $Q_{ES}^2$ used in Ref.\ \cite{ks} has, in order to match the 
conventions used elsewhere in the present work, been chosen to be 
$2p_A \cdot p_B$. The expressions for 
the functions $\psi^{(4,c)}_{m,n}$ \ and $\psi^{(6)}_{NS}$ may be found in 
Appendix B of Ref.\ \cite{ks}. The quantities $C(a_n)$ \ and $\gamma(a_n)$\ 
are given by
\[ C(a) = \left\{\begin{array}{ll}
N = 3 & \mbox{a=gluon} \\
C_F = \frac{4}{3} & \mbox{a=quark or antiquark} 
\end{array}
\right. \]   
and
\[ \gamma(a) = \left\{\begin{array}{ll}
(11 N -2 n_f)/6 & \mbox{a=gluon} \\
3 C_F/2 & \mbox{a=quark or antiquark} 
\end{array}
\right. \]   
It will be convenient for subsequent expressions to adopt the following 
notation:
\begin{equation}
{\cal F}=\left(\frac{4\pi \mu^2_R}{2 p_A \cdot p_B}\right)^{\epsilon} 
\frac{\Gamma(1-\epsilon)}{\Gamma(1-2\epsilon)}.
\end{equation}
The one loop virtual contributions can now be written as 
\begin{eqnarray}
d\sigma^v &=& \frac{1}{2x_A x_B s} \sum_{a_A, a_B, a_1, a_2} 
G_{a_A/A}(x_A) G_{a_B/B}(x_B) D_{h/a_1}(z_1) dx_A\, dx_B\, dz_1 \NO \\
& \times &\frac{(4\pi \alpha_s)^2}{w(a_A) w(a_B)} {\cal F} 
\frac{\alpha_s}{2 \pi} 
\left(\frac{A_2^v}{\epsilon^2} + \frac{A_1^v}{\epsilon} + 
A_0^v\right) d\G_2
\end{eqnarray}
where
\begin{eqnarray}
A_2^v &=& -\sum_n C(a_n) \psi^{(4)}(\vec a, \vec p)\\
A_1^v &=& -\sum_n \gamma(a_n) \psi^{(4)}(\vec a, \vec p) +\frac{1}{2}
\sum_{m,n \atop m\ne n}\ln \left(\frac{p_m \cdot p_n}{p_A \cdot p_B} \right)
\psi^{(4,c)}_{m,n}(\vec a, \vec p)\\
A_0^v &=& -\frac{\pi^2}{6}\sum_n C(a_n)\psi^{(4)}(\vec a,\vec p) + 
\psi^{(6)}_{NS}(\vec a,\vec p).
\end{eqnarray}

Next, the contributions from the $2 \rightarrow 3$ subprocesses in the 
limit where one of the final state gluons becomes soft are needed. The 
contributions of the $2 \rightarrow 3$ subprocesses may be written as 
\begin{eqnarray}
d\sigma^{2 \rightarrow 3} &=& \frac{1}{2 x_A x_B s} \frac{(4 \pi \alpha_s)^3}
{w(a_A) w(a_B)} {\cal F} \sum_{a_A, a_B, a_1, a_2, a_3} 
G_{a_A/A}(x_A) G_{a/B/B}(x_B) D_{h/a_1} (z_1) dx_A\, dx_B\, dz_1\, \NO \\
& \times & \Psi(a_A, a_B, a_1, a_2, a_3, p_A^{\mu},p_B^{\mu},p_1^{\mu},
p_2^{\mu},p_3^{\mu}) d\G_3.
\label{eqn:1PI2to3} 
\end{eqnarray}
The expressions for the $2 \rightarrow 3$ squared matrix elements appearing 
in Eq.\ (\ref{eqn:1PI2to3}) may be found in Ref.\ \cite{es}.
As noted earlier for the two-body contributions, one must include in the sum 
all possible parton to hadron fragmentations.

Consider the case where the soft gluon is parton 3. In this limit, the 
function $\Psi$ may be expanded as:
\begin{equation}
\Psi(a_A, a_B, a_1, a_2, a_3, p_A^{\mu},p_B^{\mu},p_1^{\mu},p_2^{\mu},
p_3^{\mu}) \sim \sum_{m,n \atop m < n} \delta_{a_3,g} \frac{p_m \cdot p_n}
{p_m \cdot p_3 p_n \cdot p_3} \psi^{(4,c)}_{m,n}(a_A, a_B, a_1, a_2, 
p_A^{\mu}, p_B^{\mu}, p_1^{\mu}, p_2^{\mu}).
\end{equation}

Next, one must integrate over the soft region of phase space defined by 
$E_3 < \delta_s \sqrt{2 p_A \cdot p_B}/2$.  This is easily done using the 
integrals given in the appendix. The resulting soft contribution may 
be written as 
\begin{eqnarray}
d\sigma^s &=& \frac{1}{2x_A x_B s} \sum_{a_A, a_B, a_1, a_2} 
G_{a_A/A}(x_A) G_{a_B/B}(x_B) D_{h/a_1}(z_1) dx_A\, dx_B\, dz_1 \NO \\
& \times &\frac{(4\pi \alpha_s)^2}{w(a_A) w(a_B)} {\cal F} 
\frac{\alpha_s}{2 \pi} 
\left(\frac{A_2^s}{\epsilon^2} + \frac{A_1^s}{\epsilon} + 
A_0^s\right) d\G_2
\end{eqnarray}
where
\begin{eqnarray}
A_2^s &=& \sum_n C(a_n) \psi^{(4)}(\vec a, \vec p)\\
A_1^s &=& - 2 \ln \delta_s \sum_n C(a_n) \psi^{(4)}(\vec a, \vec p) 
-\frac{1}{2} \sum_{m,n \atop m\ne n}\ln \left(\frac{p_m \cdot p_n}
{p_A \cdot p_B} \right)\psi^{(4,c)}_{m,n}(\vec a, \vec p)\\ 
A_0^s &=& 2 \ln^2 \delta_s \sum_n C(a_n) \psi^{(4)}(\vec a, \vec p) \NO \\
&+& \left( \psi^{(4,c)}_{A,1}+\psi^{(4,c)}_{B,2} \right) 
\left[ \frac{1}{2} \ln^2 
\left( \frac{p_1 \cdot p_3}{p_A \cdot p_B} \right) + \dilog \left( 
\frac{p_2 \cdot p_3}{p_A \cdot p_B} \right) + 2 \ln \delta_s \ln \left( 
\frac{p_1 \cdot p_3}{p_A \cdot p_B} \right) \right] \NO \\
&+& \left( \psi^{(4,c)}_{A,2}+\psi^{(4,c)}_{B,1} \right) 
\left[ \frac{1}{2} \ln^2 
\left( \frac{p_2 \cdot p_3}{p_A \cdot p_B} \right) + \dilog \left( 
\frac{p_1 \cdot p_3}{p_A \cdot p_B} \right) + 2 \ln \delta_s \ln \left( 
\frac{p_2 \cdot p_3}{p_A \cdot p_B} \right) \right]. 
\end{eqnarray}

After the collinear singularities associated with the two parton 
distribution functions and the fragmentation function have been factorized 
and absorbed into the corresponding bare functions, there will be 
soft-collinear terms left over due to the mismatch between the integration 
limits of the collinear singularity terms and the factorization counterterms. 
In addition, there can be collinear singularities associated with the 
non-fragmenting parton in the final state, corresponding to gluon emission or 
$q \overline q$ production. Collecting together both types of collinear 
terms, the result can be written as follows:
\begin{eqnarray}
d\sigma^{coll} &=& \frac{1}{2x_A x_B s} \sum_{a_A, a_B, a_1, a_2} 
G_{a_A/A}(x_A) G_{a_B/B}(x_B) D_{h/a_1}(z_1) dx_A\, dx_B\, dz_1 \NO \\
& \times &\frac{(4\pi \alpha_s)^2}{w(a_A) w(a_B)} {\cal F} 
\frac{\alpha_s}{2 \pi} 
\left(\frac{A_1^{coll}}{\epsilon} + A_0^{coll}\right) d\G_2
\end{eqnarray}
where
\begin{eqnarray}
A_1^{coll} &=& \sum_n [2\ln \delta_s C(a_n) + \gamma(a_n)]\\
A_0^{coll} &=& \sum_{A,B}[2\ln \delta_s C(a_n) +\gamma(a_n)]\ln \left(\frac
{2p_A \cdot p_B}{\mu^2_f}\right) +[2\ln \delta_s C(a_1) +\gamma(a_1)]\ln 
\left(\frac{2p_A \cdot p_B}{M^2_f}\right) \NO \\
&+& \gamma'(a_2).
\end{eqnarray}
Here $\mu_f \mbox{\ and \ } M_f$ are the initial and final state factorization 
scales. The function $\gamma'(a)$ is given in terms of the hard collinear 
factors of Eqs.\ (\ref{eqn:A1q}) -- (\ref{eqn:A0g}) as 
\begin{equation}
\gamma'(a) = \left\{\begin{array}{ll} A_0^{q \rightarrow qg} & 
\mbox{a= quark or antiquark} \\ A_0^{g \rightarrow gg} 
+ A_0^{g \rightarrow q\bar{q}} & \mbox{a = gluon}\end{array}\right.
\end{equation}

After the mass factorization has been performed, the bare parton distribution 
functions and fragmentation functions have been replaced by scale dependent 
$\overline {\mbox{MS}}$ functions. In addition, there are finite remainders 
involving the $\widetilde G \mbox{\ and \ } \widetilde D$ functions:
\begin{eqnarray}
d\widetilde \sigma &=& \frac{1}{2 x_A x_B s} \sum_{a_A, a_B, a_1, a_2}
\frac{(4 \pi \alpha_s)^2}{w(a_A) w(a_B)} \frac {\alpha_s}{2 \pi} 
dx_A dx_B dz_1 \j^{(4)}(\vec{a},\vec{p})
d\Gamma_2 \nonumber \\
& \times & \left[\widetilde G_{a_A/A}(x_A,\mu^2_f) G_{a_B/B}(x_B,\mu^2_f) 
D_{h/a_1}(z_1,M^2_f) \right. \nonumber \\
& + & G_{a_A/A}(x_A,\mu^2_f) \widetilde G_{a_B/B}(x_B,\mu^2_f) 
D_{h/a_1}(z_1,M^2_f) \nonumber \\ 
& + & \left. G_{a_A/A}(x_A,\mu^2_f) G_{a_B/B}(x_B,\mu^2_f) 
\widetilde D_{h/a_1}(z_1,M^2_f)\right].
\end{eqnarray}

At this point, all of the singular terms have been isolated as poles in 
$\epsilon$ or have been factorized and absorbed into the bare parton 
distribution and fragmentation functions. The $\epsilon$ dependent pole 
terms all cancel amongst each other:
\begin{eqnarray}
& & A_2^v+A_2^s = 0 \\
& & A_1^v + A_1^s + A_1^{coll} = 0.
\end{eqnarray}
The finite two-body contribution is given by
\begin{eqnarray}
d\sigma^{2\rightarrow 2} &=& d \sigma^B + d\widetilde \sigma  \NO \\
&+& \frac{1}{x_A x_B s} \sum_{a_A, a_B, a_1, a_2} 
\frac {(4 \pi \alpha_s)^2}{w(a_A) w(a_B)} G_{a_A/A}(x_A,\mu^2_f) 
G_{a_B/B}(x_B,\mu^2_f) D_{h/a_1}(z_1,M^2_f) \NO \\ 
& \times & \frac {\alpha_s}{2 \pi} \left[ A_0^v + A_0^s + A_0^{coll}\right]
dx_A\, dx_B\, dz_1\, d\Gamma_2.
\end{eqnarray}
The three-body contribution, now evaluated in four dimensions, was given in 
Eq.\ (\ref{eqn:1PI2to3}) where now the soft and collinear regions of phase 
space are excluded.

As in the previous examples, the structure of the final result is two finite 
contributions, both of which depend on the soft and collinear cutoffs -- 
one explicitly and one through the boundaries imposed on the three-body 
phase space. However, when both contributions are added while calculating an 
observable quantity, all dependence on the cutoffs cancels when sufficiantly 
small values of the cutoffs are used.

\section{Discussion and Conclusions}

A technique for performing next-to-leading-logarithm calculations
using Monte Carlo techniques was described in detail.  The method uses two
cutoff parameters which serve to separate the regions of phase space
containing the soft and collinear singularities from the non-singular
regions.  The main derivations for experimentally degenerate, tagged,
and heavy quark final states were given, as was a discussion of initial
state factorization.  We provided five illustrative examples applying
the method.

The first example was that of the QCD corrections to electron-positron
annihilation into a massive quark pair.  The quark mass serves to
regulate any would-be final state collinear singularities.  The final
state soft singular region is delineated using one cutoff.  The second
example was that of QCD corrections to electron-positron annihilation
into a massless quark pair.  In this case final state
soft and collinear singularities are encountered.  Both soft and
collinear cutoffs are therefore required.  They should be chosen such
that $\d_c \ll \d_s$.  The third example of inclusive photon
production in hadronic final states of electron-positron annihilation
was presented, illustrating the use of fragmentation functions.
Finally, the QCD corrections to lepton pair and single particle production in
hadron-hadron collisions were given.  These examples include both initial and 
final state
soft and collinear singularities.  The use of scale dependent
parton distribution and fragmentation functions was explained.

The Monte Carlo results, integrated to give an
inclusive cross section, were shown to be in complete agreement with
those available in the literature.  This is not the end of the 
utility of the method, but only the beginning.  Given the full access 
to the parton four vectors and corresponding weights, we are free to 
combine them in any way that is 
consistent with an infrared-safe measurement function, which may include 
a jet finding algorithm and experimental cuts.

The method has been applied to a wide range of hard scattering
processes and it has been found to be both simple to implement and
numerically robust.

\begin{acknowledgments}
We thank our collaborators Lewis Bergmann, Howie Baer, Jim Ohnemus, 
Bob Bailey, and Uli Baur, who have, over the years, 
applied and shaped the method presented herein, and Jack Smith 
for comments on the text.
This work was supported in part by the U.S. Department of Energy, 
High Energy Physics Division, under contracts DE-FG02-97ER41022 and 
W-31-109-Eng-38.
\end{acknowledgments}

\appendix

\section{Comparison with other methods}
The essential difference between the phase space slicing and the 
subtraction methods may be gleaned from the following 
simple example \cite{ks}.
Consider an integral to be calculated: 
\be
I = \lim_{\e \rightarrow 0^{+}} \left\{ \int_0^1 \frac{dx}{x} 
x^{\e} F(x) - \frac{1}{\e} F(0) \right\} \, ,
\ee
where $F(x)$ is a known but complicated function related to a two-to-three 
body matrix element.  
The variable $x$ represents either the energy of an emitted gluon 
or the angle between two massless partons.
In a traditional fully or single particle inclusive 
calculation the integral $I$ 
would be performed completely analytically.

In the subtraction method one simply adds and subtracts $F(0)$ under 
the integral sign.
\bea
I &=& \lim_{\e \rightarrow 0^{+}} \left\{ \int_0^1 \frac{dx}{x} 
x^\e 
\left[ F(x)-F(0)+F(0) \right] - \frac{1}{\e} F(0) \right\} 
\NO \\ 
&=& \int_0^1 \frac{dx}{x} \left[ F(x)-F(0) \right] \, ,
\eea
giving a finite and numerically calculable result.  No approximations are 
made, however in any numerical implementation there will necessarily be 
a lower limit related to machine precision below which the integral 
must be cutoff.  This is not a problem in practice.

In the phase space slicing method, the integration 
region is divided into two parts $0 < x < \d$ and $\d < x < 1$ 
with $\d \ll 1$.  A Maclaurin expansion of $F(x)$ yields 
\bea
I &=& \lim_{\e \rightarrow 0^{+}} \left\{ \int_0^{\d} \frac{dx}{x} 
x^{\e} F(x) + \int_{\d}^1 \frac{dx}{x} x^{\e} F(x) 
- \frac{1}{\e} F(0) \right\} \nonumber \\ 
&=& \int_{\d}^1 \frac{dx}{x} F(x) + F(0) \ln \d + \co(\d) \, .
\eea
Clearly, the parameter $\d$ must be chosen small enough so that 
the term linear in $\d$ may be neglected.  At the same time it must not be 
so small as to spoil the numerical convergence of the first term.

\section{Soft Integrals}
In evaluating the soft integrals we encounter angular integrals 
which may be written in the form
\be
I^{(k,l)}_n = \int_0^\p d\q_1 \sin^{n-3}\q_1 \int_0^\pi d\q_2 \sin^{n-4}\q_2
\frac{(a+b\cos\q_1)^{-k}}{(A+B\cos\q_1+C\sin\q_1\cos\q_2)^l} \, .
\ee
A large collection of these appear in the appendix of \cite{been}.  
Others may be found in the appendix of \cite{hs} or else 
computed as explained in \cite{willy}.  Here we collect together the 
results covering most of the cases encountered using the two cutoff 
slicing method.  The first two are from \cite{hs} with $A^2 \neq B^2+C^2$
\bea
I^{(0,1)}_n & = & \frac{\p}{\sqrt{B^2+C^2}} 
\left\{ \ln \left( \frac{A+\sqrt{B^2+C^2}}{A-\sqrt{B^2+C^2}} \right)
\right. \\ 
& & \left. -(n-4) \left[ \dilog\left( 
\frac{2\sqrt{B^2+C^2}}{A+\sqrt{B^2+C^2}} 
\right)+\frac{1}{4} \ln^2 \left( \frac{A+\sqrt{B^2+C^2}}{A-\sqrt{B^2+C^2}} 
\right) \right] \right\} \NO
\eea
\be
I^{(0,2)}_n = \frac{2\p}{A^2-B^2-C^2} \left[ 1 - \frac{1}{2} (n-4) 
\frac{A}{\sqrt{B^2+C^2}} \ln \left( 
\frac{A+\sqrt{B^2+C^2}}{A-\sqrt{B^2+C^2}} \right) \right] \, ,
\ee
where we drop $O((n-4)^2)$ terms.  The second two are from \cite{been} 
with $b=-a$. If $A^2 = B^2+C^2$  
\be
I^{(1,1)}_n =  2\p \frac{1}{aA} \frac{1}{n-4} \left( \frac{A+B}{2A} 
\right)^{n/2-3} \left[1+\frac{1}{4}(n-4)^2\dilog\left(\frac{A-B}{2A}\right)
\right] \, ,
\ee
whereas if $A^2 \neq B^2+C^2$ 
\bea
I^{(1,1)}_n & = & \frac{\p}{a(A+B)} \left\{ \frac{2}{n-4} + 
\ln \left[ \frac{(A+B)^2}{A^2-B^2-C^2} \right] \right. \nonumber \\
& + & \left. 
\frac{1}{2} (n-4) \left[ \ln^2 \left( \frac{A-\sqrt{B^2+C^2}}{A+B} \right) 
- \frac{1}{2} \ln^2 \left( \frac{A+\sqrt{B^2+C^2}}{A-\sqrt{B^2+C^2}} \right) 
\right. \right. \NO \\ & + & \left. \left. 2 \dilog
\left( - \frac{B+\sqrt{B^2+C^2}}{A-\sqrt{B^2+C^2}}
\right) - 2 \dilog\left(  \frac{B-\sqrt{B^2+C^2}}{A+B} \right)
\right] \right\},
\eea
again dropping $O((n-4)^2)$ terms in the second of these.  
The dilogarithm function $\dilog(x)$ is defined in \cite{dilog} and 
numerous useful properties are summarized in \cite{dd}.

\section{Recovering the ${\cal O}(\d_c/\d_s)$ terms}
In this appendix we integrate the $P_{qq}(z,\e)$ splitting kernel 
over the hard-collinear portion of phase space for the 
case of a 45 singularity as it pertains to the 
discussion given at the end of Sec.\ \ref{sec:massless_example}.  
Recall that this region is defined by
\bea
{\rm hard} &:&  \d_s \frac{\sqrt{s_{12}}}{2} \le E_5 \le 
\frac{\sqrt{s_{12}}}{2} \NO \\
{\rm collinear} &:& 0 \le s_{45} \le \d_c s_{12} \, .
\eea
From Eq.\ (\ref{eqn:coll}) we have 
$s_{34}=(p_3+p_4)^2=2p_3 \cdot p_4 \simeq (2p_3 \cdot p_{45})z$ 
and $s_{12}=(p_3+p_{45})^2 \simeq s_{45} + 2 p_3 \cdot p_{45}$
which together yield $s_{34} \simeq z(s_{12}-s_{45})$.
Using $E_5=(s_{12}-s_{34})/2\sqrt{s_{12}}$ the hard condition becomes
\be
0 \le z \le \frac{1-\d_s}{1-s_{45}/s_{12}} \, .
\ee
The approximation made in Sec.\ \ref{sec:jets} was to set $s_{45}=0$ in the 
denominator, in light of the collinear condition.  This resulted in 
a decoupling of the $z$ and $s_{45}$ integration limits in 
Eq.\ (\ref{eqn:dsHC1}).  Relaxing the $s_{45}=0$ approximation 
gives rise to terms ${\cal O}(\d_c/\d_s)$ as now described.

Keeping the $s_{45}$ dependence, the required integral is
\be
I = \int_0^{\d_c s_{12}} \frac{ds_{45}}{s_{45}} 
\left( \frac{s_{45}}{s_{12}} \right) ^{-\e}
\int_0^{\frac{1-\d_s}{1-s_{45}/s_{12}}} dz [z(1-z)]^{-\e}  P_{qq}(z,\e) \, .
\ee
We may expand $P_{qq} (z,\e)$ about $\e=0$ and make a change of variables 
$u=s_{45}/s_{12}$ giving
\be
I = C_F \int_0^{\d_c} u^{-1-\e} F(u) \, du \, ,
\ee
with
\be
\label{eqn:Fu}
F(u) = \int_0^{\frac{1-\d_s}{1-u}} dz \left\{
\frac{1+z^2}{1-z} -\e(1-z)-\e \ln\left[z(1-z)\right] 
\frac{1+z^2}{1-z} \right\} \, .
\ee
$F(u)$ may be evaluated with the help of
\bea
\int_0^a dz\, \frac{1+z^2}{1-z} &=& -a\left(1+\frac{a}{2}\right)-2\ln(1-a) \\
\int_0^a dz\, (1-z) &=& a\left(1-\frac{a}{2}\right) \\
\int_0^a dz\, \frac{\ln z}{1-z} &=& \dilog(1-a)-\frac{\p^2}{6} \\
\int_0^a dz\, \frac{\ln(1-z)}{1-z} &=& -\frac{1}{2}\ln^2(1-a) \\
\int_0^a dz\, \frac{z^2 \ln z}{1-z} &=& a\left(1+\frac{a}{4}\right)
-a\left(1+\frac{a}{2}\right)\ln a + \dilog(1-a)-\frac{\p^2}{6} \\
\int_0^a dz\, \frac{z^2 \ln(1-z)}{1-z} &=& \frac{3}{2} a 
\left( 1+\frac{a}{6} \right) + \ln(1-a) \left[ \frac{3}{2} - a 
\left( 1+\frac{a}{2} \right) -\frac{1}{2}\ln(1-a)\right]  \, .
\eea
The resulting terms in $F(u)$ may be integrated over $u$ using 
\bea
\int_0^{\d_c} du \, u^{-1-\e} \left( \frac{1-\d_s}{1-u} \right)^i 
&=& \left( -\frac{1}{\e} + \ln\d_c \right); \quad i=0,1,2\\
\int_0^{\d_c} du \, u^{-1-\e} \ln\left(1-\frac{1-\d_s}{1-u}\right) 
&=& \left( -\frac{1}{\e} + \ln\d_c \right) \ln\d_s - \dilog(\d_c/\d_s) \, ,
\eea
for the terms multiplied by ${\cal O}(\e^0)$ in Eq.\ (\ref{eqn:Fu}) and 
\bea
\int_0^{\d_c} du \, u^{-1-\e} \ln^2\left(1-\frac{1-\d_s}{1-u}\right) 
&=& - \frac{1}{\e} \ln^2 \d_s \\
\int_0^{\d_c} du \, u^{-1-\e} \left(\frac{1-\d_s}{1-u}\right)^i 
\ln\left(1-\frac{1-\d_s}{1-u}\right) &=& 
-\frac{1}{\e}\ln\d_s; \quad i=1,2 \\
\int_0^{\d_c} du \, u^{-1-\e} \left(\frac{1-\d_s}{1-u}\right)^i 
\ln\left(\frac{1-\d_s}{1-u}\right) &=& 0; \quad i=1,2 \\
\int_0^{\d_c} du \, u^{-1-\e} \dilog\left(1-\frac{1-\d_s}{1-u}\right) 
&=& 0 \, ,
\eea
for the terms multiplied by ${\cal O}(\e^1)$ in Eq.\ (\ref{eqn:Fu}).  
Terms containing or 
leading to contributions of ${\cal O}(\d_c)$ or ${\cal O}(\d_s)$ 
have been dropped.  Taking the coefficients of $1/\e$ and $\e^0$  
gives the desired result
\bea
A^{q \rightarrow qg}_1 &=& C_F \left( 3/2+2\ln\d_s \right) \\
A^{q \rightarrow qg}_0 &=& C_F \left[ 7/2 - \p^2/3 - \ln^2\d_s 
+ 2 \dilog{(\d_c/\d_s)} - \ln\d_c \left(3/2+2\ln\d_s \right) \right] \, .
\eea
The second equation is identically Eq.\ (\ref{eqn:A0q}) with the 
addition of the advertised $\dilog{(\d_c/\d_s)}$ term.  A similar analysis 
may be performed for the $P_{gg}$ splitting case with the same result: 
$\ln\d_c\ln\d_s \rightarrow \ln\d_c\ln\d_s - \dilog{(\d_c/\d_s)}$.

\section{Improving convergence of tilde terms}
We want to demonstrate how the numerical convergence of the 
$\widetilde{D}$ and $\widetilde{G}$ functions may be improved.  
To this end consider the integral
\be
F_i = \int_x^{1-\d_s} \frac{dy}{y} G_j(x/y,\mu) \widetilde{P}_{ij}(y) \, ,
\ee
with
\be
\widetilde{P}_{ij}(y) = P_{ij}(y)\ln\left(\d_c\frac{1-y}{y}
\frac{s_{12}}{\m_f^2}\right) - P_{ij}^{\prime}(y) \, .
\ee
Here a logarithm of $\d_s$ is numerically being built up.  Convergence will 
be improved if we rewrite the result in a form where the logarithmic 
dependence on $\d_s$ is manifest. To do so, use the fact that 
\be
\lim_{y \rightarrow 1} \left[ (1-y)P_{ij}(y) \right] = 2 C_i \d_{ij} \, ,
\ee
with $C_g=N$ and $C_q=C_F$.  Now add and subtract the leading singular piece 
under the integral sign:
\bea
F_i &=& \int_x^{1-\d_s} \frac{dy}{y} \left[ G_j(x/y,\mu) P_{ij}(y) 
         \ln\left(\d_c\frac{1-y}{y}\frac{s_{12}}{\m_f^2}\right) 
        - G_j(x/y,\mu) P_{ij}^{\prime}(y) \right. \NO \\
    &-& \left. G_j(x,\mu) \frac{2C_i\d_{ij}}{1-y} 
        \ln\left(\d_c\frac{1-y}{y}\frac{s_{12}}{\m_f^2}\right)
     + G_j(x,\mu) \frac{2C_i\d_{ij}}{1-y} 
        \ln\left(\d_c\frac{1-y}{y}\frac{s_{12}}{\m_f^2}\right) \right] \, .
\eea
Regrouping terms gives
\bea
F_i &=& \int_x^{1-\d_s} \frac{dy}{y} \left\{ \left[ G_j(x/y,\mu) P_{ij}(y)
       - G_j(x,\mu) \frac{2C_i\d_{ij}}{1-y} \right] 
       \ln\left(\d_c\frac{1-y}{y}\frac{s_{12}}{\m_f^2}\right) \right. \NO \\
    &-& \left. G_j(x/y,\mu) P_{ij}^{\prime}(y) + G_j(x,\mu) 
        \frac{2C_i\d_{ij}}{1-y} 
        \ln\left(\d_c\frac{1-y}{y}\frac{s_{12}}{\m_f^2}\right) \right\} \, .
\eea
The last term may be evaluated with the help of
\be
\int_x^{1-\d_s} \frac{dy}{y(1-y)}\ln\left(a\frac{1-y}{y}\right) = 
-\frac{1}{2} \ln^2\left(\frac{a\d_s}{1-\d_s}\right)
+\frac{1}{2} \ln^2\left[\frac{a(1-x)}{x}\right] \, .
\ee
The final desired expression is
\bea
F_i &=& C_i \d_{ij} G_j(x,\mu) \left[ \ln^2\left(\d_c\frac{s_{12}}{\m_f^2}
\frac{1-x}{x}\right) - \ln^2\left(\d_c\frac{s_{12}}{\m_f^2} 
\frac{\d_s}{1-\d_s} \right) \right] \NO \\
&+& \int_x^{1-\d_s} \frac{dy}{y} \Biggl\{ \left[ G_j(x/y,\mu) P_{ij}(y)
       - G_j(x,\mu) \frac{2C_i\d_{ij}}{1-y} \right] 
       \ln\left(\d_c\frac{1-y}{y}\frac{s_{12}}{\m_f^2}\right) \NO \\
& & - G_j(x/y,\mu) P_{ij}^{\prime}(y) \Biggr\} \, .
\eea
The $\ln\d_s$ is now evident in the first term, and absent from the 
second integral term.  Numerical convergence will therefore be greatly 
improved.

%
%

\end{document}